\documentclass{iopjournal}

\usepackage{physics, slashed, multirow,xcolor,booktabs,bbold}
\usepackage{cite, hyperref}

\usepackage[bitstream-charter]{mathdesign}
\hypersetup{
    colorlinks,
    linkcolor={red!50!black},
    citecolor={blue!50!black},
    urlcolor={blue!80!black}
}

\usepackage[normalem]{ulem}
\usepackage{multirow,float}
\newcommand{\bfgamma}{\mathbf{\Gamma}}
\usepackage{simpler-wick}

\renewcommand{\L}{{\text{L}}}
\newcommand{\R}{{\text{R}}}

\begin{document}

\articletype{Paper} 
\begin{flushright}DESY-25-064
\end{flushright}
\title{Perturbed symmetric-product orbifold: \\first-order mixing and puzzles for integrability}

\author{Matheus Fabri$^{a,b}$\orcid{0009-0007-0513-0655}, Alessandro Sfondrini$^{a,b,c}$\footnote{Corresponding author}\orcid{0000-0001-5930-3100} and Torben Skrzypek$^{a,d}$\orcid{0000-0003-4406-4375}}

\affil{$^a$Dipartimento di Fisica e Astronomia, Universit\`a degli Studi di Padova, via Marzolo 8, 35131
Padova, Italy}

\affil{$^b$INFN, Sezione di Padova, via Marzolo 8, 35131 Padova, Italy}

\affil{$^c$ School of Mathematics, University of Birmingham,
Watson Building, Edgbaston, Birmingham B15 2TT, UK}

\affil{$^d$Deutsches Elektronen-Synchrotron DESY, Notkestra{\ss}e 85, 22607 Hamburg, Germany}

\email{matheusaugusto.fabri@unipd.it, alessandro.sfondrini@unipd.it, torben.skrzypek@desy.de}

\begin{abstract}

We study the marginal deformation of the symmetric-product orbifold theory Sym$_N(T^4)$ which corresponds to introducing a small amount of Ramond-Ramond flux into the dual $AdS_3\times  S^3\times  T^4$ background. Already at first order in perturbation theory, the dimension of certain single-cycle operators is corrected, indicating that wrapping corrections from integrability must come into play earlier than expected. Our results provide a test for integrability computations from the mirror Thermodynamic Bethe Ansatz or Quantum Spectral Curve, akin to the computation of the Konishi anomalous dimension in $\mathcal{N}=4$ supersymmetric Yang--Mills theory.
We also discuss a flaw in the original derivation of the integrable structure of the perturbed orbifold.
Together, these observations suggest that more needs to be done to correctly identify and exploit the integrable structure of the perturbed orbifold CFT.

\end{abstract}
\vspace{\baselineskip}

\vspace{10pt}
\noindent\rule{\textwidth}{1pt}
\tableofcontents
\noindent\rule{\textwidth}{1pt}
\vspace{10pt}

\section{Introduction}
\label{sec:intro}

The symmetric-product orbifold CFT of $T^4$ plays an important role in the AdS3/CFT2 correspondence~\cite{Maldacena:1997re}. It is dual to type IIB string theory on $AdS_3\times  S^3\times  T^4$ background supported by precisely $k=1$ units of Neveu-Schwarz-Neveu-Schwarz (NSNS) flux~\cite{Giribet:2018ada,Gaberdiel:2018rqv,Eberhardt:2018ouy}. Because this is a free model, it is possible to compute in detail many of the observables of the theory. However, such computations become substantially more complicated when adding Ramond-Ramond (RR) background flux to the setup. This continuous change of the string background (which can be achieved by turning on {a combination of the RR-fields $C_0$ and $C_4$ wrapping the torus,} see e.g.~\cite{Berkovits:1999im, OhlssonSax:2018hgc})
corresponds to a specific marginal deformation in the symmetric-product orbifold CFT, see e.g.~\cite{Burrington:2012yq,Fiset:2022erp}.

On the string side, it is notoriously difficult to quantise the worldsheet theory in the presence of RR fluxes. If the Green-Schwarz (GS) action for a given background is classically integrable, one can try to quantise it in lightcone gauge and solve the theory in terms of a factorised S matrix on the worldsheet, following~\cite{Zamolodchikov:1978xm} --- an approach which has proven remarkably powerful for the study of $AdS_5\times  S^5$ superstrings, see~\cite{Arutyunov:2009ga,Beisert:2010jr}.
Remarkably, the GS action for strings on $AdS_3\times  S^3\times  T^4$ with mixed NSNS/RR flux is classically integrable~\cite{Cagnazzo:2012se}.%
\footnote{The study of classical integrability for $AdS_3$ backgrounds was initiated in~\cite{Babichenko:2009dk}.}
This spurred an active interest in constructing the quantum S~matrix on the string worldsheet by integrable bootstrap approaches, and using it to compute the spectrum of the model  (see~\cite{Sfondrini:2014via,Demulder:2023bux,Seibold:2024qkh} for reviews of various aspects of this construction).

At the same time, it was long conjectured that an integrable structure (perhaps similar to that of the $\mathcal{N}=4$ supersymmetric Yang-Mills theory~\cite{Minahan:2002ve}) should appear in the marginally-deformed symmetric-orbifold CFT~\cite{David:2008yk,Pakman:2009mi}.
A recent development was the purported identification of the symmetry algebra expected from worldsheet integrability in the symmetric-product orbifold theory~\cite{Gaberdiel:2023lco}. After correcting the identification of representations, which was done in~\cite{Frolov:2023pjw}, such an algebra reproduces the S~matrix known from the worldsheet-integrability construction~\cite{Lloyd:2014bsa}.

The purpose of this paper is to explore marginally-deformed symmetric-orbifold CFT and in particular its spectrum. More specifically, we will be interested in the anomalous dimensions of single-cycle operators (which can be thought of as the analogues of single-trace operators in $\mathcal{N}=4$ SYM) in the planar limit and at weak coupling. Calling $\lambda$ the marginal coupling, the anomalous dimension of most operators will be of order {$O(\lambda^2)$}, as it has been found by conformal perturbation theory in a number of cases, see e.g.~\cite{Gaberdiel:2015uca,Hampton:2018ygz,Guo:2019ady,Guo:2020gxm,Lima:2020boh,Lima:2020kek,Lima:2020urq,Benjamin:2021zkn}.
However, a relatively small subset of states will receive corrections of order~$O(\lambda)$, as we shall see. 
The study of these states is very important for the integrability construction, for at least two reasons:
\begin{enumerate}
    \item These operators have the leading anomalous dimension at small~$\lambda$. The aim of the integrability program is to reproduce the spectrum for all operators at any~$\lambda$. In practice, this is done numerically by solving the ``mirror'' Thermodynamic Bethe Ansatz (TBA)~\cite{Arutyunov:2007tc} or Quantum Spectral Curve (QSC)~\cite{Gromov:2013pga}, starting from small tension (here, $\lambda$) and working one's way up to larger tension.%
    \footnote{This was the strategy used in $AdS_5\times  S^5$ strings / $\mathcal{N}=4$ SYM, where the matching of the weak-tension anomalous dimension of the Konishi multiplet was one of the early successes of integrability~\cite{Gromov:2009zb,Arutyunov:2010gb,Balog:2010xa,Frolov:2010wt}; of course this requires having perturbative results against which to check the integrability result~\cite{Lukowski:2009ce}.}
    The study of the spectrum of mixed-flux $AdS_3$ backgrounds is beginning right now (the mirror TBA equations have very recently been proposed in~\cite{Frolov:2025tda}) and it appears clear that it will start from small tension, like it was done for pure-RR $AdS_3$ backgrounds~\cite{Cavaglia:2022xld,Brollo:2023pkl,Brollo:2023rgp}. In this sense, the anomalous dimensions of the operators which we study here are the most natural first test of the integrability construction.
    \item In integrability, one typically distinguishes between the ``asymptotic'' spectrum (which can be predicted from the Bethe--Yang equations) and ``wrapping'' corrections (which require the fully fledged mirror TBA/QSC approach). In $\mathcal{N}=4$ SYM, it so happened that wrapping corrections did not play a role up to $M$ loops, for an operator of bare dimension~$M$; this is why, for instance, it was important to study the Konishi multiplet up to five loops. In $AdS_3$, we expect the asymptotic corrections to the anomalous dimension to begin at order $O(\lambda^2)$; yet the states which we consider in this paper receive corrections at order~$O(\lambda)$ and are therefore ``super-sensitive'' to wrapping. In this sense, reproducing their anomalous dimension from integrability is even more of a challenge.
\end{enumerate}

Below we will work out these corrections in detail and point out their relevance to the integrability construction, and in particular to finite-volume (``wrapping'') corrections, which are their most likely explanation.
In the course of this computation we will also revisit the construction of~\cite{Gaberdiel:2023lco} which identified the integrability structure in the deformed symmetric-orbifold CFT. Surprisingly, we find an apparent flaw in that derivation, {
 specifically in how the ``large-volume'' limit of orbifold correlation functions is taken. This large-volume limit is crucial for describing asymptotic states in terms of magnons and identifying their symmetry algebra~\cite{Borsato:2012ud,Lloyd:2014bsa}. This issue is different from the mismatch in the identification of the representations already pointed out in~\cite{Frolov:2023pjw}.}

This paper is structured as follows: we start by reviewing the symmetric-product orbifold theory Sym$_N (T^4)$ in Section \ref{sec:review-section}, outlining the seed theory, the twisted-sector correlation functions, their representation in terms of covering spaces, and the marginal deformation associated to turning on RR-flux. Some details on the seed theory, the conserved currents and the chiral-ring BPS states have been collected in the Appendices \ref{app:free-field-conv}, \ref{app:currents-and-charges} and \ref{app:chiral-ring}, respectively.
In Section~\ref{sec:mixing-matrix-O(g)} we outline the main technical steps of calculating the mixing matrix and resulting spectrum at first order in perturbation theory. We sketch the various necessary ingredients for the calculation by applying them to a sample mixing-matrix element. Some detail of the computation  (the bosonic Wick contractions and the bosonisation of fermionic excitations) are presented in the Appendices \ref{app:integrals-wick} and \ref{app:bosonisation}. The full mixing matrix can be constructed using a \texttt{Wolfram Mathematica} notebook supplied as ancillary file. The resulting spectrum for states with small conformal dimensions is presented in Section~\ref{sec:results}. {
In Section~\ref{sec:integrability} we perform a scaling analysis of our results in the limit of large twist (corresponding to the large-volume regime, where integrability should be manifest), finding that $O(\lambda)$ corrections are suppressed as they should be. We then discuss their implications for integrability at finite~twist.
We present our conclusions in Section \ref{sec:conclusion}. 
Building on our discussion of the large-twist limit, we revisit the analysis of~\cite{Gaberdiel:2023lco}, highlighting an issue in the limiting procedure used in that paper, and discussing its implications as well as possible resolutions of the issue. The discussion of this point is rather technical, and therefore presented in Appendix~\ref{app:GGN}.}

\section{The (deformed) symmetric-product orbifold in a nutshell}
\label{sec:review-section}

Orbifold CFTs have played and are still playing a crucial role in string theory~\cite{Dixon:1985jw,Dixon:1986jc,Dixon:1986qv}. Over the decades, much effort has been devoted to the explicit computation of correlation functions in these theories, and while this can be done exactly when the seed theory is simple enough, the computation is nonetheless quite non-trivial in the presence of twisted sectors, see e.g.~\cite{Arutyunov:1997gi,Arutyunov:1997gt,Lunin:2000yv,Lunin:2001pw,Burrington:2012yn}.
Below we will briefly summarise the ingredients we need for the computation of first-order anomalous dimensions.

\subsection{Definition of the unperturbed orbifold}
\label{sec:orbifold-def}

The model we consider in this work is a deformation of the symmetric-product orbifold CFT which has as its seed theory the 2d $\mathcal{N}=(4,4)$ supersymmetric theory of free bosons and fermions. This seed model has four real bosons, which we indicate as $X^{A\dot{A}}(z,\bar{z})$, as well as four left- and four right-moving real fermions, which we indicate as $\psi^{\alpha \dot{A}}(z)$ and $\tilde{\psi}^{\dot{\alpha} \dot{A}}(\bar{z})$, respectively.
The Greek indices $\alpha=-,+$ and $\alpha=\dot{-},\dot{+}$ denote charge under the $SU(2)_\L$ and $SU(2)_\R$ R-symmetry, respectively. 
The Latin indices $A=1,2$ and $\dot{A}=\dot{1},\dot{2}$ encode the four bosons parameterising a target space $T^4$. More precisely, they decompose a vector of $SO(4)$ as a bispinor of $SU(2)_{\bullet}\times  SU(2)_{\circ}$; this notation is often used in the integrability literature~\cite{Borsato:2014exa} with  $SU(2)_{\bullet}$ and $SU(2)_{\circ}$ indices associated to undotted and dotted indices, respectively. For more details concerning conventions and notation for this free model we refer the reader to Appendix \ref{app:free-field-conv}.
We mention finally that $SU(2)_\bullet$ acts as an automorphism on the supercurrents, while $SU(2)_\circ$ does not act on them at all, see Appendix~\ref{app:currents-and-charges}.

The symmetric-product orbifold is defined as $N$ copies of the aforementioned free CFT where we identify all copies under the action of the symmetric group $S_N$. We denote it as
\begin{equation}
\textrm{Sym}_{N} (T^{4}) = (T^{4})^{\otimes N} / S_N \,.
\end{equation}
Here the $N\rightarrow\infty$ expansion corresponds to the large-$N$ expansion in holography and accordingly we refer to the leading order contribution as the planar limit. Let $\Phi_I$, $I=1,\dots N$, denote one of the $N$ copies of some field $\Phi$ in the seed CFT. The action of $S_N$ is realised by twist fields $\sigma_{g}(z,\bar{z})$ which introduce a branch cut and the following boundary conditions for the fields
\begin{equation}
\label{eq:810837}
\Phi_I (z+e^{2\pi i} w,\bar{z}+ e^{-2\pi i} \bar{w}) \sigma_{g}(z,\bar{z}) = \Phi_{g(I)} (z+w,\bar{z}+\bar{w}) \sigma_{g}(z,\bar{z})\,,
\end{equation}
with $g\in S_N$ being a permutation of the $N$ copies. States are defined on top of twisted vacua constructed by the corresponding twist fields. These are labelled by products of cycles (more precisely, by conjugacy classes of $S_N$ as we will discuss below). The simplest states are those where no non-trivial cycles appear, that is where $g$ is the identity. The next simplest case is the one where only one non-trivial cycle appears, involving $w$ of the $N$ copies. This set-up is already quite rich, and of great relevance in holography.\footnote{Such single-cycle states can be considered analogues of the single-trace operators of $\mathcal{N}=4$ supersymmetric Yang-Mills theory in four dimensions.} 
 Below we will consider such states defined on single-cycle permutations within $S_N$ in the large-$N$ limit, and compute their connected correlation functions.%
 \footnote{%
For a discussion of multi-cycle and disconnected correlators see~\cite{Pakman:2009zz,AlvesLima:2022elo}.
 }

As far as the spectrum is concerned, the main consequence of considering states in the twisted sectors is the existence of fractional excitations due to eq.~\eqref{eq:810837}; in the case of a single-cycle of length~$w$, the modes will be quantised in units of $1/w$. Indeed, suppose we have a left-moving field $\Phi(z)$ in the seed CFT with left and right dimensions $(h,0)$, respectively. We insert a twist field $\sigma_g$ at the origin with $g$ being a single-cycle of length $w$ ($w$ is also called the twist). We choose $g$ such that the boundary conditions \eqref{eq:810837} are now
\begin{equation}
\label{eq:905657}
\Phi_I (e^{2\pi i} z) = \Phi_{I+1} (z)\,, 
\end{equation}
with $I=1,\cdots,w$ and cyclicity condition $w+1 \equiv 1$. The action on the remaining $N-w$ copies is chosen to be trivial. Therefore the mode expansion of $\Phi(z)$ is given by
\begin{equation}
\label{eq:554705}
\Phi_I (z) = \frac{C_{\Phi}}{\sqrt{w}} \sum_{h + \frac{n}{w} \in\  \mathbb{Z}/w} \phi_{\frac{n}{w}}\ e^{-2\pi i\left(\frac{n}{w}+h\right)(I-1)} z^{-h-\frac{n}{w}}\,,
\end{equation}
where $C_{\Phi}$ is a normalisation constant and $\phi_{\frac{n}{w}}$ is the fractional mode. The nature of the mode numbers is dependent on the dimension $h$ and whether $w$ is even or odd. For the purpose of this paper, we can distinguish three cases:
\begin{align}
& h\in\mathbb{Z} : n\in\mathbb{Z}\,, \\
& h\in\mathbb{Z}+\frac{1}{2} \ \textrm{and} \ w \ \textrm{even} : n\in\mathbb{Z}\,, \\
& h\in\mathbb{Z}+\frac{1}{2} \ \textrm{and} \ w \ \textrm{odd} : n\in\mathbb{Z}+\frac{1}{2}\,.
\end{align}
Since the fermionic fields $\psi^{\alpha\dot{A}}$ and $\tilde{\psi}^{\dot{\alpha}\dot{A}}$ have conformal dimensions $h=\tfrac{1}{2}$ and $\bar{h}=\tfrac{1}{2}$, respectively, we have to take into account fermionic zero modes whenever the twist $w$ is even, resulting in a multiplet of charged vacua which will be described in Section \ref{sec:vacua-and-spectrum}. This distinction between \textit{even} and \textit{odd} twist amounts to having R or NS boundary conditions for the fermions, as we review in Section \ref{sec:correlators-lifiting}.

By inverting the mode expansion~\eqref{eq:554705} we can determine the fractional modes in terms of the original fields. For the fields considered in this work we have
\begin{align}
\label{eq:626129}
& \alpha_{\frac{n}{w}}^{A\dot{A}} = \frac{i}{\sqrt{w}} \oint_{0} \frac{dz}{2\pi i} \sum_{I=1}^{w} \partial X_I^{A\dot{A}} (z)\ e^{\frac{2\pi i n}{w} (I-1)} z^{\frac{n}{w}}\,, \\
\label{eq:626129-2}
& \psi_{\frac{n}{w} - \frac{1}{2}}^{\alpha\dot{A}} = \frac{1}{\sqrt{w}} \oint_{0} \frac{dz}{2\pi i} \sum_{I=1}^{w} \psi_I^{\alpha\dot{A}} (z)\ e^{\frac{2\pi i n}{w} (I-1)} z^{\frac{n}{w}-1}\,,
\end{align}
with similar expressions for the right-moving modes. From these we can write the currents of the model as combination of modes and derive their charges (see Appendix \ref{app:currents-and-charges}). For later convenience, we collect the charges of all modes in the theory in Table \ref{table:charges-of-modes}. 

\begin{table}[t]
	\centering
	\begin{tabular}{|c|c|c|c|c|c|c|}
\hline
Mode & $h$ & $\bar{h}$ & $j$ & $\bar{\jmath}$ & $j_{\circ}$ & $j_{\bullet}$ \\ 
\hline\hline
$\alpha_{n/w}^{A\dot{A}}$ & $-n/w$ & $0$ & $0$ & $0$ & $(-1)^{\dot{A}+1}/2$ & $(-1)^{A+1}/2$ \\ 
$\tilde{\alpha}_{n/w}^{A\dot{A}}$ & $0$ & $-n/w$ & $0$ & $0$ & $(-1)^{\dot{A}+1}/2$ & $(-1)^{A+1}/2$ \\ 
$\psi_{n/w}^{\alpha\dot{A}}$ & $-n/w$ & $0$ & $\alpha/2$ & $0$ & $(-1)^{\dot{A}+1}/2$ & $0$ \\ 
$\tilde{\psi}_{n/w}^{\dot{\alpha}\dot{A}}$ & $0$ & $-n/w$ & $0$ & $\dot{\alpha}/2$ & $(-1)^{\dot{A}+1}/2$ & $0$ \\ 
\hline 
\end{tabular} 
	\caption{Charges of the bosonic and fermionic modes. Here $h$ and $\bar{h}$ are the left- and right-moving dimensions, $j$ and $\bar{\jmath}$ are the left- and right-moving R-charges and $j_{\circ}$ and $j_{\bullet}$ are the $SU(2)_{\circ}$ and $SU(2)_{\bullet}$ charges, respectively. For more details concerning the charges see Appendix \ref{app:currents-and-charges}.}
\label{table:charges-of-modes}
\end{table}

\subsection{Correlation functions, covering maps, and lifting}
\label{sec:correlators-lifiting}

We are interested in correlation functions in the twisted sector of the symmetric-product orbifold, i.e.\ involving states defined on top of single-cycle twisted fields; this will allow us to compute mixing matrices of the perturbed theory. We want to consider local correlators. This requires some care because the twist field itself is non-local.
It is necessary to consider the conjugacy class of $g\in S_N$ defined as
\begin{equation}
[g] = \{ g' \in S_N : \exists h\in S_N \ \textrm{s.t.}\ g' = hgh^{-1} \}\,.
\end{equation}
With this we can construct the following (unit normalised) local operator 
\begin{equation}
\label{eq:811071}
\sigma_{[g]} (z,\bar{z}) = \frac{1}{\sqrt{|S_N||\textrm{stab}(g)|}} \sum_{h\in S_N} \sigma_{hgh^{-1}} (z,\bar{z})\,,
\end{equation}
with $|S_N|=N!$ being the size of the permutation group and $\textrm{stab}(g)\subset S_N$ being the stabiliser subgroup of $g$. This prefactor ensures that operators are properly normalised, and it allows us to analyse their large-$N$ behaviour. Therefore when we construct generic (excited) states in a twist-$w$ sector, we consider \eqref{eq:811071} as the vacuum on which we define the states, choosing for $g$ a particular cyclic permutation of length~$w$. 

Let us sketch the computation of an $n$-point correlation function of orbifold-invariant twisted fields. Consider operators $\sigma_{[g_j]}$ of the form \eqref{eq:811071} with $g_j$ being a single-cycle of twist $w_j$ or for short a $w_j$-cycle. All in all, the various $g_j$s will act on $c$ out of $N$ copies of the fields, and the precise value of $c$ will depend on whether and how much the cycles overlap.%
\footnote{An example which will be important later is that of a three-point function with $w_1=w$, $w_2=2$, and $w_3=w+1$. In this case we can  e.g.\ take as representatives the cycles $g_1=(12\cdots w)$, $g_2=(1,w+1)$ and $g_3=(12\cdots w+1)$, as we will see later; hence in this case we are working with $c=w+1$ copies out of~$N$.}
We are not going to describe all steps involved in computing the genus $\textrm{g}$ contribution to this correlator (the interested reader can consult \cite{Lunin:2000yv,Pakman:2009zz,Eberhardt:2019ywk,Dei:2019iym}), but we quote its final form after all $S_N$ invariance has been exploited:
 \begin{equation}
\label{eq:invariantcorrelator}
 \langle \sigma_{[g_1]} (z_1,\bar{z}_1) \cdots \sigma_{[g_n]} (z_n,\bar{z}_n) \rangle_{\textrm{g}} = \binom{N}{c} \prod_{j} \sqrt{\frac{(N-w_j)!w_j}{N!}} \  \sum_{\substack{\textrm{equiv.}\\ \textrm{classes}}}  \langle \sigma_{g_1} (z_1,\bar{z}_1) \cdots \sigma_{g_n} (z_n,\bar{z}_n) \rangle_{\textrm{g}}\,,
 \end{equation}
with the genus dependence indicated by the subscript. Let us take this expression apart term-by-term. First, the sum is realised over representatives $g_j$ of the conjugacy classes $[g_j]$, such that
\begin{enumerate}
 \item $g_1 \cdots g_n = \mathbb{I}$; this imposes the $S_N$ invariance of the correlation function.
 \item The subgroup spanned by $\{g_1 , \cdots ,g_n \}$ is acts transitively on the $c$ copies; this imposes that the correlation function is connected, which is the case of interest to us.
 \item Only one $n$-tuple of elements in the same conjugacy orbit of $\{ g_1 , \cdots , g_n \}$ is counted; other choices $\{ h g_1 h^{-1}, \cdots , h g_n h^{-1} \} $ with  $h \in S_N$ give the same correlator. This means that the sum is restricted to global $S_N$ equivalence classes, which justifies the sum notation in \eqref{eq:invariantcorrelator}.
 \end{enumerate}
Then $c\leq N$ is the number of copies on which $\{ g_1 , \cdots , g_n \}$  act (without loss of generality, we can take them to be the first $c$ copies).
Remarkably, conditions $(1)$--$(3)$ also characterise the set of inequivalent $c$-sheeted Riemann surfaces with $n$ branch points and ramifications $w_j$. The number of such surfaces is called Hurwitz number and it matches precisely with the number of terms in the sum \eqref{eq:invariantcorrelator}. 
These Riemann surfaces are the covering surfaces of $\mathbb{CP}^1$ with the cuts due to the $w_j$-cycle twist fields $\sigma_{[g_j]}$, so that the computation of the correlation function can be done on these surfaces by means of a covering map. Then the sum \eqref{eq:invariantcorrelator} can be recast as a sum over all covering maps whose branch points and ramifications are specified by the twist field insertions~\cite{Lunin:2000yv,Lunin:2001pw,Burrington:2012yn}.

Let us fix the notation for such covering maps. Consider one of the terms in the sum~\eqref{eq:invariantcorrelator}, where each $g_j$ corresponds to a $w_j$-cycle. To each such term we associate a $c$-sheeted genus-$\textrm{g}$ covering surface $\Sigma_\textrm{g}$ with coordinates which we denote $t,\bar{t}$. 
The covering map is defined as $\bfgamma:\Sigma_{\textrm{g}} \rightarrow \mathbb{CP}^1$ such that in the vicinity of the ramification point $t_j$ it satisfies\footnote{Here we follow the notation of \cite{Gaberdiel:2023lco} and use $\bfgamma$ for the covering map and $\Gamma$ for the standard Gamma function.}
\begin{equation}
\label{eq:coveringmap}
\bfgamma(t) = z_j + a_j (t-t_j)^{w_j}+ O((t-t_j)^{w_j +1}) \qquad \textrm{for} \qquad t\sim t_j\,, 
\end{equation}
where $t_j \in\Sigma_g$ is the pre-image of the point $z_j$ where the $w_j$-cycle is inserted. The genus of the covering surface is given by the Riemann-Hurwitz formula
\begin{equation}
\label{eq:RiemannRoch}
\textrm{g} = 1-c+\sum_{j=1}^{n} \frac{(w_j - 1)}{2}\,.
\end{equation}
This can also be found from the large-$N$ analysis of the correlator prefactor in~\eqref{eq:invariantcorrelator} by using the general fact that a genus-$\textrm{g}$ $n$-point function goes as $N^{1-\textrm{g}-n/2}$. The advantage of the covering space is that the collection of fields on the $c$ copies appearing in the correlator, $\Phi_I (z,\bar{z})$ with $I=1,\dots, c$\,, can be expressed as a unique single-valued ``lifted'' field $\Phi (t,\bar{t})$ on the covering surface.

Let us see how this lifting works for our fields --- bosons and fermions. Consider a twist-$w$ insertion at the origin $z=\bar{z}=0$ and choose the covering map such that $\Gamma(0)=0$ and the local behaviour near this point takes the form
\begin{equation}
\label{eq:285212}
\bfgamma(t) = a t^w +O(t^{w+1})\,.
\end{equation} 
 For bosonic operators such as $\partial X^{A\dot{A}}(t)$,  we find periodic fields in the covering surface. However for the fermions  $\psi^{\alpha\dot{A}} (t)$, one finds after the coordinate change that
\begin{equation}
\psi^{\alpha\dot{A}} (t e^{2\pi i}) = (-1)^{w-1}\psi^{A\dot{A}} (t)\,.
\end{equation}
Then for even and odd twist $w$ we get R and NS boundary conditions in the covering surface, respectively.
The fractional modes of the fields introduced above can also be expressed in terms of the lifted fields. A change of coordinates from \eqref{eq:626129} and \eqref{eq:626129-2} results in%
\footnote{By virtue of the form of the covering map near the points $t_k$, these integrals are well-defined; for the fermions, the square-root term accounts for the NS/R-sector periodicity.}
\begin{align}
\label{eq:667038-1}
& \alpha_{\frac{n}{w}}^{A\dot{A}} = \frac{i}{\sqrt{w}} \oint_{t_k} \frac{dt}{2\pi i}\  (\bfgamma(t) - z_k)^{\frac{n}{w}} \ \partial X^{A\dot{A}} (t) \,, \\
\label{eq:6548651687}
& \psi_{\frac{n}{w} - \frac{1}{2}}^{\alpha\dot{A}} = \frac{1}{\sqrt{w}} \oint_{t_k} \frac{dt}{2\pi i}\ \left(\frac{\partial\bfgamma(t)}{\partial t}\right)^{1/2} \ (\bfgamma(t) - z_k)^{\frac{n}{w}-1} \ \psi^{\alpha\dot{A}} (t) \,, 
\end{align}
with analogous expressions for right-moving modes. Here we defined the integrals around a generic insertion point $z_k$ satisfying $z_k = \bfgamma(t_k)$. 

Even though on the covering space we are left with a free CFT of bosons and fermions, this does not mean that the twist field contribution completely drops out. In fact their presence induces a conformal anomaly when going to covering coordinates and their contribution can be computed through a Liouville action approach as detailed in \cite{Lunin:2000yv,Lunin:2001pw,Burrington:2012yn} for three-point functions. The result is universal and depends on the seed theory only through its central charge. In summary, when one goes to the covering surface the contribution of the original twist insertions factorises in a universal prefactor as, schematically,%
\footnote{For three-point functions, which is the only case used in this work, this expression is valid as it is since there is an unique genus $\textrm{g}=0$ covering surface. However for higher-point correlators there are multiple covering maps and it is necessary to sum over all of them \cite{Dei:2019iym,Pakman:2009zz}.}
\begin{equation}
    \label{eq:schematically-lift}
    \langle\textrm{Sym}_N(T^4)\textrm{ correlator}\rangle = \langle\sigma_{g_j}\ \textrm{correlation}\rangle\ \times \ \langle\textrm{free correlator in }\Sigma_g\rangle\,,
\end{equation}
which was discussed in more detail in~\cite{Lunin:2001pw}.

\subsection{Spectrum before the perturbation}
\label{sec:vacua-and-spectrum}

\begin{table}[t]
	\centering
	\begin{tabular}{|c|c|c|c|}
\hline
State & $j$ & $\bar{\jmath}$ & $j_{\circ}$ \\ 
\hline\hline
$|0\rangle^{\alpha\dot{\beta}}_{w}$ & $\alpha/2$ & $\dot{\beta}/2$ & $0$  \\ 

$|0\rangle^{\alpha\dot{A}}_{w}$ & $\alpha/2$ & $0$ & $(-1)^{\dot{A}+1}/2$  \\ 

$|0\rangle^{\dot{A}\dot{\alpha}}_{w}$ & $0$ & $\dot{\alpha}/2$ & $(-1)^{\dot{A}+1}/2$  \\ 

$|0\rangle^{\dot{A}\dot{B}}_{w}$ & $0$ & $0$ & $((-1)^{\dot{A}+1} + (-1)^{\dot{B}+1}) /2$  \\ 
\hline 
\end{tabular} 
	\caption{$SU(2)_{L}\times  SU(2)_{R}$ and $SU(2)_{\circ}$ charges of the even $w$ vacua.}
\label{table:even-vacuum-charges}
\end{table}

We now describe the spectrum of the theory before the perturbation. We first start with the vacuum of each $w$-cycle sector. From the Virasoro generators in Appendix \ref{app:currents-and-charges} we see that the twisted vacuum has different conformal dimensions for odd and even twist. Below we give the dimensions and notation for each case:
\begin{align}
        \label{eq: vacuum-dim-odd}
        & |0\rangle_w:&&h_w=\bar{h}_w= \frac{1}{4}\left(w-\frac{1}{w}\right),\qquad &&\textrm{for odd w}\ ,\\
        \label{eq:vacuum-dim-even}
        &|0\rangle_w^{xy}: &&h_w=\bar{h}_w=\frac{w}{4},\ \qquad&&\textrm{for even w}\ .
\end{align}
The untwisted $w=1$ sector yields the CFT vacuum. Any even-$w$ vacuum is charged under $SU(2)_{\circ}$ and/or $SU(2)_{L}\times  SU(2)_{R}$ due to the fermionic zero modes which exist in this case, with  $x$ and $y$ in \eqref{eq:vacuum-dim-even} being placeholders for the appropriate indices. The full set of even vacua is given in Table \ref{table:even-vacuum-charges} together with their charges. As usual the fermionic zero modes allow us to cycle through all vacua:
\begin{equation}
\label{eq:997119}
 \psi_0^{\alpha\dot{A}} |0\rangle^{\beta y}_{w}  = -\epsilon^{\alpha \beta} |0\rangle_{w}^{\dot{A} y}\,, \qquad
 \psi_0^{\alpha\dot{A}} |0\rangle_{w}^{\dot{B}y} = \epsilon^{\dot{A}\dot{B}} |0\rangle^{\alpha y}_w\,,
\end{equation}
with analogous relations for the right-moving fermionic zero modes acting on the second index $y$. As described above, in the lift to the covering surface the odd twisted vacuum is lifted to the NS vacuum and the even twist one to the R vacuum, that is upon lifting
\begin{align}
    & |0\rangle_w \rightarrow |0\rangle_{\textrm{NS}} && \textrm{for odd}\ w\,, \\
    & |0\rangle_w^{xy} \rightarrow |0\rangle_{\textrm{R}}^{{xy}} && \textrm{for even}\ w\,.
\end{align}
We can also write
\begin{equation}
\mathcal{S}^{xy}(0)\,|0\rangle_{\textrm{NS}} = |0\rangle^{xy}_{\textrm{R}}\,,
\end{equation}
where $\mathcal{S}^{xy}(t)$ is the spin field that gives antiperiodic boundary conditions to the fermions. Therefore the computation of correlation functions in covering space may involve not only free bosons and fermions, but also spin fields \cite{Lunin:2001pw}. 

The rest of the states are then built on top of the twisted vacua for even/odd $w$ by acting with the usual creation and annihilation operators, which in the orbifold theory are fractionally moded. We have then the usual highest-weight conditions\footnote{We are considering states with no winding or momentum along $T^4$.}
\begin{equation}
\begin{aligned}
& \alpha_{\frac{n}{w}}^{A\dot{A}} |0\rangle_w = \tilde{\alpha}_{\frac{\bar{n}}{w}}^{A\dot{A}} |0\rangle_w = 0\,, &&\textrm{for} \quad n,\bar{n}\geq 0\,, \\
& \psi_{\frac{n}{w}-\frac{1}{2}}^{\alpha\dot{A}} |0\rangle_w = \tilde{\psi}_{\frac{\bar{n}}{w}-\frac{1}{2}}^{\dot{\alpha}\dot{A}} |0\rangle_w = 0\,, &&\textrm{for}\quad n\,,\bar{n} > \frac{w}{2} \ ,
\end{aligned}
\end{equation}
for odd $w$ and analogous expressions for even $w$. By using the mode algebra 
\begin{equation}
\label{eq: commutator}
\left[\alpha_{\frac{n}{w}}^{A\dot{A}},\,\alpha_{\frac{m}{w}}^{B\dot{B}}\right] = \frac{n}{w} \delta_{n+m,0} \ \epsilon^{AB} \epsilon^{\dot{A}\dot{B}}\,, \qquad
 \left\{ \psi_{\frac{n}{w}-\frac{1}{2}}^{\alpha\dot{A}} ,\, \psi_{\frac{m}{w}+\frac{1}{2}}^{\beta\dot{B}} \right\} = \delta_{n+m,0} \ \epsilon^{\alpha\beta} \epsilon^{\dot{A}\dot{B}}\,,
\end{equation}
we see that excited states are obtained by acting with the negative modes on the twisted vacuum. For physical states, the total mode number of left- and right-moving excitations must vanish modulo~$w$,
\begin{equation}
\label{eq:orbifoldinvariance}
    \sum_j n_j-\sum_j\bar{n}_j=0\quad \text{mod}~w\,,
\end{equation}
so that the spin $h-\bar{h}$ of the operator is appropriately quantised.
Due to the fractional nature of the modes we have high levels of degeneracy even for low-lying  states. Indeed, as an example we show in Table~\ref{table:881263} a list of all states in the sector with $h=\bar{h}=1$ and $j = \bar{\jmath} = 0$. The remaining charges are found by simply adding the excitation and vacuum charges of Table~\ref{table:charges-of-modes} and Table~\ref{table:even-vacuum-charges}, respectively. For simplicity we are going to use the notation in Table~\ref{table:881263} for all states henceforth. 
Note that, among these states, some sit in short (half-BPS) multiplets of the $\mathcal{N}=(4,4)$ algebra. Their dimensions and three-point functions are protected in conformal perturbation theory, as they descend from highest-weight states satisfying  $h=j$ and $\bar{h}=\bar{\jmath}$ \cite{Baggio:2012rr}. Such states can be constructed by dressing the twist fields by fermion modes, see Appendix \ref{app:chiral-ring} for details. A generic excited state can be described with reference to the twisted vacuum $|0\rangle_w$ or to a reference BPS state in the $w$-sector as done in~\cite{Gaberdiel:2023lco}. This choice is immaterial of course and, in this work, we find it easier to work with reference to $|0\rangle_w$. 

\begin{table}[t]
	\centering
	\begin{tabular}{|c|c|}
        \hline
		Twist $w$ & States \\
        \hline\hline
		$1\phantom{\bigg(}$ & $|\alpha_{-1}^{A\dot{A}}\tilde{\alpha}_{-1}^{B\dot{B}}\rangle_1$\,,\,\,$|\alpha_{-1}^{A\dot{A}}\tilde{\psi}_{-\frac{1}{2}}^{\dot{+}\dot{B}}\tilde{\psi}_{-\frac{1}{2}}^{\dot{-}\dot{C}}\rangle_1$\,,\,\,$|\psi_{-\frac{1}{2}}^{+\dot{A}}\psi_{-\frac{1}{2}}^{-\dot{B}}\tilde{\alpha}_{-1}^{C\dot{C}}\rangle_1$\,,\,\,$|\psi_{-\frac{1}{2}}^{+\dot{A}}\psi_{-\frac{1}{2}}^{-\dot{B}}\tilde{\psi}_{-\frac{1}{2}}^{\dot{+}\dot{C}}\tilde{\psi}_{-\frac{1}{2}}^{\dot{-}\dot{D}}\rangle_1$\\
		$2\phantom{\bigg(}$ & 
		$|\psi_{-\frac12}^{\pm\dot{A}}\tilde{\psi}_{-\frac12}^{\dot{\pm}\dot{B}}\rangle_2^{\mp\dot{\mp}}$\,,\,\,$|\alpha_{-\frac12}^{A\dot{A}}\tilde{\psi}_{-\frac12}^{\dot{\pm}\dot{B}}\rangle_2^{\dot{C}\dot{\mp}}$\,,\,\,$|\psi_{-\frac12}^{\pm\dot{A}}\tilde{\alpha}_{-\frac12}^{B\dot{B}}\rangle_2^{\mp\dot{C}}$\,,\,\,$|\alpha_{-\frac12}^{A\dot{A}}\tilde{\alpha}_{-\frac12}^{B\dot{B}}\rangle_2^{\dot{C}\dot{D}}$, \\
		$3\phantom{\bigg(}$ & $|\alpha_{-\frac{1}{3}}^{A\dot{A}}\tilde{\alpha}_{-\frac{1}{3}}^{B\dot{B}}\rangle_3$\,,\,\,$|\alpha_{-\frac{1}{3}}^{A\dot{A}}\tilde{\psi}_{-\frac{1}{6}}^{\dot{+}\dot{B}}\tilde{\psi}_{-\frac{1}{6}}^{\dot{-}\dot{C}}\rangle_3$\,,\,\,$|\psi_{-\frac{1}{6}}^{+\dot{A}}\psi_{-\frac{1}{6}}^{-\dot{B}}\tilde{\alpha}_{-\frac{1}{3}}^{C\dot{C}}\rangle_3$\,,\,\,$|\psi_{-\frac{1}{6}}^{+\dot{A}}\psi_{-\frac{1}{6}}^{-\dot{B}}\tilde{\psi}_{-\frac{1}{6}}^{\dot{+}\dot{C}}\tilde{\psi}_{-\frac{1}{6}}^{\dot{-}\dot{D}}\rangle_3$\\
		$4\phantom{\bigg(}$ & $|0\rangle_{4}^{\dot{A}\dot{B}}$ \\
        \hline
	\end{tabular}
	\caption{All states in the sector with $h=\bar{h}=1$ and $j = \bar{\jmath} = 0$ separated by twist. Taking into account all possible index values, we have a total amount of 276 states.}
\label{table:881263}
\end{table}

\subsection{Marginal deformation}
\label{sec:RR-deformation}

Thanks to $\mathcal{N}=(4,4)$ supersymmetry, there are exactly marginal operators in this model (see~\cite{David:2002wn} for a description of them). In particular, one such operator preserves the full superconformal symmetry, is neutral under $SU(2)_{\circ}$ and $ SU(2)_{\bullet}$, and can be constructed from the $w=2$ sector. This operator is built as a supersymmetric descendant of a half-BPS state. More specifically the deformation $\mathcal{D}$ considered in this work is given by
\begin{equation}
|\mathcal{D} \rangle = \frac{1}{2\sqrt{2}} \ \epsilon_{AB} \epsilon_{\alpha\beta}\epsilon_{\dot{\alpha}\dot{\beta}}\ G^{\alpha A}_{-1/2} \tilde{G}^{\dot{\alpha} B}_{-1/2} |0\rangle_{2}^{\beta\dot{\beta}}\,,
\end{equation}
with the normalisation chosen such that its two-point function is unit-normalised and with $G^{\alpha A}_{-1/2}$ and $\tilde{G}^{\dot{\alpha} B}_{-1/2}$ being the supercharges described in Appendix~\ref{app:currents-and-charges}. This state has $h=\bar{h}=1$ and the resulting local operator, once integrated, is exactly marginal. This is the same deformation operator considered in earlier perturbative computations \cite{Gava:2002xb,Gaberdiel:2015uca,Hampton:2018ygz,Guo:2019ady,Guo:2020gxm,Lima:2020boh,Lima:2020kek,Lima:2020urq,Lima:2020nnx,Lima:2021wrz,Benjamin:2021zkn,Guo:2022ifr,Apolo:2022fya,Hughes:2023apl,Gaberdiel:2023lco,Gaberdiel:2024nge,Gaberdiel:2024dfw}. By using the identities \eqref{eq:648952-1} and \eqref{eq:648952-2} we can rewrite it as 
\begin{equation}
\label{eq:17944}
|\mathcal{D} \rangle  = \frac{1}{\sqrt{2}} \ \epsilon_{AB} \epsilon_{\dot{N}\dot{M}}\epsilon_{\dot{Q}\dot{P}}\ |\alpha^{A\dot{Q}}_{-1/2} \tilde{\alpha}^{B\dot{N}}_{-1/2} \rangle_{2}^{\dot{P}\dot{M}}\,,
\end{equation}
or as its lift to the covering surface
\begin{equation}
\label{eq:425268}
|\mathcal{D} \rangle  = \frac{1}{\sqrt{2}} \ \epsilon_{AB} \epsilon_{\dot{N}\dot{M}}\epsilon_{\dot{Q}\dot{P}}\ \alpha^{A\dot{Q}}_{-1/2} \tilde{\alpha}^{B\dot{N}}_{-1/2} \mathcal{S}^{\dot{P}\dot{M}} (0) | 0 \rangle_{\textrm{NS}} \,.
\end{equation}
Expression~\eqref{eq:425268} will be the most useful form for the mixing matrix computations in Section~\ref{sec:mixing-matrix-O(g)}. In the holographic picture this deformation corresponds to deforming the $AdS_3 \times  S^3 \times  T^4$ background with one unit of NSNS flux (which itself is dual to $\textrm{Sym}_N(T^4)$) by turning on an axion in the background, which sources the RR flux~\cite{OhlssonSax:2018hgc}. An analysis from the worldsheet point-of-view of this deformation can be found in \cite{Fiset:2022erp}.

\section{First-order perturbation theory and mixing matrices}
\label{sec:mixing-matrix-O(g)}
Having laid out the basic premises of the theory at hand, we now want to study the effect of the RR-deformation at first order in perturbation theory. The deformation changes the spectrum of the theory, but affects only certain subsectors of the full Hilbert space. In these sectors, the basis of independent (but a priori degenerate) states which we constructed above is reorganised into mixed states which diagonalise the so-called mixing matrix generated by the three-point function of states with the deformation. The eigenvalues of this mixing matrix constitute the anomalous conformal dimensions of the corresponding eigenstates.

We will first lay out the technical details of the calculation and discuss which sectors are affected. We then present some explicit results for light operators, i.e., operators that have conformal dimension $\Delta=h+\bar{h}\leq 3$ in the unperturbed theory.
\subsection{Conformal perturbation theory and Sym\texorpdfstring{$_N (T^4)$}{N(T4)}}
\label{sec:mixing-matrix-def}

Let us briefly review the logic behind conformal perturbation theory. Consider fields $\phi_{j} (z,\bar{z})$ with \textit{unperturbed} dimensions $(h_{j} ,\bar{h}_{j})$ in a generic CFT$_2$ perturbed by a marginal deformation operator $\mathcal{D}(z,\bar{z})$. Then a generic correlator has the expansion
\begin{equation}
\label{eq:PT-def}
\langle \phi_{1} (z_1 , \bar{z}_1 ) \cdots \phi_{n} (z_n , \bar{z}_n ) \rangle_{\xi} =
\sum_{\ell=0}^{\infty} \xi^\ell \langle \phi_{1} (z_1 , \bar{z}_1 ) \cdots \phi_{n} (z_n , \bar{z}_n ) \rangle^{(\ell)}\,,
\end{equation}
where $\xi$ is  the marginal coupling. The correlators on the right-hand side are computed in the unperturbed theory and are defined as
\begin{multline}
\langle \phi_{1} (z_1 , \bar{z}_1 ) \cdots \phi_{n} (z_n , \bar{z}_n ) \rangle^{(\ell)} = \\
\frac{1}{\ell!} \int_{D} d^2 y_1  \cdots d^2 y_l \  \langle \phi_{1} (z_1 , \bar{z}_1 ) \cdots \phi_{n} (z_n , \bar{z}_n ) \mathcal{D}(y_1 , \bar{y}_1 ) \cdots \mathcal{D}(y_\ell, \bar{y}_\ell )\rangle\,.
\end{multline}
The main problem then becomes constructing a proper regularisation scheme and choosing the integration domain $D$ such that one finds consistent physical results, before finally performing the integration, which is difficult in general. In this work we are interested in the anomalous dimension at first order in perturbation theory, for which there are closed-form expressions for the integrated correlator in terms of OPE data.

As usual, to compute anomalous dimensions we focus on two-point functions. The first order correction is given by
\begin{equation}
\langle \phi_n (z_1 , \bar{z}_1) \phi_m (z_2 , \bar{z}_2) \rangle^{(1)} = \int\limits_{\mathbb{C}\setminus\text{discs}_{\varepsilon}} d^2 y\ \langle \phi_n (z_1 , \bar{z}_1) \mathcal{D}(y,\bar{y})  \phi_m (z_2 , \bar{z}_2) \rangle\,,
\end{equation}
where the integration region is the complex plane with $\varepsilon$-discs cut around the insertions at $(z_1 , \bar{z}_1)$ and $(z_2 , \bar{z}_2)$ to avoid singularities. We will not reproduce in full details how to compute this integral and refer the reader to the thorough discussion of ref.~\cite{Keller:2019yrr}. The main point is that the leading $\log\varepsilon^2$ behaviour has to be cancelled by an appropriate renormalisation scheme. In the end, one finds the following renormalised two-point function
\begin{equation}
\langle \phi_n (z_1 , \bar{z}_1) \phi_m (z_2 , \bar{z}_2) \rangle_{\textrm{ren}} = \frac{\delta_{nm}}{z_{12}^{2h} \bar{z}_{12}^{2\bar{h}}} + 2\pi \xi\frac{ C_{n\mathcal{D}m}}{z_{12}^{2h} \bar{z}_{12}^{2\bar{h}}}\ \log\left(\frac{|z_{12}|^2 }{\Lambda^2}\right) + O(\xi^2)\,,
\end{equation}
with $h_{n} = h_{m} = h$ due to conformal symmetry, $z_{12}=z_1-z_2$ as usual, $\Lambda$ an appropriate length scale and $C_{n\mathcal{D}m}$ denoting the structure constant for the fields $\phi_n$, $\phi_m$ and the deformation operator $\mathcal{D}$. We can see that the mixing matrix in this case is given by
\begin{equation}
    \label{eq:mixing-mat-gen-def}
    \mathcal{M}_{nm} = -2\pi \xi \,C_{n\mathcal{D}m}\,.
\end{equation}
The logarithmic $z_{12}$-dependence allows us to reabsorb the $O(\xi)$ perturbation in a shift of the conformal dimensions $h$ and $\bar{h}$ by the eigenvalues of $\mathcal{M}_{nm}$.

Hence, our computation boils down to computing a three-point function involving the marginal operator~$\mathcal{D}$.
We shall see that the mixing matrix separates into blocks, corresponding to degenerate sectors which do not interact with each other. This block structure follows broadly from the charges of the states.

In the case at hand, we will pick a basis of states of the undeformed theory written in terms of (fractional) oscillators of the free fields, see Section~\ref{sec:vacua-and-spectrum}. 
Due to conformal symmetry and R-symmetry, the mixing matrix splits into blocks with definite $(h,\bar{h},j,\bar{\jmath})$ that can be independently diagonalised.\footnote{Here $h$ and $\bar{h}$ denote the unperturbed left- and right-moving dimensions, i.e., up to $O(g)$ corrections.} Each block is a finite (but large) matrix which can be computed and then, exactly or numerically, diagonalised. For instance, in the block with $(h,\bar{h},j,\bar{\jmath})=(1,1,0,0)$ given in Table \ref{table:881263} we have a $276\times 276$ matrix. 

To understand the structure of the three-point function under consideration, let us look at eq.~\eqref{eq:invariantcorrelator} and without loss of generality consider as $w$-cycle the permutation $g_1=(12\cdots w)$ on the first $w$ copies. If we want a connected correlator, we have two possible choices for the 2-cycle appearing in $\mathcal{D}$: either it involves two copies among the first $w$ --- for instance, $g_2=(12)$, or it involves one copy among the first $w$ and a different copy --- for instance, $g_2=(1,w+1)$. In the former case, we see that to satisfy the orbifold invariance condition $g_1g_2g_3=1$, the permutation $g_3$ must be a $(w-1)$-cycle, $g_3=(23\cdots w)$; in the latter case, $g_3$ must be a $(w+1)$-cycle, $g_3=(12\cdots w+1)$.
Hence, we are interested in mixing the states involving a $w$-cycle with those involving a $(w\pm1)$-cycle.  

\begin{figure}[t]
	\centering
	\includegraphics[width=0.7\textwidth]{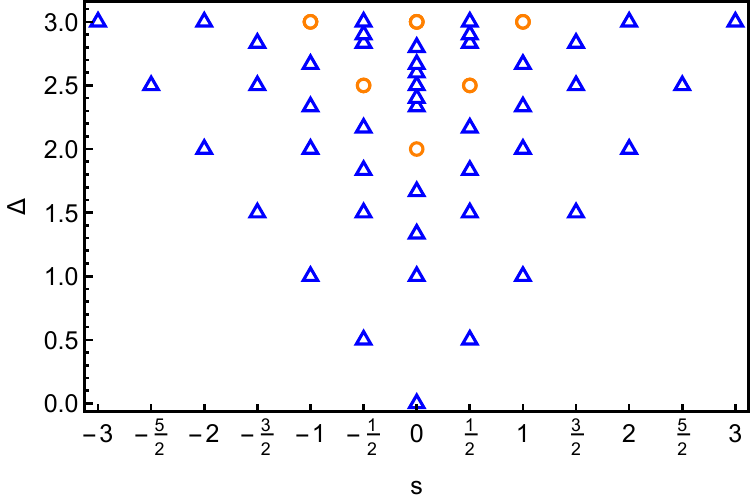}
	\caption{Free conformal dimensions $\Delta=h+\bar{h}$ and spins $s=h-\bar{h}$ of light ($\Delta\leq3$) states in the symmetric-product orbifold Sym$_N(T^4)$. We marked sectors with non-trivial mixing under first-order RR-deformation as orange circles. We can observe the relative scarcity in comparison with the states that are corrected only at higher orders in perturbation theory which are denoted by the blue triangles.}\label{fig:sectors}
\end{figure}

The conformal dimensions $h,\bar{h}$ of such states are quantised as (half-)integer multiples of $1/w$ and $1/(w\pm1)$, respectively. Therefore, for them to mix it is necessary that the scaling dimensions are \textit{integer} or \textit{half-integer}. This is a relatively rare occurrence, but such states can be constructed for any  $h,\bar{h}\in\mathbb{Z}/2$. Figure~\ref{fig:sectors} illustrates the relative scarcity of such states as $h,\bar{h}$ increases.

As in usual large-$N$ expansions, we define an effective coupling in perturbation theory so that non-planar and perturbative corrections are easily separable. We remember, as discussed in Section \ref{sec:correlators-lifiting}, that a generic genus-$\textrm{g}$ contribution to an $n$-point function in the symmetric-product orbifold has leading $N$ behaviour given by $N^{1-\textrm{g}-n/2}$. Applying this to the planar mixing matrix in consideration here we find $N^{-1/2}$. Then it is useful to define the following effective coupling in perturbation theory%
\footnote{The overall sign of the coupling \textcolor{red}{$\lambda$} is not so crucial: as we will see explicitly, at $O(\xi)$ the anomalous dimensions come in pairs of opposite sign. This can be understood by noting that correlation functions are invariant under the parity operator $g\to\text{sgn}[g]g$  (where $\text{sgn}[g]$ is the sign of the permutation $g$) and  $\mathcal{D}\to-\mathcal{D}$ under such a transformation.}
\begin{equation}
\label{eq:eff-coupling-def}
\lambda = -\frac{2 \pi \xi}{\sqrt{N}}\,,
\end{equation}
where $-2\pi$ and $N^{-1/2}$ absorb the factor in \eqref{eq:mixing-mat-gen-def} and the planar large-$N$ factor, respectively. One can see that $\lambda$ is indeed a good expansion parameter. Consider the $k$-th order in perturbation theory. From \eqref{eq:PT-def} we see that this contribution is proportional to
\begin{equation}
\xi^k N^{1-\textrm{g}-\frac{(k+2)}{2}} \sim \frac{\lambda^k}{N^{\textrm{g}}}\,,
\end{equation}
because in the $O(\xi^k)$ contribution one computes an integrated $(k+2)$-point function. Upon using $\lambda$, the non-planar corrections are suppressed by powers $N^{-\textrm{g}}$, as expected from a 't Hooft-like expansion.

Let us now introduce an explicit notation for the  matrix elements of $\mathcal{M}$, namely\footnote{To exemplify we assumed $w$ even, thus the right state vacuum is charged and the left $w\pm 1$ twisted vacuum is uncharged. The opposite case works analogously.}
\begin{equation}
\label{eq:338781}
_{w\pm 1}\langle Y_{-m_1 /(w\pm 1)}^{(1)} \cdots Y_{-m_{m'} /(w\pm 1)}^{(m')} | \mathcal{M} |X_{-n_1 /w}^{(1)} \cdots X_{-n_m /w}^{(m)} \rangle^{xy}_{w} \,,
\end{equation}
with $X_{-n_k /w}^{(k)}$ and $Y_{-m_k /(w\pm 1)}^{(k)}$ representing generic bosonic and/or fermionic excitations detailed in Section \ref{sec:review-section}. Using expressions \eqref{eq:mixing-mat-gen-def} for the mixing matrix and \eqref{eq:eff-coupling-def} for the effective coupling we find that a generic mixing matrix element is given by
\begin{multline}
\label{eq:970115}
_{w\pm 1}\langle Y_{-m_1 /(w\pm 1)}^{(1)} \cdots Y_{-m_{m'} /(w\pm 1)}^{(m')} |\mathcal{M} |X_{-n_1 /w}^{(1)} \cdots X_{-n_m /w}^{(m)} \rangle^{xy}_{w} = \frac{\lambda}{\sqrt{2}} \ C^{\sigma}_{w\pm 1,2,w}\  \times  \\  
\epsilon_{AB} \epsilon_{\dot{N}\dot{M}}\epsilon_{\dot{Q}\dot{P}} \ _{\textrm{NS}}\langle Y_{-m_1 /(w\pm 1)}^{(1)} \cdots Y_{-m_{m'} /(w\pm 1)}^{(m')} | (\alpha^{A\dot{Q}}_{-1/2} \tilde{\alpha}^{B\dot{N}}_{-1/2} \mathcal{S}^{\dot{P}\dot{M}})(1) |X_{-n_1 /w}^{(1)} \cdots X_{-n_m /w}^{(m)} \rangle^{xy}_{\textrm{R}} \,.
\end{multline}
In the above formula we lifted the correlator on the right-hand side of \eqref{eq:mixing-mat-gen-def} to the covering surface using the prescription of~\cite{Lunin:2001pw} (and given schematically in \eqref{eq:schematically-lift}). Here $C^{\sigma}_{w\pm 1,2,w}$ denotes the twist field contribution as in \eqref{eq:invariantcorrelator} (with twists $w$, $w\pm 1$, and $2$). The correlator in the second line of \eqref{eq:970115} involves free bosons and fermions on the covering surface. The computation of the mixing matrix in each sector $(h,\bar{h},j,\bar{\jmath})$ then reduces to the calculation of these matrix elements. In the next section we show how to compute them. 

\subsection{Constructing the mixing matrix elements}\label{Sec: 3.2}

Here we show the necessary steps to compute the mixing matrix elements of the type \eqref{eq:970115}. The main idea is to break the lifted structure constant on the right-hand side into twist field, boson, and fermion contributions and compute each term independently. We are going to detail how each contribution works in general and also exemplify these using the following matrix element
\begin{equation}
\label{eq:example-comp}
_{2}^{\dot{2}\dot{2}}\langle \alpha_{-1/2}^{1 \dot{2}} \tilde{\alpha}_{-1/2}^{2 \dot{2}} | \mathcal{M} | \psi^{-\dot{2}}_{-1/2} \psi^{+\dot{2}}_{-1/2} \tilde{\psi}^{\dot{-}\dot{2}}_{-1/2} \tilde{\psi}^{\dot{+}\dot{2}}_{-1/2}  \rangle_{1}  = \frac{\lambda}{\sqrt{2}} \ C^{\sigma}_{2,2,1}\ \epsilon_{AB} \epsilon_{\dot{N}\dot{M}}\epsilon_{\dot{Q}\dot{P}} \ \mathcal{B}^{A\dot{Q}B\dot{N}}\ \mathcal{F}^{\dot{P}\dot{M}}\,,
\end{equation}
where $C^{\sigma}_{2,2,1}$, $\mathcal{B}^{A\dot{Q}B\dot{N}}$, and $\mathcal{F}^{\dot{P}\dot{M}}$ are the twist, bosonic, and fermionic contributions, respectively. Their explicit forms will be given as we compute each one. Note that this is an example of mixing occurring in the sector $(1,1,0,0)$ described in Table \ref{table:881263}.

The last ingredient we need is the covering map used to lift the mixing matrix element computations to the covering surface and also to define the bosonic and fermionic modes in \eqref{eq:667038-1} and \eqref{eq:6548651687}, which are
\begin{equation}
\label{eq:930576}
\bfgamma_{+} (t) = (w+1)t^{w} -wt^{w+1}\,,  \qquad
 \bfgamma_{-} (t) = \frac{t^w}{wt-w+1}\,,
\end{equation}
for $w$-$(w+1)$ and $w$-$(w-1)$ mixings, respectively. It can be checked that these follow the expansion \eqref{eq:coveringmap} around the insertion points as expected.

\subsubsection{Twist field contribution}

The twist field contribution amounts to computing the three-point function of normalised twist operators and it was found in \cite{Lunin:2000yv}. We write this result below as\footnote{More specifically, this is the large-$N$ result \cite{Lunin:2000yv}, where we stripped out the $N^{-1/2}$ factor since we absorbed it into the planar coupling~$\lambda$.}
\begin{equation}
C^{\sigma}_{n,m,q} = \sqrt{nmq}\  |\mathcal{C}_{nmq}|^{\mathbf{c}}\,,
\end{equation}
where $\mathbf{c}$ is the central charge of the seed theory and $n$, $m$, and $q$ are the twists of the operator insertions. The $|\mathcal{C}_{nmq}|$ is given by the following rather unwieldy expression
\begin{multline}
\log |\mathcal{C}_{nmq}|^2 = \frac{1}{6} \log\left(\frac{q}{nm}\right) - \frac{n-1}{12} \log n - \frac{m-1}{12} \log m + \frac{q-1}{12} \log q \\
- \frac{n-1}{12n} \log\left( \frac{d_1!d_2!}{n!(n-1)!} \frac{(d_1-m)!}{(d_1-n)!} \right) - \frac{m-1}{12m} \log\left( \frac{d_1!d_2!}{m!(m-1)!} \frac{(d_1-n)!}{(d_1-m)!} \right) \\
+ \frac{q-1}{12q} \log\left( \frac{(q-1)! d_2!}{(d_1 -n)!(d_1 -m)!} \frac{(d_1-d_2)!}{d_1!} \right) + \frac{d_2 (d_2 -1)}{3} \log 2 -\frac{d_2}{6} \log n + \frac{\log\mathcal{D}}{3}\\
+ \frac{(3d_2 -4)}{6} \log d_2! - \frac{d_2}{6} \log\left( \frac{d_1!}{n!(d_1 - n)!} \right)-\frac{(n+d_2 -1)}{6} \log\left( \frac{(n-1)!}{(n-d_2 -1)!} \right) \\
 - \frac{(d_1 - d_2 +3)}{6} \log\left( \frac{(d_1 -d_2)!}{d_1!} \right) - \frac{(d_1 + d_2 -n)}{6} \log\left( \frac{(d_1+d_2-n)!}{(d_1-n)!} \right)\,,
\end{multline}
with $d_1$, $d_2$, and $\mathcal{D}$ being
\begin{align}
& d_1 = \frac{n+m+q-1}{2}\,, \\
& d_2 = \frac{n+m-q-1}{2}\,, \\
& \mathcal{D} = \frac{1}{2^{d_2 (d_2 -1)}} \prod_{k=1}^{d_2} k^{k+2-2d_2} (k-n)^{k-1} (k-d_1-d_2+n-1)^{k-1} (k-d_1 -1)^{d_2-k}\ .
\end{align}
Note that this expression is universal and is dependent on the seed theory only through the central charge $\mathbf{c}$. Although it is not manifest, this expression is fully symmetric in $n$, $m$, and $q$.

For the mixing matrix applications, we are interested in the case with $\mathbf{c}=6$ (see Appendix~\ref{app:free-field-conv}) and fixed twists. We then only need the expressions
\begin{equation}
\label{eq:7696}
 C^{\sigma}_{w+1,2,w} = \frac{w^{-\frac{w (2 w+3)+3}{8 (w+1)}} (w+1)^{\frac{2 w^2+w+2}{8
   w}}}{2^{5/8}}\,,\qquad
C^{\sigma}_{w-1,2,w} = \frac{(w-1)^{\frac{-2 w^2+w-2}{8 w}} w^{\frac{2 w^2-3 w+3}{8
   (w-1)}}}{2^{5/8}}\,.
\end{equation}
For the example \eqref{eq:example-comp} the twist contribution is simply
\begin{equation}
\label{eq:twist-example}
    C^{\sigma}_{2,2,1} =1\,,
\end{equation}
which is expected since the twist operators are unit normalised and in this case the structure constant reduces to the norm of the two-point function of twist-2 operators.

\subsubsection{Boson contribution}
\label{sec:boson-contribution}

This contribution involves only the bosonic modes $\alpha^{A\dot{A}}_{-n/w}$ and $\tilde{\alpha}^{A\dot{A}}_{-n/w}$ of the excited states in the mixing matrix elements. To compute it we first convert the modes into fields by using their definition \eqref{eq:667038-1} and end up with an integrated free boson correlator. This correlator is computed by usual Wick contractions. However, due to the factorisation of Wick contractions we may equally factorise the  integrals and compute them contraction-by-contraction. Using this we can analytically compute all Wick contractions involving bosonic modes around $0$ (right state in \eqref{eq:970115}), around $1$ (deformation mode), and around $\infty$ (left state in \eqref{eq:970115}). For instance, an integrated Wick contraction involving bosonic modes of the left and right states is given generically by
\begin{equation}
\label{eq:example-wick}
    \wick{ (\c1 \alpha^{B \dot{B}}_{m/(w\pm 1)})_{\infty} ( \c1 \alpha^{A \dot{A}}_{-n/w})_{0}} = 
    \frac{-\epsilon^{AB} \epsilon^{\dot{A}\dot{B}}}{\sqrt{w(w\pm 1)}} \oint_{\infty} \frac{dt'}{2\pi i} \oint_0 \frac{dt}{2\pi i}\ \bfgamma_{\pm} (t')^{m/(w\pm 1)} \bfgamma_{\pm} (t)^{-n/w} \frac{1}{(t-t')^2}\,, 
\end{equation}
for a mixing involving twists $w$ and $w\pm 1$. Similar expressions can be found for integrated Wick contractions involving all other modes. Note that the $SU(2)_{\circ}$ and $SU(2)_{\bullet}$ indices are simply contracted with the $SU(2)$-invariant structure $\epsilon$ and the integrated Wick contractions can thus be written as
\begin{align}
    \label{eq:bos-wick1}
    & \wick{ (\c1 \alpha^{B \dot{B}}_{-m/w})_{0} ( \c1 \alpha^{A \dot{A}}_{-n/w})_{0} } = \epsilon^{AB} \epsilon^{\dot{A}\dot{B}}\ \mathcal{W}_{00}^{\pm}(n,m)\,, \\
    & \wick{ (\c1 \alpha^{B \dot{B}}_{-1/2})_{1} ( \c1 \alpha^{A \dot{A}}_{-n/w})_{0} } = \epsilon^{AB} \epsilon^{\dot{A}\dot{B}}\ \mathcal{W}_{01}^{\pm}(n)\,, \\
    & \wick{ (\c1 \alpha^{B \dot{B}}_{m/(w\pm 1)})_{\infty} ( \c1 \alpha^{A \dot{A}}_{-n/w})_{0} } = \epsilon^{AB} \epsilon^{\dot{A}\dot{B}}\ \mathcal{W}_{0\infty}^{\pm}(n,m)\,, \\
    \label{eq:6435468}
    & \wick{ (\c1 \alpha^{B \dot{B}}_{m/(w\pm 1)})_{\infty} ( \c1 \alpha^{A \dot{A}}_{-1/2})_{1} } = \epsilon^{AB} \epsilon^{\dot{A}\dot{B}}\ \mathcal{W}_{1\infty}^{\pm}(m)\,, \\
    \label{eq:bos-wick5}
    & \wick{ (\c1 \alpha^{B \dot{B}}_{m/(w\pm 1)})_{\infty} ( \c1 \alpha^{A \dot{A}}_{n/(w\pm 1)})_{\infty} } = \epsilon^{AB} \epsilon^{\dot{A}\dot{B}}\ \mathcal{W}_{\infty\infty}^{\pm}(n,m)\,,
\end{align}
with similar expressions for the right-moving modes. The right-hand side factors correspond to integrals similar to \eqref{eq:example-wick}, which can be computed analytically and are given in Appendix \ref{app:integrals-wick}. 

With these tools we can compute the bosonic contribution in our example \eqref{eq:example-comp}. It is explicitly given by 
\begin{equation}
    \mathcal{B}^{A\dot{Q}B\dot{N}} = - \langle 0 | (\alpha_{1/2}^{2 \dot{1}} \tilde{\alpha}_{1/2}^{1 \dot{1}})_{\infty} (\alpha^{A\dot{Q}}_{-1/2} \tilde{\alpha}^{B\dot{N}}_{-1/2})_{1} | 0  \rangle\,,
\end{equation}
where the modes are grouped by the point (indicated in the subscript $\infty$ and $1$ in this example) around which we integrate the boson fields using the definition \eqref{eq:667038-1}. Note that we have a minus sign coming from conjugation of the modes in the state using the conjugation rule \eqref{eq:bosons-conj}. As described before, we could write explicitly the integrated boson correlator, compute the integrand using Wick contractions and then integrate the latter. However, by using the previously described factorisation of Wick contractions and relation \eqref{eq:6435468} we find
\begin{equation}
\label{eq:boson-example}
    \mathcal{B}^{A\dot{Q}B\dot{N}} = \frac{1}{2} \ \delta^{A}_{1} \delta^{B}_{2} \delta^{\dot{Q}}_{\dot{2}} \delta^{\dot{N}}_{\dot{2}} \ ,
\end{equation}
which completes the computation of the bosonic contribution to \eqref{eq:example-comp}.

\subsubsection{Fermion contribution and bosonisation}
\label{sec:fermion-contribution}

The fermion contribution is computed analogously to the bosonic term. That is, we write the fermionic modes using \eqref{eq:6548651687} and then we calculate integrated fermionic correlators. To compute these fermion correlators we are going to use bosonisation, which is useful when spin fields are present. For this we introduce left- and right-moving bosons $\phi_a (t)$ and $\tilde{\phi}_a (\bar{t})$ with $a=1,2$. The fermions and spin fields are then given by
\begin{equation}
\begin{gathered}
 \psi^{\alpha\dot{A}}(t) = C^{\alpha\dot{A}} :e^{i \boldsymbol{q}^{\alpha\dot{A}}\cdot\phi(t)}:\,,\qquad
 \tilde{\psi}^{\dot{\alpha}\dot{A}}(\bar{t}) = \tilde{C}^{\dot{\alpha}\dot{A}} :e^{i \boldsymbol{q}^{\dot{\alpha}\dot{A}}\cdot\tilde{\phi}(\bar{t})}: \,, \\
 \mathcal{S}^{xy}(t,\bar{t}) = Q^{xy} :e^{i \boldsymbol{\xi}^{x}\cdot\phi(t)} e^{i \boldsymbol{\xi}^{y}\cdot\tilde{\phi}(\bar{t})}: \,,
\end{gathered}
\end{equation}
with the prefactors $C^{\alpha\dot{A}}$, $\tilde{C}^{\dot{\alpha}\dot{A}}$, and $Q^{xy}$ being cocycles which ensure the correct statistics of the fields. The definition of $\boldsymbol{q}^{\alpha\dot{A}}$, $\boldsymbol{q}^{\dot{\alpha}\dot{A}}$, and $\boldsymbol{\xi}^{x}$ and a detailed description of the bosonisation procedure used here are reported in Appendix~\ref{app:bosonisation}. The main point of introducing bosonisation is that the fermion correlators boil down to vertex-operator correlators which are known to evaluate to \cite{polchinski1998string}
\begin{equation}
\label{eq:702796asas}
\langle :e^{i\boldsymbol{\xi}_1 \cdot \phi(t_1)}: \cdots :e^{i\boldsymbol{\xi}_n \cdot \phi(t_n)}: \rangle = \prod_{i<j} t_{ji}^{\boldsymbol{\xi}_i \cdot \boldsymbol{\xi}_j} \ \ \ \textrm{with} \ \ \ \sum_i \boldsymbol{\xi}_i = 0\
,
\end{equation}
where radial ordering ($|t_{j+1}| > |t_{j}|$) is assumed. For right-moving operators we have a similar contribution. The vanishing of the sum over all polarisations is just the statement of charge neutrality of the correlator.\footnote{This implies that if some operator is defined at infinity, its contribution is cancelled by the inversion factor used to move it to infinity and thus it does not enter \eqref{eq:702796asas} directly, only through charge neutrality.}

We now compute the fermionic contribution of our example mixing matrix element \eqref{eq:example-comp}. First we note that by combining the index structure of \eqref{eq:example-comp} with the one coming from the bosonic contribution \eqref{eq:boson-example} we can fix the indices of the fermionic contribution. It is given as
\begin{equation}
    \mathcal{F}^{\dot{P}\dot{M}} = \delta_{\dot{1}}^{\dot{P}} \delta_{\dot{1}}^{\dot{M}} \ _{\textrm{NS}}\langle 0 |(\mathcal{S}^{\dot{1}\dot{1}})_{\infty} ( \mathcal{S}^{\dot{1}\dot{1}})_{1} ( \psi^{-\dot{2}}_{-1/2} \psi^{+\dot{2}}_{-1/2} \tilde{\psi}^{\dot{-}\dot{2}}_{-1/2} \tilde{\psi}^{\dot{+}\dot{2}}_{-1/2})_0 |0 \rangle_{\textrm{NS}}\,,
\end{equation}
where the conjugation rules like \eqref{eq:fermions-conj} were used. By using the mode definition \eqref{eq:6548651687} it is possible to rewrite it as 
\begin{multline}
\label{eq:example-integral}
 \mathcal{F}^{\dot{P}\dot{M}} = \delta_{\dot{1}}^{\dot{P}} \delta_{\dot{1}}^{\dot{M}} \oint_0 \frac{dt_1}{2\pi i } \oint_0 \frac{dt_2}{2\pi i } \oint_0 \frac{d\bar{t}_1}{2\pi i } \oint_0 \frac{d\bar{t}_2}{2\pi i }\ \frac{(\partial\bfgamma_+ (t_1))^{1/2}}{\bfgamma_+(t_1)} \frac{(\partial\bfgamma_+(t_2))^{1/2}}{\bfgamma_+(t_2)} \frac{(\partial\bfgamma_+(\bar{t}_1))^{1/2}}{\bfgamma_+(\bar{t}_1)} \times  \\  \frac{(\partial\bfgamma_+(\bar{t}_2))^{1/2}}{\bfgamma_+(\bar{t}_2)}\ 
 _{\textrm{NS}}\langle 0 |\mathcal{S}^{\dot{1}\dot{1}} (\infty) \mathcal{S}^{\dot{1}\dot{1}} (1) \psi^{-\dot{2}}(t_1) \psi^{+\dot{2}}(t_2) \tilde{\psi}^{\dot{-}\dot{2}}(\bar{t}_1) \tilde{\psi}^{\dot{+}\dot{2}}(\bar{t}_2)  |0 \rangle_{\textrm{NS}}\,.
\end{multline}
The correlator in the integrand is found by using the bosonisation prescription (see Appendix~\ref{app:bosonisation}) combined with formula~\eqref{eq:702796asas}. In this particular case it is simply
\begin{multline}
_{\textrm{NS}}\langle 0 |\mathcal{S}^{\dot{1}\dot{1}} (\infty) \mathcal{S}^{\dot{1}\dot{1}} (1) \psi^{-\dot{2}}(t_1) \psi^{+\dot{2}}(t_2) \tilde{\psi}^{\dot{-}\dot{2}}(\bar{t}_1) \tilde{\psi}^{\dot{+}\dot{2}}(\bar{t}_2)  |0 \rangle_{\textrm{NS}} = \\
(1-t_1 )^{-1/2} (1-t_2 )^{-1/2} (1-\bar{t}_1 )^{-1/2} (1-\bar{t}_2 )^{-1/2} .
\end{multline}
Integrating the above result as in \eqref{eq:example-integral} we find the following fermionic contribution
\begin{equation}
\label{eq:fermion-example}
    \mathcal{F}^{\dot{P}\dot{M}} =\frac{1}{4} \ \delta_{\dot{1}}^{\dot{P}} \delta_{\dot{1}}^{\dot{M}}\,.
\end{equation}

Finally, combining the twist field \eqref{eq:twist-example}, the bosonic \eqref{eq:boson-example}, and the fermionic \eqref{eq:fermion-example} contributions with expression \eqref{eq:example-comp} we can construct the desired matrix element
\begin{equation}
\label{eq:final-example}
_{2}^{\dot{2}\dot{2}}\langle \alpha_{-1/2}^{1 \dot{2}} \tilde{\alpha}_{-1/2}^{2 \dot{2}} | \mathcal{M} | \psi^{-\dot{2}}_{-1/2} \psi^{+\dot{2}}_{-1/2} \tilde{\psi}^{\dot{-}\dot{2}}_{-1/2} \tilde{\psi}^{\dot{+}\dot{2}}_{-1/2}  \rangle_{1}
=\frac{\lambda}{4\sqrt{2}}\,.
\end{equation}
We remark that expression \eqref{eq:970115} for generic mixing matrix elements is valid for normalised states, thus \eqref{eq:final-example} is found after dividing by the norms of the left and right states in \eqref{eq:example-comp}.

\subsection{Anomalous dimensions for light states}\label{sec:results}

Having established the necessary techniques to compute mixing matrices, we now move on to the analysis of low-lying states with bare dimension $\Delta\leq3$. As exemplified in Table \ref{table:881263} the number of such states in each mixing sector is large and so we automated the mixing matrix computation in \texttt{Wolfram Mathematica}. The interested reader can consult the ancillary file in the \texttt{arXiv} submission.

As discussed above, the mixing problem can be broken down into sectors labelled by conformal dimensions and R-charges $(h,\bar{h},j,\bar{\jmath})$, which remain independent under first-order RR-deformation. However these are not the only restrictions to mixing. Indeed, due to the bosonic contribution, detailed in Section~\ref{sec:boson-contribution}, we observe that non-vanishing mixing only occurs when the overall number of bosonic excitations in in- and out-state is odd in both left-moving and right-moving sector. This is such that one can absorb the bosonic modes of the deformation seen in eq.~\eqref{eq:970115}. One immediate consequence of this is that half-BPS operators are not corrected, as expected (see Appendix~\ref{app:chiral-ring}). Combining this restriction on bosonic oscillators with the fact that mixing occurs only between  twist $w$ and $w\pm 1$ sectors we see that numerous sectors are excluded, such as those with $h,\bar{h}\notin\mathbb{Z}/2$ as illustrated in Figure~\ref{fig:sectors}.

With all these restrictions we are left with 15 non-trivial sectors for bare dimension $\Delta\leq3$ which fit into five groups
\begin{equation}\begin{split}
    \text{Group 1:}& \quad (1,1,0,0)\,,\\
    \text{Group 2:}& \quad (1,\tfrac{3}{2},0,\tfrac{1}{2})\,,\quad (1,\tfrac{3}{2},0,-\tfrac{1}{2})\,,\quad(\tfrac{3}{2},1,\tfrac{1}{2},0)\,,\quad(\tfrac{3}{2},1,-\tfrac{1}{2},0)\,,\\
    \text{Group 3:}& \quad (1,2,0, 1)\,,\quad(1,2,0, -1)\,,\quad (2,1,1,0)\,,\quad (2,1,-1,0)\,,\\
    \text{Group 4:}& \quad (\tfrac{3}{2},\tfrac{3}{2},\tfrac{1}{2},\tfrac{1}{2})\,,\quad(\tfrac{3}{2},\tfrac{3}{2},\tfrac{1}{2},-\tfrac{1}{2})\,,\quad (\tfrac{3}{2},\tfrac{3}{2},-\tfrac{1}{2},\tfrac{1}{2})\,,\quad(\tfrac{3}{2},\tfrac{3}{2},-\tfrac{1}{2},-\tfrac{1}{2})\,,\\
    \text{Group 5:}& \quad (1,2,0,0)\,,\quad (2,1,0,0)\,.
\end{split}\label{eq:mixingsecs}\end{equation}
The sectors within each group are related by exchanging left- and right-moving fields or inverting the R-charge and thus have the same spectrum of anomalous dimensions. It is therefore enough to compute the mixing matrix in the first sector of each group.

One could further decompose each sector in irreducible representations of $SU(2)_\bullet\times  SU(2)_\circ$ which mix strictly among themselves. However, this was not necessary for our purposes where the full mixing matrix could already be generated in a reasonable amount of time. We simply determined the $SU(2)_\bullet\times  SU(2)_\circ$ representations retrospectively to verify that mixing does not occur between different irreducible representations.

For the lowest sector $(1,1,0,0)$  we found that 74 of the 276 states are mixed by the RR-deformation and acquire an anomalous dimension. We present the results of the diagonalisation of the mixing matrix in Table \ref{table:firstsector}. Although the eigenvalues can be computed analytically, they do not follow any obvious pattern apart from appearing in pairs of opposite sign. We also presented which untwisted states partake in mixing. For example, one eigenstate in the first eigenspace of Table \ref{table:firstsector} is approximately given by the combination
\begin{equation}
   (0.3-0.5i)\big|\psi_{-\frac{1}{2}}^{+\dot{1}}\psi_{-\frac{1}{2}}^{-\dot{1}}\tilde{\psi}_{-\frac{1}{2}}^{+\dot{1}}\tilde{\psi}_{-\frac{1}{2}}^{-\dot{1}}\big\rangle_{1}+
   (0.3-0.4i)\epsilon_{AB}\big|\alpha_{-\frac{1}{2}}^{A\dot{1}}\tilde{\alpha}_{-\frac{1}{2}}^{B\dot{1}}\big\rangle_{2}^{\dot{1}\dot{1}}+0.4\big|\psi_{-\frac{1}{2}}^{+\dot{1}}\psi_{-\frac{1}{2}}^{-\dot{1}}\tilde{\psi}_{-\frac{1}{6}}^{+\dot{1}}\tilde{\psi}_{-\frac{1}{6}}^{-\dot{1}}\big\rangle_{3}\,.
   \end{equation}
It can also be seen that the eigenstates organise themselves in irreducible representations of $SU(2)_\bullet\times  SU(2)_\circ$ (see Table~\ref{table:sectors1and2}) so the index structure of the constituents is necessarily aligned. Note that the sector $(1,1,0,0)$ contains the RR-deformation operator \eqref{eq:425268} itself, but our analysis shows that it does not participate in first order mixing (i.e. it is part of ``all other states" in Table \ref{table:firstsector}). This is an important consistency check because it tell us that the deformation considered remains exactly marginal at first order in perturbation theory.

\begin{table}[t]
	\centering
	\renewcommand{\arraystretch}{1.5}
	\begin{tabular}{|c|c|c|c|}
    \hline
		 $\gamma/\lambda$ & Numerical values & \# states & Constituent states \\\hline\hline
		$\pm \frac{1}{12}\sqrt{9+2^\frac{8}{3}3^{-\frac{1}{2}}}$
		&$\pm0.2966$ &9 & \begin{tabular}{c}
			$\big|\psi_{-\frac{1}{2}}^{+(\dot{A}}\psi_{-\frac{1}{2}}^{-\dot{B})}\tilde{\psi}_{-\frac{1}{2}}^{+(\dot{C}}\tilde{\psi}_{-\frac{1}{2}}^{-\dot{D})}\big\rangle_1$\\
			$\epsilon_{AB}\big|\alpha_{-\frac{1}{2}}^{A(\dot{A}}\tilde{\alpha}_{-\frac{1}{2}}^{B(\dot{C}}\big\rangle_{2}^{\dot{B})\dot{D})}$\\$\big|\psi_{-\frac{1}{2}}^{+(\dot{A}}\psi_{-\frac{1}{2}}^{-\dot{B})}\tilde{\psi}_{-\frac{1}{6}}^{+(\dot{C}}\tilde{\psi}_{-\frac{1}{6}}^{-\dot{D})}\big\rangle_{3}$
			\end{tabular}\\\hline
		$\pm2^{-\frac{25}{24}}3^{-\frac{55}{16}}\sqrt{27+2^{\frac{27}{4}}3^{\frac{3}{8}}}$
		&$\pm0.1532$ &4 & \begin{tabular}{c}
			$\epsilon_{\pm\mp}{\epsilon}_{\dot{\pm}\dot{\mp}}\big|\psi_{-\frac{1}{2}}^{\pm\dot{A}}\tilde{\psi}_{-\frac{1}{2}}^{\pm'\dot{B}}\big\rangle_{2}^{\mp\dot{\mp}}$\\$\epsilon_{AB}\big|\alpha_{-\frac{1}{3}}^{A\dot{A}}\tilde{\alpha}_{-\frac{1}{3}}^{B\dot{B}}\big\rangle_{3}$\\
			$|0\rangle_{4}^{\dot{A}\dot{B}}$
		\end{tabular} \\\hline
		$\pm 2^{\frac{1}{3}}3^{-\frac{9}{4}}$
		&$\pm0.1064$ & 24 (12+12) & \begin{tabular}{c}
			$\epsilon_{\pm\mp}\big|\psi_{-\frac{1}{2}}^{\pm\dot{A}}\tilde{\alpha}_{-\frac{1}{2}}^{A(\dot{B}}\big\rangle_{2}^{\mp\dot{C})}$\\$\big|\alpha_{-\frac{1}{3}}^{A\dot{A}}\tilde{\psi}_{-\frac{1}{2}}^{+(\dot{B}}\tilde{\psi}_{-\frac{1}{2}}^{-\dot{C})}\big\rangle_{3}$\\\hline
			$\epsilon_{\pm\mp}\big|\alpha_{-\frac{1}{2}}^{A(\dot{B}}\tilde{\psi}_{-\frac{1}{2}}^{\pm\dot{A}}\rangle_{2}^{\dot{C})\mp}$\\$\big|\psi_{-\frac{1}{2}}^{+(\dot{B}}\psi_{-\frac{1}{2}}^{-\dot{C})}\tilde{\alpha}_{-\frac{1}{3}}^{A\dot{A}}\big\rangle_{3}$
		\end{tabular} \\\hline
		$0$
		&$0$ & 202 & All other states \\\hline
	\end{tabular}
	\caption{Anomalous dimensions $\gamma$ and constituent fields in sector $(1,1,0,0)$. The parentheses in the index structure denote symmetrisation and it only affects the two indices right next to them. In the third row we observe two closed degenerate subsectors related by left-right symmetry.} \label{table:firstsector}\end{table}

The following sectors in \eqref{eq:mixingsecs} have a growing number of distinct eigenvalues. Crucially, the anomalous dimensions found in sector $(1,1,0,0)$ reappear. This is expected since the RR-deformation does not break the superconformal symmetry of the theory. As a result the deformed states in $(1,1,0,0)$ generate long superconformal multiplets, such that their descendants will reappear in other sectors and have the same anomalous dimension. We demonstrate this effect by listing the anomalous dimensions and $SU(2)_\bullet\times  SU(2)_\circ$-representations of the sectors $(1,1,0,0)$ and $(1,\tfrac{3}{2},0,\tfrac{1}{2})$ in Table \ref{table:sectors1and2}. Acting with the supercharges $\tilde{G}_{-1/2}^{\dot{+}A}$ (defined in \eqref{eq: supercharge}) on states in sector $(1,1,0,0)$ will generate an $SU(2)_\bullet$-doublet of states in sector $(1,\tfrac{3}{2},0,\tfrac{1}{2})$. We find precisely these states in an explicit diagonalisation. 

\begin{figure}[h]
	\centering
	\includegraphics[width=1\textwidth]{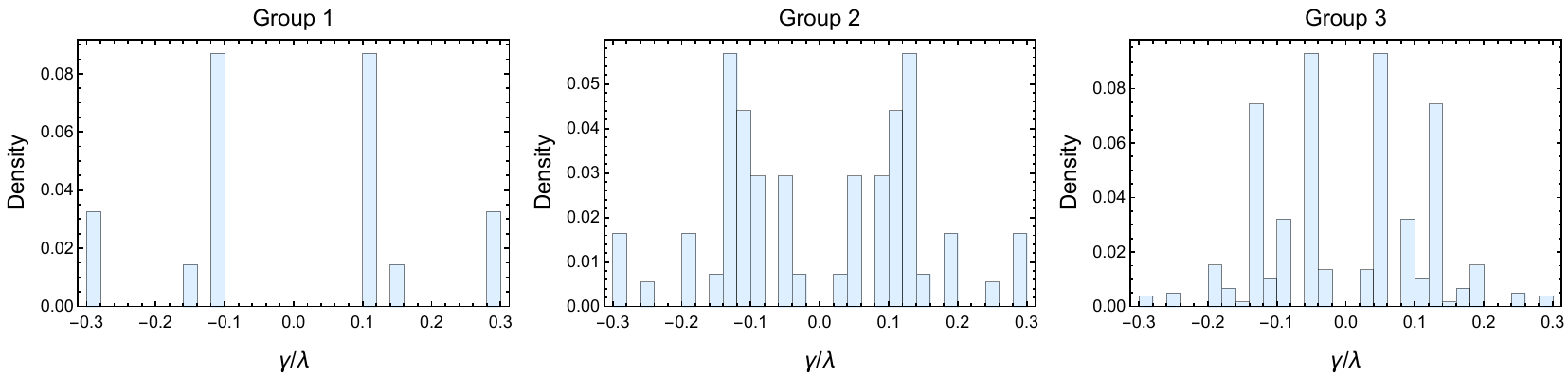}
    \includegraphics[width=0.666\textwidth]{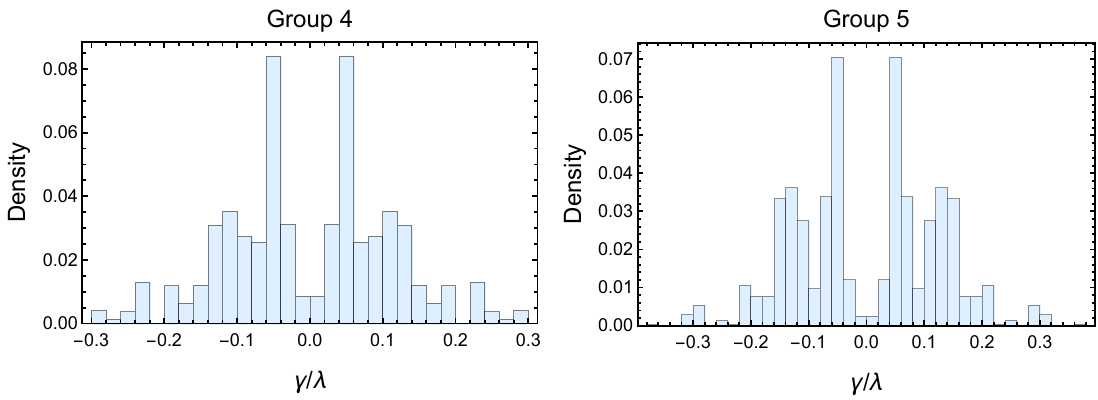}
	\caption{\label{fig:plot-eigenvalues}
    Density of non-vanishing eigenvalues of the mixing matrix, for each of the  five groups  of states listed in~\eqref{eq:mixingsecs}.}
\end{figure}

\begin{table}[t]
	\centering
    \small
	\renewcommand{\arraystretch}{1.5}
	\begin{tabular}{|c|c|c|c|}
    \hline
		Sector&$\gamma/\lambda$  & \# states & $SU(2)_\bullet\times SU(2)_\circ$ representation \\\hline\hline
		\multirow{4}{*}{\begin{tabular}{c}$(1,1	,0,0)$\end{tabular}}&$\pm 0.2966$
		&9 & $ (0,2)\oplus (0,1)\oplus (0,0)$ \\ \cline{2-4}
		&$\pm 0.1532$
		&4 & $  (0,1)\oplus (0,0)$ \\  \cline{2-4}
		&$\pm 0.1064$
		&24 & $ 2\times\left[(\frac{1}{2},\frac{3}{2})\oplus (\frac{1}{2},\frac{1}{2}) \right]$ \\ \cline{2-4}
		&$0$
		&202& Various representations \\ \hline\hline
		\multirow{15}{*}{\begin{tabular}{c}$(1,\tfrac{3}{2},0,\pm\tfrac{1}{2})$\\and\\$(\tfrac{3}{2},1,\pm\tfrac{1}{2},0)$\end{tabular}}&$\mathbf{\underline{\pm 0.2966}}$
		& \textbf{\underline{18}} & $ (\tfrac{1}{2},2)\oplus (\tfrac{1}{2},1)\oplus (\tfrac{1}{2},0)$ \\ \cline{2-4}
		&$\pm 0.2568$ 
		&6 & $(0,\tfrac{3}{2})\oplus (0,\tfrac{1}{2}) $ \\ \cline{2-4}
		&$\pm 0.1876$
		&18 & $ (\tfrac{1}{2},2)\oplus (\tfrac{1}{2},1)\oplus (\tfrac{1}{2},0)$ \\ \cline{2-4}
		&$\mathbf{\underline{\pm 0.1532}}$
		&\textbf{\underline{8}} & $  (\tfrac{1}{2},1)\oplus (\tfrac{1}{2},0)$ \\ \cline{2-4}
		&$\pm 0.1398$
		&6 & $(0,\tfrac{3}{2})\oplus (0,\tfrac{1}{2}) $ \\ \cline{2-4}
		&$\pm 0.1303$
		&54&$(1,\tfrac{5}{2}) \oplus(1,\tfrac{3}{2}) \oplus(1,\tfrac{1}{2}) \oplus (\tfrac{1}{2},1)\oplus (0,\tfrac{5}{2}) \oplus(0,\tfrac{3}{2}) \oplus(0,\tfrac{1}{2}) $ \\ \cline{2-4}
		&$\pm 0.1251$
		&2 & $(0,\tfrac{1}{2 }) $ \\ \cline{2-4}
		&$\mathbf{\underline{\pm 0.1064}}$
		&\textbf{\underline{48}} & $ 2\times\left[(1,\tfrac{3}{2})\oplus (1,\tfrac{1}{2}) \oplus (0,\tfrac{3}{2})\oplus(0,\tfrac{1}{2}) \right]$ \\ \cline{2-4}
		&$\pm 0.0971$
		&18 & $(2,\tfrac{1}{2})\oplus(1,\tfrac{1}{2})\oplus(0,\tfrac{1}{2}) $ \\ \cline{2-4}
		&$\pm 0.0921$
		&8& $ (\tfrac{1}{2},1)\oplus(\tfrac{1}{2},0)$ \\ \cline{2-4}
		&$\pm 0.0869$
		&6 & $ (\tfrac{1}{2},1)$ \\ \cline{2-4}
		&$\pm 0.0485$
		&24 & $(1,\tfrac{3}{2})\oplus(1,\tfrac{1}{2})\oplus(0,\tfrac{3}{2})\oplus(0,\tfrac{1}{2})$ \\ \cline{2-4}
		&$\pm 0.0483$
		&8 & $ (\tfrac{1}{2},1)\oplus(\tfrac{1}{2},0)$ \\ \cline{2-4}
		&$\pm 0.0357$
		&8& $(\tfrac{1}{2},1)\oplus(\tfrac{1}{2},0) $ \\ \cline{2-4}
		&$0$
		&626& Various representations \\ \hline
	\end{tabular}
    \normalsize
	\caption{Anomalous dimensions $\gamma$ and $SU(2)_\bullet\times SU(2)_\circ$ representations of states in the lightest non-trivial sectors. The highlighted states are descendants from the $(1,1,0,0)$ sector. Note that their representation is the tensor product of the $(1,1,0,0)$ states with an additional $SU(2)_{\bullet}$-doublet.  \label{table:sectors1and2}} 
    \end{table}

We present the full spectrum of the five distinct groups \eqref{eq:mixingsecs} in Figure~\ref{fig:plot-eigenvalues}.%
\footnote{Let us mention that already at $\Delta=3$ the numerical diagonalisation becomes quite challenging due to the large size of the blocks in the mixing matrix, introducing numerical errors in the eigenvalues. It is likely that this can be avoided by more carefully constructing smaller mixing blocks and using more sophisticated numerical techniques.}
For simplicity we are not going to show all the numerical value of the eigenvalues in a table, however we give them in the ancillary file in the \texttt{arXiv} submission. The numbers of perturbed states are presented in Table \ref{table:accounting}. We note that a  fraction of 12.4\% of physical states with bare dimension $\Delta\leq 3$ receive first-order corrections. We expect this fraction to decrease as $\Delta = h+\bar{h}$ increases --- the size of the mixing sectors grows with $\Delta$, but parametrically slower than the size of the Fock subspace with given $\Delta$ (see Figure~\ref{fig:sectors}). Nonetheless, infinitely many such $O(\lambda)$ mixing sectors appear as $\Delta$~grows.

\begin{table}[h!]
	\centering
	\renewcommand{\arraystretch}{1.5}
	\begin{tabular}{|c|c||c|c|c|}\hline
	group & $\Delta $&\# states &  \# deformed states & fraction \\\hline\hline
	 1 &2& $276$ & $74$ & 26.8\%\\ 		2 & $\frac{5}{2}$&$4\times  1090 $&$ 4\times  464$&42.6\% \\
	3 & 3&$4\times 2368 $& $4\times  1210$ & 51.1\%\\
		4 & 3& $4\times 6467$ &$4\times  3828$ & 59.2\%\\
		5 & 3& $2\times 8280 $&$2\times  4342$ & 52.4\%\\\hline\hline
	 All&$\leq3$ & $248778$ & $30766$ &12.4\% \\\hline
 	\end{tabular}
	\caption{Numbers of total and deformed states in sectors grouped as in \eqref{eq:mixingsecs}. We also give the total number of physical states that do  (or do not) receive corrections  at order $O(\lambda)$ with bare dimension $\Delta\leq3$.\label{table:accounting}}\end{table}

\section{Relation to integrability}
\label{sec:integrability}

The integrability approach to AdS3/CFT2 is most easily explained by starting from the worldsheet of the type IIB string on $AdS_3 \times  S^3 \times  T^4$ background, where it was first developed (see~\cite{Sfondrini:2014via,Demulder:2023bux,Seibold:2024qkh} for reviews). By fixing a suitable lightcone gauge along coordinates $X^\pm$ (whereby the density of the conjugate momentum~$P_-$ is constant), one obtains a non-conformal QFT in two dimensions, whose fields are the transverse excitations of a string of finite length~$R$. If the choice of the gauge fixing condition preserves as much supersymmetry as possible (half, in this case), the length can be identified with the  R-charge of a half-BPS state~$|\text{BPS}\rangle_{R}$. Integrability is manifest in the $R\to\infty$ limit of the theory, whereby the worldsheet decompactifies to a plane and  one can construct a factorised S-matrix. The two-to-two scattering matrix $S(p_1,p_2)$ is bootstrapped from the symmetries, with the exception of an overall scalar pre-factor (the ``dressing factor'') for which one can make an Ansatz based on unitarity, crossing symmetry, analyticity, and comparison with perturbative computations~\cite{Frolov:2021fmj,Frolov:2024pkz,Frolov:2025uwz,Frolov:2025tda}.

On the orbifold side of the duality, it is natural to identify $|\text{BPS}\rangle_{R}$ with one of the half-BPS states in Appendix \ref{app:chiral-ring}, constructed out of the $w$-cycle sector of the model with $w=R$. Hence, it is expected that integrability becomes manifest in the $w\to\infty$ limit. By representation theory it is clear that the transverse fields of the strings in lightcone gauge must be identified with the fields of the orbifold~\cite{Frolov:2023pjw}, and more specifically that worldsheet excitations (i.e., particles of momentum $p$) above the lightcone vacuum must be identified  with the creation operators on the $w$-cycle ``vacuum'' (with fractional modes $n/w$).

It is therefore natural to explore the $w\to\infty$ limit of the $O(\lambda)$ mixing matrix and see how its behaviour fits in the integrability picture. This will also give us an opportunity to revisit some of the arguments of~\cite{Gaberdiel:2023lco} and point out some important issues that must be addressed to link the orbifold model to the integrability structure. Finally, a more difficult (but crucial) question is what integrability predicts at finite~$w$ and $O(\lambda)$, and whether this is compatible with our results.

\subsection{Order-\texorpdfstring{$\lambda$}{lambda} mixing matrix in the \texorpdfstring{$w\to\infty$}{w->infty} limit}

Since the dimensions of the twist vacuum (\ref{eq: vacuum-dim-odd}--\ref{eq:vacuum-dim-even}) increase with $w$, the  large-$w$ limit requires the bare dimension $\Delta$ to be large; in turn, this implies dealing with a very large mixing problem, making explicit computations along the lines of Section~\ref{sec:results} unfeasible.
Of course while the number of states which may mix at $O(\lambda)$ grows, the total number of states grows even faster, see Figure~\ref{fig:sectors}. 

\begin{figure}[t]
  \centering
  \includegraphics[width=0.8\linewidth]{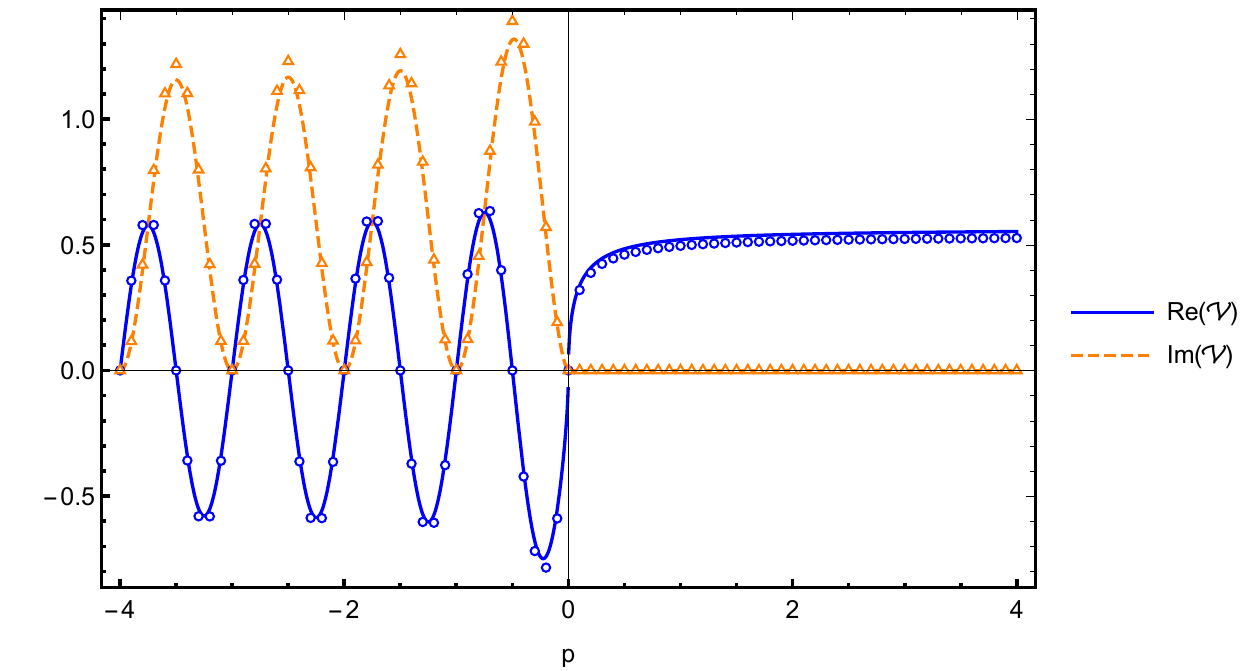}
 \caption{\label{fig:large-w} The asymptotic function $\mathcal{V}_\infty (p)$ as well as the $w=10$ real (circles) and imaginary parts (triangles) of appropriately rescaled $\mathcal{V}_{w}(n)$, respectively. We notice a good approximation already at relatively small twist $w$. The zeros at negative integer values are needed to cancel the poles in the propagator.}
\end{figure}

Let us now consider the large-$w$ behaviour of a generic mixing matrix element of the type~\eqref{eq:970115}. The most straightforward part is the twist-field structure function $C_{w\pm1,2,w}^\sigma$ which behaves as
\begin{equation}
    \lim_{w\to\infty} C_{w\pm1,2,w}=e^{1/4}2^{-5/8}\,.
\end{equation}
Consider now the boson contributions of Section~\ref{sec:boson-contribution}. 
In this case it is useful to work in terms of ``momenta'' $p_j = n_j/w$, rather than mode numbers $n_j$; this will allow  us to consider the scaling limit where $w\to\infty$ with $p_j$ fixed, which is natural in integrability (and more generally, at large-$w$).
The commutation relation \eqref{eq: commutator} suggests that the normalisation of the bosonic oscillators requires the introduction of factors $1/\sqrt{\lvert{p}\rvert}$ which are generically finite in the large-$w$ limit.\footnote{One may worry about the behaviour around $p\sim0$ but it will turn out that the correlators of interest cancel this divergence naturally.} With this normalisation we can compute the limit of the bosonic Wick contractions \eqref{eq:bos-wick1}-\eqref{eq:bos-wick5}. To do so we use the building blocks $\mathcal{V}_{w}(n)$ and $\mathcal{V}_{-(w+1)}(m)$ defined in~\eqref{eq:calVn} in Appendix~\ref{app:integrals-wick} so that the Wick contractions are  given by (\ref{eq:first-wick}--\ref{eq:last-wick}). Without loss of generality we focus on $w$ and $w+1$ mixing. By assuming unit-normalised bosonic oscillators we can write these blocks as a function of the momentum as:
\begin{equation}
    \mathcal{V}_w(p)=\sqrt{\frac{2}{\lvert{(w+1)p}\rvert}}\left(\frac{w}{w+1}\right)^{p (w+1)-1}w^{-p}\frac{\Gamma(p(w+1))}{\Gamma(p w)\Gamma(p)}\,,
\end{equation}
with a similar expression for $\mathcal{V}_{-(w+1)}(q)$ where $q=m/(w+1)$. By making use of the following elementary limits
\begin{equation}
    \lim_{w\to\infty}\left(\frac{w}{w+1}\right)^{p(w+1)}= e^{-p}\,,\qquad \lim_{w\to\infty}w^{-p}\frac{\Gamma(p(w+1))}{\Gamma(p w)}=p^p\,,
\end{equation}
one then finds the following asymptotic behaviour for the Wick contraction building blocks:
\begin{equation}
\label{eq:aympt1}
     \mathcal{V}_w(n)\ \rightarrow\ \frac{1}{\sqrt{w}}\, \mathcal{V}_\infty(p)\,,\qquad  \mathcal{V}_{-(w+1)}(m)\ \rightarrow\ \frac{1}{\sqrt{w}}\,{\mathcal{V}}_\infty(-q)\,, 
\end{equation}
where
\begin{equation}
    \mathcal{V}_\infty(p)=\sqrt{\frac{2}{\lvert{p}\rvert}}\frac{e^{-p}p^p}{\Gamma(p)}\,. 
\end{equation}
We can see in Figure \ref{fig:large-w} that the asymptotic function $\mathcal{V}_\infty$ is a good approximation of the finite result already for $w=10$.

The final ingredients which we need are the ``propagators'' appearing in Wick-contractions among in- and out-state excitations \eqref{eq:first-wick},\eqref{eq:third-wick},\eqref{eq:fifth-wick}. While contractions of two in-state or two out-state excitations among themselves only result in a finite factor, a contraction of one in- and one out-state excitation results in a contribution like
\begin{equation}\label{pole}
    \frac{1}{2}\frac{1}{\frac{n}{w}-\frac{m}{w+1}}\qquad\rightarrow\qquad\frac{1}{2}\frac{1}{p-q}\,,
\end{equation}
 which generates a pole at coinciding incoming and outgoing momenta. At finite $w$ these momenta are fractions with co-prime denominators, so they can only truly coincide at integer values of $p$ and $q$; in this case however,  the accompanying factor of $\mathcal{V}_\infty(-q)$ vanishes and cancels the pole (see fig.\ref{fig:large-w}). At large $w$ however, non-coincident values can get parametrically close, e.g. 
\begin{equation}
\label{eq:orderwcontraction}
    p=\frac{n}{w}\,,\quad q=\frac{n}{w+1}\,:\qquad \frac{1}{p-q}=\frac{1}{\frac{n}{w}-\frac{n}{w+1}}=\frac{w(w+1)}{n}=\frac{w}{q}\,,
\end{equation}
generating a resonance that contributes with another (large) factor $w$. However, despite such factors the correlation function \textit{remains finite} as $w\to\infty$. To see this, let us now discuss which $O(\lambda)$-processes are least suppressed at large-$w$. We start with states involving only two bosonic excitations. Schematically the boson contribution to three-point functions on the right-hand side of \eqref{eq:970115} behaves as
\begin{equation}\label{eq:largewprocess}
\begin{gathered}
     _{w+1}\langle{0\phantom{_j}\hspace{-2 pt}}\vert \mathcal{D}  \vert{\alpha_{-p} \tilde{\alpha}_{-\bar{p} }}\rangle_{w} =O(w^{-1})\,,\qquad
    _{w+1}\langle{\alpha_{-q} \tilde{\alpha}_{-\bar{q} }}\vert  \mathcal{D}  \vert{\,0\phantom{_j}\hspace{-3 pt}}\rangle_{w} =O(w^{-1})\,,\\
    _{w+1}\langle{\alpha_{-q}}\vert \mathcal{D} \vert{\tilde{\alpha}_{-\bar{p} }}\rangle_{w} =O(w^{-1})\,,
\end{gathered}\end{equation}
for finite momenta $p,\bar{p}, q,\bar{q}$. In this case both bosonic modes contract with the deformation operator, yielding a $1/w$ suppression. 
Moving on to states with more than two bosonic excitations, we notice that each additional excitation carries a normalisation factor of $1/\sqrt{w}$ from~\eqref{eq:aympt1}, generically suppressing the process at hand. This suppression is avoided for pairs of additional modes, one in the in- and one in the out-state, with approximately equal momentum like in~\eqref{eq:orderwcontraction} which instead give an overall $O(w^0)$ contribution. We are thus left with (at best) an overall $1/w$ behaviour. Hence, even on the ``poles''~\eqref{eq:orderwcontraction} the correlation function is finite as $w\to\infty$.

The final ingredient for our large-$w$ analysis are the fermion contributions of the mixing-matrix elements. The fermionic excitations will also be parameterised by asymptotically continuous momenta $p$ now defined as 
\begin{equation}
    p=\frac{n}{w}-\frac{1}{2}\,.
\end{equation}
As described in Section~\ref{sec:fermion-contribution}, we used bosonisation to compute fermion correlators, which slightly obscures the large-$w$ analysis. Suffice to say that the vertex-operator correlators of the type~\eqref{eq:702796asas} do not depend on $w$, neither do the cocycle contributions detailed in Appendix~\ref{app:bosonisation}. The only $w$-dependence arises from the mode integrals
\eqref{eq:6548651687}, which behave very similar to their bosonic counterparts. In fact, we could have employed a similar contraction scheme as for the bosons and found very similar expressions for fermion-fermion contractions~\cite{Gaberdiel:2023lco}. This suggests that excited fermion modes come with the same $1/\sqrt{w}$ suppression as in the bosonic case, which is only ameliorated if two fermionic modes in in- and out-state have parametrically close momenta and therefore hit a resonance. The leading fermionic contribution is therefore of order $O(w^0)$ if only such resonant pairs are excited. 

Collecting all ingredients, we have shown that the large-$w$ limit of a generic mixing-matrix element at $O(\lambda)$ is suppressed at least like $1/w$, and that this comes from bosonic contractions with the deformation operator. It is harder to estimate the precise scaling of the \textit{eigenvalues} of the mixing matrix, as the number of non-vanishing elements in each column is itself of order~$O(w)$. Generically, this would yield a scaling of the eigenvalues as ~$O(w^{-1/2})$, but it is likely that this estimate can be strengthened. Indeed, if we want to interpret these effects as the finite-volume (``wrapping''~\cite{Ambjorn:2005wa}) corrections of some integrable QFT, we would expect a scaling of $1/w$, see for instance~\cite{Brollo:2023pkl}. We will discuss such potential interpretations below.

{
It is also worth mentioning that our discussion of the $w\to\infty$ limit shares many similarities with the one used in~\cite{Gaberdiel:2023lco} for the purpose of deriving the symmetry algebra expected from integrability~\cite{Borsato:2012ud,Lloyd:2014bsa} for ``off-shell'' states (states that do not satisfy the orbifold-invariance condition, or in string theory the level-matching condition). Crucially, the algebra of~\cite{Borsato:2012ud,Lloyd:2014bsa} features a central extension (similar to Beisert's~\cite{Beisert:2005tm} for $\mathcal{N}=4$ SYM) which vanishes on on-shell states.
This is important because, in turn, this algebra yields the S-matrix of worldsheet excitations~\cite{Lloyd:2014bsa}, up to a dressing factor.%
\footnote{The derivation of the dressing factors is more involved, and was recently carried out in~\cite{Frolov:2024pkz,Frolov:2025uwz,Frolov:2025tda} for mixed-flux backgrounds.}
In repeating the computations of~\cite{Gaberdiel:2023lco}, we match the results for finite-$w$ physical (on-shell) states, but we cannot reproduce those for off-shell states at $w\to\infty$. 
Starting from the finite-$w$ results of~\cite{Gaberdiel:2023lco}, which are on-shell, we find that the limit $w\to \infty$ can be taken unambiguously and that it yields a trivial central extension. The treatment in~\cite{Gaberdiel:2023lco} instead combines the large-$w$ limit with a particular analytic continuation (in $q=m/(w+1)$, where $m\in\mathbb{N}$ is a mode number). 
This procedure should take the states off-shell and match the integrability results of~\cite{Borsato:2012ud,Lloyd:2014bsa}. Looking carefully at those steps, we could not find a mathematical justification for the prescription used in~\cite{Gaberdiel:2023lco}. It is therefore not clear to us that it provides a consistent way to perform off-shell computations. Rather, we believe that the definition of off-shell states requires a more substantial redefinition of the correlation functions --- more specifically, of the orbifold projection. This is likely to be necessary both to convincingly derive the off-shell symmetry algebra at $w\to\infty$ and, importantly, to perform more complicated computations for novel quantities, such as the perturbative computation of the dressing factors at small~$\lambda$. To elucidate these points, we present a rather thorough discussion of the results of~\cite{Gaberdiel:2023lco} in Appendix~\ref{app:GGN}. }

\subsection{Order-\texorpdfstring{$\lambda$}{lambda} anomalous dimensions at finite~\texorpdfstring{$w$}{w} from integrability}

Notwithstanding the issue with the derivation of~\cite{Gaberdiel:2023lco} described above, there is ample reason to believe that the perturbed symmetric-orbifold CFT may be described by integrability. Here we want to discuss the role of the $O(\lambda)$ corrections within the integrability picture.

In the limit $w\to\infty$ we expect that the $O(\lambda)$ mixing can be overlooked. The main contribution should then come from an ``S~matrix'' scattering the magnons on a $w=\infty$ vacuum. Once this S~matrix is known, one can use it to reverse-engineer the corrections to the finite-$w$ spectrum of the model, similarly to what was done in $\mathcal{N}=4$ SYM. In fact, the S-matrix is already known from the string worldsheet~\cite{Lloyd:2014bsa}, including the dressing factors which have been very recently proposed~\cite{Frolov:2024pkz,Frolov:2025uwz,Frolov:2025tda}. When $w$ is finite, the spectrum of excitation should become discrete. The momenta / mode numbers should be quantised according to the Bethe-Yang equations which very schematically are%
\footnote{Usually, in the integrability literature, the equations involve a term of the form~$e^{i p R}$ with $R=w$; here we are using a non-standard definition of the momenta, rescaled by $2\pi$.}
\begin{equation}
    1=e^{2\pi i p_j w}\prod_{k=1}^M S(p_j,p_k)\,,\qquad j=1,\dots,M\,,
\end{equation}
which can indeed be solved by the free-orbifold excitations plus corrections%
\footnote{More precisely, chiral and antichiral mode excitations should be matched with positive and negative momenta, respectively, see~\cite{Frolov:2023pjw}.}
\begin{equation}
    p_j = \frac{n_j}{w}+O(\lambda^2)\,,
\end{equation}
where we used that, as it is suggested from the orbifold perturbation theory, the S~matrix should have the small-$\lambda$ expansion $S(p,q)=1+O(\lambda^2)$.
The contribution to the lightcone energy (that is to $E=h-j+\bar{h}-\bar{\jmath}$) is then found from~\cite{Hoare:2013lja,Lloyd:2014bsa}
\begin{equation}
\label{eq:dispersion}
    E=\sum_{j=1}^M E(p_j)\,,\qquad
    E(p)=\sqrt{p^2+4\lambda^2\sin^2(\pi p)}=|p|+2\lambda^2\frac{\sin(\pi p)^2}{|p|}+O(\lambda^4)\,.
\end{equation}
We see that \textit{generically} the dimension should then be
\begin{equation}
    E=\sum_{j=1}^M \frac{|n_j|}{w}+O(\lambda^2)\,,
\end{equation}
where we recognise in the first term the free orbifold result. It seems difficult, in this setup, to generate $O(\lambda)$ corrections.

A possible way out may be to consider magnons with momentum $p=0+O(\lambda^2)$; however, bosonic zero-modes annihilate the vacuum and fermionic zero-modes just move us within the Clifford module of the even-$w$ vacuum. It is not obvious to us that this mechanism can explain the anomalous dimensions which we computed, especially as they must come with both positive and negative sign --- while~\eqref{eq:dispersion} would yield positive corrections only.

A different and more likely explanation may lie in the observation that the Bethe-Yang equations themselves must be corrected by finite-volume effects~\cite{Ambjorn:2005wa}, which for a theory involving gapless modes (such as this one) are expected to yield $O(1/w)$ corrections to the spectrum~\cite{Brollo:2023pkl}. Strictly speaking, these corrections must be found from the mirror thermodynamic Bethe ansatz or quantum spectral curve, which have been derived for pure-NSNS~\cite{Dei:2018mfl} and pure-RR~\cite{Ekhammar:2021pys,Cavaglia:2021eqr,Frolov:2021bwp} backgrounds, and they have been proposed only very recently for the case at hand~\cite{Frolov:2025tda}.
However, to illustrate the behaviour we might expect, let us consider a typical finite-volume correction of the type first derived by L\"uscher~\cite{Luscher:1985dn,Luscher:1986pf} (see~\cite{Janik:2010kd} for a pedagogical introduction). Very schematically, we may expect contributions of the form
\begin{equation}
\label{eq:luscher}
    \delta E \sim
    \int\limits_{-\infty}^{+\infty}d \tilde{q}\,e^{-w\tilde{H}(\tilde{q})}\, \tilde{S}(\tilde{q},p_1)\cdots \tilde{S}(\tilde{q},p_M)\,,
\end{equation}
where the excitation of momentum $\tilde{q}$ is in the \textit{mirror kinematics}~\cite{Arutyunov:2007tc} and the S~matrix has been analytically continued so that it has one ``leg'' in the mirror kinematics and the other in the physical (string) kinematics.%
\footnote{To be precise, we should consider the string-mirror transfer matrix~\cite{Seibold:2022mgg} and sum over all possible types of virtual particles.}
The mirror dispersion relation~\cite{Arutyunov:2007tc} in this case takes the form
\begin{equation}
    \tilde{H}(\tilde{q})=|\tilde{q}|+2\lambda^2\frac{\sinh^2(\pi \tilde{q})}{|\tilde{q}|}+O(\lambda^4)\,.
\end{equation}
We see that the order of $\delta E$ hinges on the precise form of mirror-string S~matrix; more specifically, we expect that the dependence should come through the dressing factors, which have been investigated only very recently~\cite{Frolov:2024pkz,Frolov:2025uwz,Frolov:2025tda}.%
\footnote{It is worth stressing that, even if the physical (string-string) S-matrix were of order $O(\lambda^2)$, this need not be the case for the mirror-string one, as the analytic continuation of the dressing factors is very non-trivial and can change the weak-tension scaling; this is what happens for $AdS_5\times  S^5$~\cite{Arutyunov:2009kf}.}
Regardless, it is very peculiar that this should be the case for a relatively small fraction of states, whose common feature is to have integer dimensions at order $O(\lambda^0)$; we see that the correction $\delta E$ does not really depend on the \textit{total} mode number of a state, but rather on the mode numbers of individual excitations. It might be possible to explain this behaviour if it turned out that all of the states that mix at order $O(\lambda)$ involve, say, some particle with special values of the momenta, so that $\tilde{S}(\tilde{q},p)$ is singular. This is what happens for the so-called exceptional operators of $\mathcal{N}=4$ SYM, see~\cite{Arutyunov:2012tx}, where the momenta of the physical excitations are configured such that a double pole pinches the real line in~\eqref{eq:luscher}. It will be important to revisit this question once the mirror-string S~matrix (and its singularities) are worked out.

\section{Conclusions and outlook}
\label{sec:conclusion}

We have studied the mixing matrix of the marginally deformed symmetric-product orbifold CFT of~$T^4$ at first order in conformal perturbation theory, and computed the relative anomalous dimensions by direct diagonalisation.
Only a small fraction of states receive corrections at $O(\lambda)$, while the vast majority of them is corrected at $O(\lambda^2)$. Nonetheless, we expect infinitely many states to receive $O(\lambda)$ corrections. In fact, the dimension of the mixing matrix grows quite fast with the bare scaling dimension $\Delta$, and we have therefore restricted our analysis to $\Delta\leq3$ as the explicit diagonalisation becomes computationally challenging otherwise.
To our knowledge, this is the first analysis of such $O(\lambda)$ corrections and it is likely that the analysis can be further refined by projecting out of the mixing matrix some uninteresting states (for instance, symmetry descendants of states whose anomalous dimension has been determined at smaller~$\Delta$). It would be interesting to see how much further such a computation can be pushed.

Aside from the general interest of such a perturbative computation for orbifold-CFT practitioners, our investigation is motivated by the importance of this model within AdS3/CFT2. More specifically, our aim is to better understand how integrability, which is present on the string worldsheet, should manifest itself in the perturbed orbifold-CFT. 
A first puzzle for integrability  comes from he fact that some states receive corrections at $O(\lambda)$ rather than $O(\lambda^2)$. Indeed this may appear surprising from integrability; the form of the dispersion relation~\eqref{eq:dispersion} would suggest that the natural perturbation parameter is $\lambda^2$, not $\lambda$. The probable resolution of this puzzle lies in the finite-volume (``wrapping'') corrections to the anomalous dimension. A full analysis would require knowing the mirror TBA equations for the model, which in turn would require knowing the dressing factors of the model.%
\footnote{The construction of the S~matrix~\cite{Borsato:2014hja,Baggio:2018gct}, dressing factors~\cite{Frolov:2021fmj}, and mirror TBA / QSC~\cite{Ekhammar:2021pys,Cavaglia:2021eqr,Frolov:2021bwp} has so far been completed for pure-NSNS and pure-RR models, but here we would need the mixed-flux case which is currently under development. Specifically, the S-matrix is known~\cite{Lloyd:2014bsa} up to the dressing factors,whose construction is being completed at the time of writing~\cite{Frolov:2024pkz,Frolov:2025uwz,Frolov:2025tda}.}
It is not impossible that some special states receive wrapping corrections at a lower order than expected, similar to the case of ``exceptional operators'' in $\mathcal{N}=4$ SYM~\cite{Arutyunov:2012tx}. However, reproducing the pattern of lifting that we have observed (not to mention, the explicit values of the anomalous dimensions) would be an absolutely remarkable test of the integrability construction.

{
In the course of our investigation, we have also revisited the large-$w$ limit performed in~\cite{Gaberdiel:2023lco}, see Appendix~\ref{app:GGN} for a thorough discussion. In fact, the identification of the integrability structure performed in~\cite{Gaberdiel:2023lco} and subsequently corrected by~\cite{Frolov:2023pjw} relies on perturbative computations similar to the ones performed here, but in the $w\to\infty$ limit and for off-shell states (i.e., states that do not satisfy the orbifold-invariance condition)\footnote{For comparison, in $\mathcal{N}=4$ this would be non-cyclic, i.e.\ non-gauge-invariant states.}.
At finite twist~$w$ and for physical states, our results reproduce those of~\cite{Gaberdiel:2023lco}. However we cannot reproduce their $w\to\infty$ limit for off-shell states, and the resulting symmetry algebra.%
\footnote{This symmetry algebra in turn can be used to determine the S-matrix of orbifold excitations, as proposed in~\cite{Gaberdiel:2023lco} and corrected in~\cite{Frolov:2023pjw}.}
The expectation from worldsheet integrability, which~\cite{Gaberdiel:2023lco} claims to reproduce, is the presence of a central extension in that algebra~\cite{Borsato:2012ud,Lloyd:2014bsa}, similar to Beisert's~\cite{Beisert:2005tm}. This central extension is necessarily zero on physical states --- this is why it is necessary to consider off-shell states.
As we discuss at some length in Appendix~\ref{app:GGN},  in~\cite{Gaberdiel:2023lco} the large-$w$ limit is combined with an analytic continuation of sorts --- a procedure which according to~\cite{Gaberdiel:2023lco} yields the expected central extensions, but for which  we cannot find a mathematical justification. 
We find instead that, starting like~\cite{Gaberdiel:2023lco} from finite-$w$ expressions for physical states, the $w\to\infty$ limit can be taken unambiguously and consistently, and that doing so yields no central extension.
To our mind, this apparent mismatch is the sign of a fundamental issue.
We believe that to convincingly reproduce the central extension of~\cite{Borsato:2012ud,Lloyd:2014bsa} it is necessary to modify the structure of the orbifold correlation functions in a more fundamental way to account for the possibility of states being off-shell --- most likely, the invariance under the cyclic group $\mathbb{Z}_w\subset S_N$ should be relaxed in an appropriate way, before taking $w\to\infty$.\footnote{Moreover, it may also be necessary to first perform the $w\to\infty$ limit and only later to integrate over the conformal perturbation, as that too requires a regularisation, see Appendix~\ref{app:GGN} for a further details.}}
While it might be possible to circumvent 
the proper definition of off-shell excitations (as done in~\cite{Gaberdiel:2024dfw}), it is unclear to us to what extent any such argument may be trusted or extended to the computation of more general off-shell quantities. 
Thus far, this treatment has been used to reproduce the known algebraic structure of integrability, which in itself is quite robust; a first-principle derivation of integrability, especially for more subtle properties (such as the small-$\lambda$ expansion of the dressing factor, which in principle should be obtainable from the deformed orbifold CFT) is likely to require a very careful definition of the off-shell states.
We hope to return to these questions in the near future.

\section*{Acknowledgements}

We are grateful to Sebastian Harris, Volker Schomerus and Roberto Volpato for helpful comments and discussions, and especially to Sergey Frolov for many helpful discussions and collaboration at early stages of this project. We also would like to thank Beat Nairz for related correspondence. We furthermore thank the participants of the Workshop \textit{Integrability in Low Supersymmetry Theories} in Trani (Italy) in 2024 for stimulating discussions that initiated and furthered this project.

\paragraph{Funding information}

MF, AS and TS acknowledge support from the EU -- NextGenerationEU, program STARS@UNIPD, under project \textit{Exact-Holography}, and from the PRIN Project n.~2022ABPBEY. MF and AS also acknowledge support from the CARIPLO Foundation under grant n.~2022-1886, and from the CARIPARO Foundation Grant under grant n.~68079.
AS\ thanks the MATRIX Institute in Creswick \& Melbourne, Australia, for support through a MATRIX Simons Fellowship in conjunction with the program \textit{New Deformations of Quantum Field and Gravity Theories}, as well as the IAS in Princeton for hospitality. TS would like to thank the Cluster of Excellence EXC 2121 Quantum Universe 390833306 and the Collaborative Research Center SFB1624 for creating a productive research environment at DESY.
\pagebreak
\begin{appendix}
\numberwithin{equation}{section}

\section{Conventions for the seed theory}
\label{app:free-field-conv}

The seed theory of the orbifold defined in Section \ref{sec:orbifold-def} is the free CFT of four real bosons $X^{i} (z,\bar{z})$, four real left-moving fermions $\psi^{i} (z)$ and four real right-moving fermions $\tilde{\psi}^{i} (\bar{z})$ (with $i\in\{1,\cdots,4\}$). This model has $\mathcal{N}=(4,4)$ supersymmetry with the central charge being $\mathbf{c}=6$. These fields have the usual OPEs
\begin{align}
    \label{eq:OPE-free-1}
    & \partial X^{i} (z) \partial X^{j} (w) \sim -\frac{\delta^{ij}}{(z-w)^2}\,,  \\
    \label{eq:OPE-free-2}
    & \psi^{i} (z) \psi^{j} (w) \sim \frac{\delta^{ij}}{z-w}\,,
\end{align}
with analogous expressions for the right-moving fields. We have an $SO(4)$ global symmetry which we split as $SO(4)=SU(2)_{\bullet}\times  SU(2)_{\circ}$. Then we can convert the fields to a bi-spinor notation by defining
\begin{align}
    & X^{A\dot{A}} (z,\bar{z}) = \frac{i}{\sqrt{2}} \sum_{k=1}^{4} \sigma_k^{A\dot{A}}\ X^{k}(z,\bar{z}) \,, \\
    & \psi^{\alpha\dot{A}} (z) = \frac{i}{\sqrt{2}} \sum_{k=1}^{4} \sigma_k^{\alpha\dot{A}}\ \psi^{k}(z)\,, \\
    & \tilde{\psi}^{\dot{\alpha}\dot{A}} (\bar{z}) = \frac{i}{\sqrt{2}} \sum_{k=1}^{4}   \sigma_k^{\dot{\alpha}\dot{A}}\ \tilde{\psi}^{k}(\bar{z})\,,
\end{align}
with $\sigma_k$ being the Pauli matrices and $\sigma_4 = i\mathbb{I}_2$. We denote $A=1,2$ as an $SU(2)_{\bullet}$-index, $\dot{A}=\dot{1},\dot{2}$ as an $SU(2)_{\circ}$-index, $\alpha=+,-$ as an $SU(2)_{L}$-index, and $\dot{\alpha}=\dot{+},\dot{-}$ as an $SU(2)_{R}$-index, where $SU(2)_{L} \times  SU(2)_{R}$ is the R-symmetry.  Using the identity 
\begin{equation}
\sum_{k=1}^{4} \sigma_k^{A\dot{A}} \sigma_k^{B\dot{B}} = - 2 \epsilon^{AB} \epsilon^{\dot{A}\dot{B}}\,,
\end{equation}
the OPEs \eqref{eq:OPE-free-1} and \eqref{eq:OPE-free-2} become
\begin{align}
& \partial X^{A\dot{A}} (z) \partial X^{B\dot{B}} (w) \sim -\frac{\epsilon^{AB} \epsilon^{\dot{A}\dot{B}}}{(z-w)^2}\,, \\
& \psi^{\alpha \dot{A}} (z) \psi^{\beta \dot{B}} (w) \sim \frac{\epsilon^{\alpha\beta} \epsilon^{\dot{A}\dot{B}}}{z-w}\,.
\end{align}
Note that the fermions are defined in the NS sector (periodic in the plane). We can also define bosonic left- and right-moving modes $\alpha^{A\dot{A}}_{n}$ and $\tilde{\alpha}^{A\dot{A}}_{n}$ and fermionic modes $\psi^{\alpha\dot{A}}_{n}$ and $\tilde{\psi}^{\dot{\alpha}\dot{A}}_{n}$. We finish this section with the conjugation properties of these modes
\begin{align}
\label{eq:bosons-conj}
& (\alpha^{A \dot{A}}_{n})^\dagger = \epsilon^{AB} \epsilon^{\dot{A}\dot{B}} \alpha^{B \dot{B}}_{-n}\,, \\
\label{eq:fermions-conj}
& (\psi^{\alpha \dot{A}}_n)^\dagger = \epsilon^{\alpha\beta} \epsilon^{\dot{A}\dot{B}} \psi_{-n}^{\beta \dot{B}}\,,
\end{align}
with an implicit sum of $B$, $\dot{B}$, and $\beta$ on the right-hand side. We have analogous expressions for the right-moving modes. Note that these relations ensure the positivity of two-point functions and they are inherited by the orbifold theory.

\section{Currents and charges of the symmetric-product orbifold}
\label{app:currents-and-charges}

The $\mathcal{N}=(4,4)$ currents of $\textrm{Sym}_N(T^4)$ are found by taking the diagonal element of $N$ copies of the usual $\mathcal{N}=(4,4)$ currents, yielding
\begin{align}
\label{eq:currents-1}
& T(z) = - \frac{1}{2} \epsilon_{AB}\epsilon_{\dot{A}\dot{B}} \sum_{I=1}^{N} : \partial X_I^{A\dot{A}} \partial X_I^{B\dot{B}} : + \frac{1}{2} \epsilon_{\alpha\beta}\epsilon_{\dot{A}\dot{B}} \sum_{I=1}^{N} : \psi_I^{\alpha \dot{A}} \partial\psi_I^{\beta \dot{B}} : \,, \\
\label{eq:currents-2}
&  J^{\alpha\beta}(z) = \frac{1}{2} \epsilon_{\dot{A}\dot{B}} \sum_{I=1}^{N} : \psi_I^{\alpha \dot{A}} \psi_I^{\beta \dot{B}} : \,, \\
\label{eq:currents-3}
& G^{\alpha A}(z) = \frac{i}{\sqrt{2}} \epsilon_{\dot{A}\dot{B}} \sum_{I=1}^{N} : \psi_I^{\alpha\dot{A}} \partial X_I^{A\dot{B}} : \,,
\end{align}
for the left-moving energy-momentum tensor, $SU(2)_L$ current and supercurrent, respectively, with analogous expressions for the right-moving currents. The $SU(2)_{\circ}$ and $SU(2)_{\bullet}$ currents do not separate into left- and right-moving components so we just write a total current for each
\begin{align}
    & J^{\dot{A}\dot{B}}(z,\bar{z}) = \frac{1}{2}  \sum_{I=1}^{N} \left(\epsilon_{\alpha\beta} : \psi_I^{\alpha \dot{A}} \psi_I^{\beta \dot{B}} : + \epsilon_{\dot{\alpha}\dot{\beta}} : \tilde{\psi}_I^{\dot{\alpha} \dot{A}} \tilde{\psi}_I^{\dot{\beta} \dot{B}} : + \epsilon_{AB} : X_I^{A\dot{A}}\bar{\partial} X_I^{B\dot{B}}: + \epsilon_{AB} : X_I^{A\dot{A}} \partial X_I^{B\dot{B}}: \right)\,, \\
    & J^{AB}(z,\bar{z}) = \frac{1}{2} \epsilon_{\dot{A}\dot{B}} \sum_{I=1}^{N} \left( :X_I^{A\dot{A}} \partial X_I^{B\dot{B}}: +  : X_I^{A\dot{A}} \partial X_I^{B\dot{B}} : \right)\,.
\end{align}
Due to their failure to split into holomorphic and anti-holomorphic parts and the appearance of the non-conformal field $X^{A\dot{A}}(z,\bar{z})$, these cannot be considered conformal currents. Nevertheless, we can assign $SU(2)_{\circ}$ and $SU(2)_{\bullet}$ charges for the modes as done in Table \ref{table:charges-of-modes}. More specifically, $SU(2)_{\circ}$ commutes with the superalgebra while $SU(2)_{\bullet}$ is an outer automorphism acting on $G^{\alpha A}$ as can be seen from the orbifold currents.

From (\ref{eq:currents-1}--\ref{eq:currents-3}) we can derive the appropriate mode decompositions in a twisted sector corresponding to a $w$-cycle. However, their explicit expressions are dependent on whether we act on a sector with even or odd twist $w$. Let us start from the simpler odd-$w$ case. For the left-moving Virasoro generators and R-charges we find
\begin{align}
    & L_k = \frac{1}{2} \sum_{\frac{m+n}{w} = k} \left( \epsilon_{AB}\epsilon_{\dot{A}\dot{B}} : \alpha_{\frac{n}{w}}^{A\dot{A}} \alpha_{\frac{m}{w}}^{B\dot{B}} : + \epsilon_{\alpha\beta}\epsilon_{\dot{A}\dot{B}} \frac{n}{w} : \psi_{\frac{m}{w}+\frac{1}{2}}^{\alpha \dot{A}} \psi_{\frac{n}{w}-\frac{1}{2}}^{\beta \dot{B}} : \right) + \frac{1}{4} \left(w-\frac{1}{w}\right) \delta_{k,0}\ , \\
    & J^{\alpha \beta}_k = \frac{1}{2} \sum_{\frac{m+n}{w} = k}  \epsilon_{\dot{A}\dot{B}} : \psi_{\frac{m}{w}+\frac{1}{2}}^{\alpha \dot{A}} \psi_{\frac{n}{w}-\frac{1}{2}}^{\beta \dot{B}} :\ ,
\end{align}
respectively. We have similar expressions for the right-moving generators. From these it becomes clear that the odd-$w$ twisted vacuum has dimensions given in \eqref{eq: vacuum-dim-odd} and also is uncharged under $SU(2)_L$ and $SU(2)_R$. However for even twist $w$ the Virasoro and R-symmetry generators are modified as
\begin{align}
    & L_k = \frac{1}{2}\sum_{\frac{m+n}{w} = k} \left( \epsilon_{AB}\epsilon_{\dot{A}\dot{B}} : \alpha_{\frac{n}{w}}^{A\dot{A}} \alpha_{\frac{m}{w}}^{B\dot{B}} : + \epsilon_{\alpha\beta}\epsilon_{\dot{A}\dot{B}} \frac{n}{w} : \psi_{\frac{m}{w}+\frac{1}{2}}^{\alpha \dot{A}} \psi_{\frac{n}{w}-\frac{1}{2}}^{\beta \dot{B}} : \right) + \frac{w}{4} \delta_{k,0}\ , \\
    & J^{\alpha \beta}_k = \frac{1}{2} \sum_{\frac{m+n}{w} = k} \epsilon_{\dot{A}\dot{B}} : \psi_{\frac{m}{w}+\frac{1}{2}}^{\alpha \dot{A}} \psi_{\frac{n}{w}-\frac{1}{2}}^{\beta \dot{B}} : + \frac{1}{4} \epsilon_{\dot{A}\dot{B}} \left(\psi_{0}^{\alpha \dot{A}} \psi_{0}^{\beta \dot{B}} -\psi_{0}^{\beta \dot{B}} \psi_{0}^{\alpha \dot{A}}\right)\delta_{k,0}\ ,
\end{align}
respectively. This means that the twisted vacuum now has dimensions \eqref{eq:vacuum-dim-even} and is charged under R-symmetry and $SU(2)_{\circ}$ due to fermionic zero modes (with its charges given in Table~\ref{table:even-vacuum-charges}). 

The supercharges have a common expression for odd and even twist $w$:
\begin{equation}\label{eq: supercharge}
    G_{r}^{\alpha A} = \frac{1}{\sqrt{2}}  \sum_{\frac{m+n}{w} = r + \frac{1}{2}} \epsilon_{\dot{A}\dot{B}} : \psi_{\frac{m}{w}-\frac{1}{2}}^{\alpha \dot{A}} \alpha_{\frac{n}{w}}^{A \dot{B}} :\ ,
\end{equation}
and analogously for the right-moving supercharges. The following two identities involving these charges are useful
\begin{align}
\label{eq:648952-1}
& \tilde{G}_{-1/2}^{\dot{\alpha}B} |0\rangle^{\beta\dot{\beta}}_{2} = -\frac{1}{\sqrt{2}} \epsilon_{\dot{M}\dot{N}} \epsilon^{\dot{\alpha}\dot{\beta}} \tilde{\alpha}^{B\dot{N}}_{-1/2} |0\rangle^{\beta\dot{M}}_{2}\,, \\
\label{eq:648952-2}
& G_{-1/2}^{\alpha A} |0\rangle^{\beta\dot{M}}_{2} = -\frac{1}{\sqrt{2}} \epsilon_{\dot{P}\dot{Q}} \epsilon^{\alpha\beta} \alpha^{A\dot{Q}}_{-1/2} |0\rangle^{\dot{P}\dot{M}}_{2}\,.
\end{align}
These are found by using the zero-mode action \eqref{eq:997119} and yield expression \eqref{eq:17944} for the deformation.

\section{Details on the chiral ring}
\label{app:chiral-ring}

\begin{figure}[t]
\centering
  \includegraphics[width=1\linewidth]{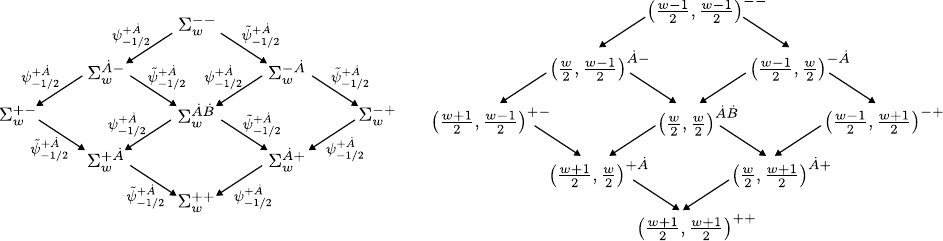}
\caption{Left: Structure of half-BPS operators and their definitions starting from $\Sigma^{--}_w$. Right: Dimensions $(h,\bar{h})$ (or equivalently R-charges $(j,\bar{\jmath})$) of the corresponding operators in the left figure. The indices $\dot{A}$ and $\dot{B}$ are $SU(2)_{\circ}$-indices.}
\label{fig:hal-BPS-multiplet}
\end{figure}

The chiral ring is a set of sixteen excited protected operators per twist $w$ defined as the states saturating the BPS bound
\begin{equation}
h = j \ \ \textrm{and} \ \ \bar{h} = \bar{\jmath}\,.
\end{equation}
Operators satisfying this constraint are half-BPS operators and have protected dimensions and protected structure constants in the entire moduli space of $AdS_3 \times  S^3 \times  T^4$ \cite{Baggio:2012rr}.
Their explicit definition depends on $w$ being even or odd. Let us start with the simpler odd-$w$ case. The main idea behind their construction is to add fermionic modes to the twisted vacuum to increase its R-charge and dimension to then saturate the BPS bound \cite{Lunin:2001pw}. The first half-BPS operator can be constructed as\footnote{We follow the notation of \cite{Pakman:2007hn}. However we remark that the $\pm$ indices here are not R-charge indices and are just labels for the half-BPS states.}
\begin{equation}
|\Sigma^{--}_{w}\rangle = \prod_{j=1}^{\frac{w-1}{2}} \left( \psi_{-\frac{1}{2}+\frac{j}{w}}^{+\dot{1}} \psi_{-\frac{1}{2}+\frac{j}{w}}^{+\dot{2}} \right) \prod_{j=1}^{\frac{w-1}{2}} \left( \tilde{\psi}_{-\frac{1}{2}+\frac{j}{w}}^{\dot{+}\dot{1}} \tilde{\psi}_{-\frac{1}{2}+\frac{j}{w}}^{\dot{+}\dot{2}} \right) |0\rangle_w\,,
\end{equation}
which satisfies the BPS bound and has dimensions
\begin{equation}
\Sigma^{--}_{w} : h=\bar{h}= \frac{w-1}{2}\,.
\end{equation}
To construct the remaining fifteen operators we can add the modes $\psi_{-1/2}^{+\dot{A}}$ and $\tilde{\psi}_{-1/2}^{\dot{+}\dot{A}}$ to $\Sigma^{--}_{w}$ and since these have equal R-charge and dimension they do not violate the BPS condition.  The operators of the full chiral ring are defined in Figure \ref{fig:hal-BPS-multiplet} together with their dimensions. For $w$ even, the definition changes slightly since now the twisted vacuum is charged. One then starts from the vacuum $|0\rangle^{+\dot{+}}_w$ and builds the lowest half-BPS operator as
\begin{equation}
|\Sigma^{--}_{w}\rangle = \prod_{j=1}^{\frac{w}{2} -1} \left( \psi_{-\frac{1}{2}+\frac{j}{w}}^{+\dot{1}} \psi_{-\frac{1}{2}+\frac{j}{w}}^{+\dot{2}} \right) \prod_{j=1}^{\frac{w}{2} -1} \left( \tilde{\psi}_{-\frac{1}{2}+\frac{j}{w}}^{\dot{+}\dot{1}} \tilde{\psi}_{-\frac{1}{2}+\frac{j}{w}}^{\dot{+}\dot{2}} \right) |0\rangle^{+\dot{+}}_w\,,
\end{equation}
which then has the same dimensions as in the odd-$w$ case. The rest of the chiral ring is obtained in the same way by adding fermionic modes.

From the form of the half-BPS operators we can derive that they are indeed protected at first order in perturbation theory. Charge conservation in the mixing matrix suggests that a half-BPS state could potentially mix only with another half-BPS state. However, as explicitly shown in this section, no half-BPS state has bosonic excitations and thus the bosonic contribution to their mixing matrix elements vanishes since one can never absorb the bosonic excitations of the deformation operator. We note that this proof works for all odd orders in perturbation theory.

\section{Integrated Wick contractions}
\label{app:integrals-wick}

Here we detail the computation of integrated Wick contractions described in Section \ref{sec:boson-contribution}. In general we have to compute the following integrals
\begin{align}
\label{eq:wickbb1}
& \mathcal{W}^{\pm}_{00}(n,m) = -\frac{1}{w} \oint_0 \frac{dt'}{2\pi i} \oint_0 \frac{dt}{2\pi i}\ \bfgamma_{\pm} (t')^{-m/w} \bfgamma_{\pm} (t)^{-n/w} \frac{1}{(t-t')^2}\,, \\
\label{eq:wickbb2}
& \mathcal{W}^{\pm}_{01}(n) = -\frac{1}{\sqrt{2w}} \oint_1 \frac{dt'}{2\pi i} \oint_0 \frac{dt}{2\pi i}\ (\bfgamma_{\pm} (t')-1)^{-1/2} \bfgamma_{\pm} (t)^{-n/w} \frac{1}{(t-t')^2}\,, \\
\label{eq:wickbb3}
& \mathcal{W}^{\pm}_{1\infty}(n) = -\frac{1}{\sqrt{2(w\pm 1)}} \oint_1 \frac{dt'}{2\pi i} \oint_{\infty} \frac{dt}{2\pi i}\ (\bfgamma_{\pm} (t')-1)^{-1/2} \bfgamma_{\pm} (t)^{n/(w\pm 1)} \frac{1}{(t-t')^2}\,, \\
\label{eq:wickbb4}
& \mathcal{W}^{\pm}_{0\infty}(n,m) =  -\frac{1}{\sqrt{w(w\pm 1)}} \oint_{\infty} \frac{dt'}{2\pi i} \oint_0 \frac{dt}{2\pi i}\ \bfgamma_{\pm} (t')^{m/(w\pm 1)} \bfgamma_{\pm} (t)^{-n/w} \frac{1}{(t-t')^2}\,, \\
\label{eq:wickbb5}
& \mathcal{W}^{\pm}_{\infty\infty}(n,m) = -\frac{1}{w\pm 1} \oint_{\infty} \frac{dt'}{2\pi i} \oint_{\infty} \frac{dt}{2\pi i}\ \bfgamma_{\pm} (t')^{m/(w\pm 1)} \bfgamma_{\pm} (t)^{n/(w\pm 1)} \frac{1}{(t-t')^2}\,. 
\end{align}
We will not show the computation of all these expressions, but instead focus on the more complicated case of $\mathcal{W}^{+}_{0\infty}(n,m)$ with the remaining ones being calculated in an analogous way. 

Let us consider \eqref{eq:wickbb4} for the $w$ and $w+1$ mixing, the integral around infinity is given by
\begin{multline}
\label{eq:87737}
\oint_{\infty} \frac{dt}{2\pi i}  \frac{\bfgamma_{+}(t)^{m/(w+1)}}{(t-t')^2} = (-w)^{m/(w+1)} (-1)^{m-1} \left( \frac{w+1}{w}\right)^{m-1} \times  \\
\sum_{a=0}^{m-1} (-1)^a (a+1) \binom{\frac{m}{w+1}}{m-a-1} \left( \frac{w}{w+1}\right)^{a} (t')^a\,.
\end{multline}
To find this result we combined binomial expansions with the residue theorem. Then by using the integral
\begin{equation}
\label{eq:5619872}
\oint_{0} \frac{dt'}{2\pi i}\ \bfgamma_{+}(t')^{-n/w} (t')^a = \frac{(-1)^{n-1-a}}{(w+1)^{n/w}} \left( \frac{w}{w+1} \right)^{n-1-a} \binom{-n/w}{n-1-a}\,,
\end{equation}
combined with \eqref{eq:87737} we find that
\begin{multline}
\mathcal{W}^{+}_{0\infty}(n,m) = \frac{1}{\sqrt{w(w+1)}} \frac{(-w)^{m/(w+1)}}{(w+1)^{n/w}} \frac{(-1)^{n-m}}{(n-1)!(m-1)!}\frac{1}{\left( \frac{n}{w} - \frac{m}{w+1} \right)} \left( \frac{w}{w+1} \right)^{n-m} \times  \\ \frac{\Gamma\left( 1-\frac{n}{w} \right) \Gamma\left( 1+\frac{m}{w+1} \right) }{\Gamma\left( 1-\frac{mw}{w+1} \right) \Gamma\left( 1-\frac{n(w+1)}{w} \right)}\,.
\end{multline}
For the remaining cases one has to compute analogues of \eqref{eq:87737} and \eqref{eq:5619872} around other insertions, which are found in the same way by combining binomial expansions and the residue theorem. It turns out that the various integrals lead to similar expressions in terms of the function 
\begin{equation}\label{eq:calVn}
    \mathcal{V}_w(n)=\sqrt{\frac{2}{\lvert{w+1}\rvert}}\left(\frac{w}{w+1}\right)^{n\frac{w+1}{w}-1}w^{-\frac{n}{w}}\frac{\Gamma(n+\frac{n}{w})}{\Gamma(n)\Gamma(\frac{n}{w})}\,.
\end{equation}
The $w\to w+1$ contractions of \eqref{eq:wickbb1}-\eqref{eq:wickbb5} are given by 
\begin{align}\label{eq:first-wick}
&\mathcal{W}^{+}_{00}(n,m) =-\frac{1}{2}\frac{\mathcal{V}_w(n)\mathcal{V}_w(m)}{\frac{n}{w}+\frac{m}{w}}\,,\\
&\mathcal{W}^{+}_{01}(n) =  i \mathcal{V}_w(n)\,,\\  
&\mathcal{W}^{+}_{0\infty}(n,m) =\frac{1}{2}\frac{\mathcal{V}_w(n)\mathcal{V}_{-(w+1)}(m)}{\frac{n}{w}-\frac{m}{w+1}}\,,\label{eq:third-wick}\\
&\mathcal{W}^{+}_{1\infty}(n) = -i  \mathcal{V}_{-(w+1)}(n)\,,\\ 
\label{eq:fifth-wick}
&\mathcal{W}^{+}_{\infty\infty}(n,m) = \frac{1}{2}\frac{\mathcal{V}_{-(w+1)}(n)\mathcal{V}_{-(w+1)}(m)}{\frac{n}{w+1}+\frac{m}{w+1}}\,.
\end{align}

The $w\to w-1$ contractions can be deduced from the $w\to w+1$ case by exchanging the points $0\leftrightarrow\infty$, shifting $w\rightarrow w-1$ and a complex conjugation
\begin{align}
& \mathcal{W}^{-}_{00}(n,m) = (\mathcal{W}^{+}_{\infty\infty}(n,m)|_{w\rightarrow w-1})^*\,,  \\
& \mathcal{W}^{-}_{01}(n) = (\mathcal{W}^{+}_{1\infty}(n)|_{w\rightarrow w-1})^*\,,  \\
& \mathcal{W}^{-}_{1\infty}(n) = (\mathcal{W}^{+}_{01}(n)|_{w\rightarrow w-1})^*\,,  \\
& \mathcal{W}^{-}_{0\infty}(n,m) = (\mathcal{W}^{+}_{0\infty}(m,n)|_{w\rightarrow w-1})^*\,, \\
& \mathcal{W}^{-}_{\infty\infty}(n,m) = (\mathcal{W}^{+}_{00}(n,m)|_{w\rightarrow w-1})^*\,.\label{eq:last-wick}
\end{align}
An explicit evaluation of the respective integrals confirms this matching.

\section{Bosonisation in covering space}
\label{app:bosonisation}

As explained in Section \ref{sec:fermion-contribution} the bosonisation procedure consists in introducing the left- and right-moving bosons $\phi_a (t)$ and $\tilde{\phi}_a (\bar{t})$ with $a=1,2$, respectively. These are unit normalised so that vertex operators have the usual dimensions
\begin{equation}
:e^{i\boldsymbol{\xi}\cdot\phi(t)}e^{i\boldsymbol{\chi}\cdot\tilde{\phi}(\bar{t})}:\  \rightarrow h=\frac{\boldsymbol{\xi}^2}{2}\,,\ \bar{h}=\frac{\boldsymbol{\chi}^2}{2}\,.
\end{equation}
The fermions and spin fields are then given by the following expressions 
\begin{align}
& \psi^{\alpha\dot{A}}(t) = C^{\alpha\dot{A}} :e^{i \boldsymbol{q}^{\alpha\dot{A}}\cdot\phi(t)}:\ , \\
& \tilde{\psi}^{\dot{\alpha}\dot{A}}(\bar{t}) = \tilde{C}^{\dot{\alpha}\dot{A}} :e^{i \boldsymbol{q}^{\dot{\alpha}\dot{A}}\cdot\tilde{\phi}(\bar{t})}: \ , \\
& \mathcal{S}^{xy}(t,\bar{t}) = Q^{xy} :e^{i \boldsymbol{\xi}^{x}\cdot\phi(t)} e^{i \boldsymbol{\xi}^{y}\cdot\tilde{\phi}(\bar{t})}:\ .
\end{align}
The polarisations are chosen as
\begin{align}
\label{eq:pol-1}
& \boldsymbol{q}^{\pm \dot{A}} = \frac{(\pm 1, (-1)^{\dot{A}+1})}{\sqrt{2}}\,, \\
\label{eq:pol-2}
& \boldsymbol{\xi}^{\pm} = \frac{(\pm 1, 0)}{\sqrt{2}}\,, \\
\label{eq:pol-3}
& \boldsymbol{\xi}^{\dot{A}} = \frac{(0, (-1)^{\dot{A}+1})}{\sqrt{2}}\,.
\end{align}
To check the charges of fermions and spin fields we need the bosonised currents. These are found to be\footnote{For the $SU(2)_{\circ}$  current we display here only the fermionic contribution to it.} 
\begin{align}
& J^{+-}(t) = \frac{i}{\sqrt{2}} \partial\phi_1 (t)\,, \\
& \tilde{J}^{+-}(\bar{t}) = \frac{i}{\sqrt{2}} \bar{\partial}\tilde{\phi}_1 (\bar{t})\,, \\
& J^{\dot{1}\dot{2}}(t,\bar{t}) = \frac{i}{\sqrt{2}} (\partial\phi_2 (t) + \bar{\partial}\tilde{\phi}_2 (\bar{t}) )\,.
\end{align}
One then finds the following charges for the bosonised operators
\begin{equation}
:e^{i\boldsymbol{\xi}\cdot\phi(t)}e^{i\boldsymbol{\chi}\cdot\tilde{\phi}(\bar{t})}: \rightarrow j = \frac{\xi_1}{\sqrt{2}}\,,\ \bar{\jmath}=\frac{\chi_1}{\sqrt{2}} ,\ j_{\circ} = \frac{\xi_2}{\sqrt{2}} + \frac{\chi_2}{\sqrt{2}}\,.
\end{equation}
With this we can confirm that the fermions and spin fields with polarisations \eqref{eq:pol-1}-\eqref{eq:pol-3} possess the correct charges in covering space.

To define the cocycle algebra we must specify the statistics of the fields. Note that fermionic zero modes can be used to flip between different Ramond vacua, which is just the lifted version of the relations \eqref{eq:997119}. It follows that eight of the sixteen twisted vacua are fermionic. Let $F_{xy}$ be the fermion number of the spin field $\mathcal{S}^{xy}$, we use the following fermion number assignment
\begin{align}
& F_{\alpha \dot{A}} = F_{\dot{A}\dot{\alpha}} = 1\,, \\
& F_{\dot{A}\dot{B}} = F_{\alpha\dot{\beta}} = 0\,,
\end{align}
which means that fermionic and bosonic Ramond vacua carry half-integer and integer total R-charge, respectively. Then we impose the following statistics 
\begin{align}
	\label{eq:422998}
	& \psi^{\alpha\dot{A}} (z)\mathcal{S}^{xy}(w,\bar{w}) = (-1)^{F_{xy}}\ \mathcal{S}^{xy}(w,\bar{w}) \psi^{\alpha\dot{A}} (z)\,, \\
	& \tilde{\psi}^{\dot{\alpha}\dot{A}} (\bar{z})\mathcal{S}^{xy}(w,\bar{w}) = (-1)^{F_{xy}}\  \mathcal{S}^{xy}(w,\bar{w}) \tilde{\psi}^{\dot{\alpha}\dot{A}} (\bar{z})\,,  \\
	\label{eq:72894}
	& \psi^{\alpha\dot{A}} (z) \psi^{\beta\dot{B}} (w) = - \psi^{\beta\dot{B}} (w) \psi^{\alpha\dot{A}} (z)\,, \\
	& \tilde{\psi}^{\dot{\alpha}\dot{A}} (\bar{z}) \tilde{\psi}^{\dot{\beta}\dot{B}} (\bar{w}) = - \tilde{\psi}^{\dot{\beta}\dot{B}} (\bar{w}) \tilde{\psi}^{\dot{\alpha}\dot{A}} (\bar{z})\,,  \\
	\label{eq:164508}
	& \psi^{\alpha\dot{A}} (z) \tilde{\psi}^{\dot{\beta}\dot{B}} (\bar{w}) = - \tilde{\psi}^{\dot{\beta}\dot{B}} (\bar{w}) \psi^{\alpha\dot{A}} (z)\,.
\end{align}
To ensure these, we require the following cocycle algebra
\begin{align}
\label{eq:cocycle-com1}
	& C^{\alpha\dot{A}} Q^{xy} = (-1)^{F_{xy}} i^{2 \boldsymbol{q}^{\alpha\dot{A}} \cdot \boldsymbol{\xi}^{x}} Q^{xy} C^{\alpha \dot{A}}\,, \\
	\label{eq:cocycle-com3}
	& \tilde{C}^{\dot{\alpha}\dot{A}} Q^{xy} = (-1)^{F_{xy}} i^{2\boldsymbol{q}^{\dot{\alpha}\dot{A}} \cdot \boldsymbol{\xi}^{y}} Q^{xy} \tilde{C}^{\dot{\alpha} \dot{A}}\,, \\
	\label{eq:cocycle-com4}
	& C^{\alpha\dot{A}} C^{\beta\dot{B}} = -(-1)^{\boldsymbol{q}^{\alpha \dot{A}} \cdot \boldsymbol{q}^{\beta \dot{B}}} C^{\beta\dot{B}}C^{\alpha\dot{A}}\,, \\
	\label{eq:cocycle-com5}
	& \tilde{C}^{\dot{\alpha}\dot{A}} \tilde{C}^{\dot{\beta}\dot{B}} = -(-1)^{\boldsymbol{q}^{\dot{\alpha} \dot{A}} \cdot \boldsymbol{q}^{\dot{\beta} \dot{B}}} \tilde{C}^{\dot{\beta}\dot{B}}\tilde{C}^{\dot{\alpha}\dot{A}}\,, \\
	\label{eq:cocycle-com6}
	& C^{\alpha\dot{A}} \tilde{C}^{\dot{\beta}\dot{B}} = - \tilde{C}^{\dot{\beta}\dot{B}} C^{\alpha\dot{A}} \,. 
\end{align}
Not all cocycles are fully independent since from the two-point functions we can derive annihilation conditions
\begin{align}
\label{eq:anni-cocycle-1}
& (C^{\alpha\dot{A}})^{\dagger} C^{\alpha\dot{A}} = 1\,, \\
\label{eq:anni-cocycle-2}
& (Q^{xy})^{\dagger} Q^{xy} = 1\,.
\end{align}
Since fermion conjugation was determined in \eqref{eq:fermions-conj}, we further require
\begin{align}
\label{eq:conj-cocycle-1}
& (C^{\alpha\dot{A}})^{\dagger} = \epsilon^{\alpha\beta}\epsilon^{\dot{A}\dot{B}} C^{\beta\dot{B}}\,, \\
\label{eq:conj-cocycle-2}
& (\tilde{C}^{\dot{\alpha}\dot{A}})^{\dagger} = \epsilon^{\dot{\alpha}\dot{\beta}}\epsilon^{\dot{A}\dot{B}} \tilde{C}^{\dot{B}\dot{B}}\,, \\
\label{eq:conj-cocycle-3}
& (Q^{xy})^{\dagger} = \epsilon^{xz}\epsilon^{yw} Q^{zw}\,.
\end{align}
The final set of relations that these cocycle factors have to satisfy comes from the zero-mode relations \eqref{eq:997119}. From these we infer
\begin{equation}
\label{eq:591402}
\begin{split}
& C^{\alpha\dot{A}} Q^{\beta y} = -\epsilon^{\alpha \beta} Q^{\dot{A}y}\,, \\
& C^{\alpha\dot{A}} Q^{\dot{B}y} = \epsilon^{\dot{A}\dot{B}} Q^{\alpha y}\,,
\end{split}
\end{equation}
with similar relations for right-moving cocycles. These relations allow us to connect distinct cocycles of spin fields by annihilating indices with fermion cocycles. 

This set of conditions is sufficient for all computations done in this work. For instance, we can derive that the cocycle contribution of the fermion factor \eqref{eq:fermion-example} in the example \eqref{eq:example-comp} is simply
\begin{equation}
Q^{\dot{1}\dot{1}} Q^{\dot{1}\dot{1}} C^{-\dot{2}} C^{+\dot{2}} \tilde{C}^{\dot{-}\dot{2}} \tilde{C}^{\dot{+}\dot{2}} = 1\,.
\end{equation}

\section{Comments on the large-\texorpdfstring{$w$}{w} analysis of Gaberdiel-Gopakumar-Nairz}
\label{app:GGN}

{
As discussed in Section~\ref{sec:integrability}, our results can be taken to large~$w$ without ambiguity. A similar large-$w$ analysis was performed in~\cite{Gaberdiel:2023lco}. }
At finite $w$, our expressions for Wick contractions match the ones of~\cite{Gaberdiel:2023lco}. However, we cannot reproduce their large-$w$ limit, e.g., for the Wick contraction \eqref{eq:example-wick} and the results which follow from it. We would like to comment on that result at some length.%
\footnote{We thank the referees for suggesting to expand this discussion.}

\paragraph{Generalities of the computation.}
Ref.~\cite{Gaberdiel:2023lco} aims at determining the action of supersymmetry generators on states with e.g.\ two bosonic excitations on top of the BPS operators defined in Appendix~\ref{app:chiral-ring}. The form of this representation is what then fixes (up to the dressing factors) the two-to-two S~matrix, and it is known from the worldsheet analysis of~\cite{Sfondrini:2014via} (see also~\cite{Hoare:2013lja}).
A crucial expectation from string theory~\cite{Lloyd:2014bsa} is that, due to the RR deformation, the symmetries of the lightcone gauge-fixed model involve a non-trivial anticommutator between chiral and anti-chiral supercharges (similarly to what was famously found by Beisert in $\mathcal{N}=4$ SYM~\cite{Beisert:2005tm} and correspondingly by Arutyunov, Frolov, Plefka and Zamaklar in $AdS_5\times  S^5$~\cite{Arutyunov:2006ak}).
This central extension is meant to vanish on physical states, i.e.\ on states which satisfy the string theory level-matching condition or, in the case at hand, the orbifold-invariance condition~\eqref{eq:orbifoldinvariance}.
The presence of such a non-vanishing anticommutator (e.g.\ $\{S_1,\tilde{S}_2\}$ in \cite{Gaberdiel:2023lco}) is necessary to match the single-magnon dispersion relation known from integrability~\cite{Hoare:2013lja,Lloyd:2014bsa}, and eventually the two-to-two S~matrix. To determine the action of these symmetries on the orbifold excitations, the authors of \cite{Gaberdiel:2023lco} compute a quantity which we dub~$\mathcal{C}(n_i,m_i,w)$ and which (in their notation) is defined as%
\footnote{The mode number used in~\cite{Gaberdiel:2023lco} are shifted with respect to those we defined above.}
\begin{equation}
\label{eq:centralex}
\mathcal{C}(n_i,m_i,w)=\langle{\textrm{BPS}_{w+1}}\vert\alpha^2\left(-\frac{m_1}{w+1}\right)\alpha^2\left(-\frac{m_2}{w+1}\right)\ \big\{S_1,\,\tilde S_2\big\}\  \alpha^2\left(\frac{n_1}{w}\right)\alpha^2\left(\frac{n_2}{w}\right)\vert{\textrm{BPS}_w}\rangle\,.
\end{equation}
The interpretation is that one acts with the central extension $\{S_1,\,\tilde S_2\big\}$ on a two-excitation state of the $w$-twist sector (labelled by $n_1,n_2$), which yields a number of possible states in the $(w+1)$-twist sector, labelled by $m_1,m_2$, over which one has to sum.
Explicitly,~\eqref{eq:centralex} takes the form
\begin{equation}\label{eq:centr}
\mathcal{C}(n_i,m_i,w)=\delta_{\frac{n_1+n_2}{w},\frac{m_1+m_2}{w+1}}
    \left(\frac{1}{\frac{n_1}{w}-\frac{m_1}{w+1}}+\frac{1}{\frac{n_1}{w}-\frac{m_2}{w+1}}+\frac{1}{\frac{n_2}{w}-\frac{m_1}{w+1}}+\frac{1}{\frac{n_2}{w}-\frac{m_2}{w+1}}\right)\,\mathcal{F}(n_i,m_i,w)\,,
\end{equation}
where $\mathcal{F}(n_i,m_i,w)$ is a regular function of the mode numbers and of~$w$, which is known explicitly and contains the functions~$\mathcal{V}_w(n)$ introduced above,  but is too bulky to report here.
Here, the Kronecker $\delta$ is a consequence of $SL(2)$ invariance of the correlation function under consideration (i.e., conservation of the scaling dimension). Technically, it arises by performing the integral over the position~$(y,\bar{y})$ of the deforming operator $\mathcal{D}(y,\bar{y})$, as discussed in Section~3.4 of~\cite{Gaberdiel:2023lco}.

\paragraph{Vanishing of the central extension at finite~$w$.}
The correlator is hence supported on the space where
\begin{equation}
\label{eq:constrain}
    \frac{n_1}{w}=-\frac{n_2}{w}+\frac{m_1+m_2}{w+1}\,.
\end{equation}
There is no way to relax this constraint in the correlation function computation, as it is necessary for the action of the supercharges to be compatible with $SL(2)$. 
For finite~$w$, eq.~\eqref{eq:constrain} also implies the orbifold-invariance condition (or in string theory, the level matching condition)
\begin{equation}
\label{levelmatch}
   n_1=-n_2~\text{mod}~w\,, 
\end{equation}
because $w$ and $w+1$ are coprime and we are solving the constraint \textit{over the integers}.
Hence, at finite $w$ imposing $SL(2)$ invariance means imposing level matching, which in turns means that the central extension vanishes. Indeed, on the constraint~\eqref{eq:constrain}, the bracket in~\eqref{eq:centr} vanishes identically, so that
\begin{equation}
\label{eq:Cn-is-zero}
\mathcal{C}(n_i,m_i,w)=0\,,\qquad \forall n_i,m_i,w\in\mathbb{N}\,,
\end{equation}
as also pointed out in appendix C in \cite{Gaberdiel:2023lco}.

\paragraph{Looking for a non-vanishing central extensions at infinite~$w$.}
In general, a crucial requirement of the AdS/CFT integrability construction is to deal with \textit{off-shell} states, which do not satisfy the level-matching condition --- that is the cyclicity (gauge-invariance) condition in $\mathcal{N}=4$ SYM, or in this case the orbifold-invariance condition \eqref{eq:orbifoldinvariance}. This is because we are interested in building the two-to-two S-matrix $S(p_1,p_2)$ from representation theory, and use it to construct a factorised multi-excitation scattering process. Even if we are dealing with $M$ excitations $p_1,\dots p_M$ which constitute a physical state (and all together satisfy level matching), pairs of excitations will not satisfy level matching.
In this case, there is clearly a difficulty in imposing the $SL(2)$ invariance condition \textit{without imposing level matching}. As discussed at length in~\cite{Arutyunov:2006ak}, the off-shell algebra is expected to be manifest in the infinite-volume of the lightcone gauge-fixed worldsheet model, which here corresponds to taking
\begin{equation}
\label{eq:largewlimit}
    n_i =p_i\,w\,,\qquad m_i=q_i(w+1)\,,\qquad p_i,\ q_i\quad\text{fixed}\,,\qquad w\to\infty\,.
\end{equation}
Quite rightly, Ref.~\cite{Gaberdiel:2023lco} considers this limit; what we cannot reproduce is \textit{how} the limit is taken at the level of~\eqref{eq:centralex} or~\eqref{eq:centr}. Namely, we find that the limit~\eqref{eq:largewlimit} turns~\eqref{eq:centr} into
\begin{equation}
\label{eq:limitresult}
\begin{aligned}
\mathcal{C}(n_i,m_i,w)\to\mathcal{C}(p_i,q_i)=&\,\delta(p_1+p_2-q_1-q_2)\left(\frac{1}{p_1-q_1}+\frac{1}{p_1-q_2}+\frac{1}{p_2-q_1}+\frac{1}{p_2-q_2}\right)\\
&\times \frac{e^{p_1+p_2-q_1-q_2}}{\pi^4}e^{i\pi(p_1+p_2)}\frac{q_1^{q_1}q_2^{q_2}}{p_1^{p_1}p_2^{p_2}}\sqrt{\frac{p_1p_2}{q_1q_2}}\Gamma(p_1)\Gamma(p_2)\\
&\times \Gamma(1-q_1)\Gamma(1-q_2)\sin(\pi p_1) \sin(\pi p_2) \sin(\pi q_1) \sin(\pi q_2)\,,
\end{aligned}
\end{equation}
where we spelled out the limit of~$\mathcal{F}(n_i,m_i,w)$ to make the expression completely explicit, though bulky. After the limit, the sums over $m_1,m_2$ turn into integrals over~$q_1,q_2$ (with $q_i\in[0,+\infty]$), and the Kronecker~$\delta$ into a Dirac~$\delta$; the relative measure factors have already been accounted for in~\eqref{eq:limitresult}.
The expression is regular everywhere, and in fact it vanishes on the support of the Dirac $\delta$-function, except at the poles~$p_i=q_j$, where it is not defined and necessitates a regularisation. 

\paragraph{Regularising the large-$w$ limit.}
The simplest way to regularise the poles is the principal value prescription,
\begin{equation}
    \frac{1}{p_i-q_j}\quad\to\quad PV\left(\frac{1}{p_i-q_j}\right)\,.
\end{equation}
By using the symmetry properties of the integrand under exchange of the integration variables~$q_i$, it is easy to check that with this prescription
\begin{equation}
    \mathcal{C}(p_i,q_j)=0\,.
\end{equation}
This is quite natural, given that~\eqref{eq:Cn-is-zero} held for any $n_i,m_i,w$, however large.
One could have attempted to give an $i\varepsilon$-prescription, and in fact we could pick different choices for each of the poles%
\footnote{We thank Beat Nairz for discussions around this point in private correspondence.}
\begin{equation}
    \frac{1}{p_i-q_j}\quad\to\quad \frac{1}{p_i-q_j+i\varsigma_{ij}\varepsilon}\,,\qquad
    \varsigma_{ij}=\pm1\,,\qquad \varepsilon>0\,.
\end{equation}
Splitting the pole into its principal value and a Dirac $\delta$, we find that only the latter contributes
\begin{equation}
\begin{aligned}
    \mathcal{C}(p_i,q_j)=\,&\left[(\varsigma_{12}+\varsigma_{21})\delta(p_1-q_2)\delta(p_2-q_1)
    +(\varsigma_{11}+\varsigma_{22})\delta(p_1-q_1)\delta(p_2-q_2)\right]\\
    &\times \frac{e^{i\pi(q_1+q_2)}}{i\pi}\sin(\pi q_1)\sin(\pi q_2)\,,
\end{aligned}
\end{equation}
which is not necessarily zero, but anyway does not match the integrability result~\cite{Lloyd:2014bsa}. Moreover, if $\varsigma_{ij}$ are chosen so that the result is non-vanishing, it is not consistent with the requirement that $\mathcal{C}=0$ on all on-shell states, i.e.\ whenever $p_1+p_2=0$ mod~$1$. 
We conclude that a standard principal-value regularisation procedure yields $\mathcal{C}=0$, while an $i\varepsilon$-prescription which yields a non-vanishing result is not physical. Of course one could explore other regularisations, but we limit ourselves to discussing the approach of~\cite{Gaberdiel:2023lco}.

\paragraph{The manipulations of~\cite{Gaberdiel:2023lco}.}
The main claim of \cite{Gaberdiel:2023lco} is that the quantity~\eqref{eq:centr} yields the wanted result if a special large-$w$ limit is employed. That procedure is tantamount to identifying  expressions involving $q_j=m_j/(w+1)$ and treating $q_j$ as a real numbers \textit{before taking the limit} $w\to\infty$. Although this procedure may disentangle~\eqref{levelmatch} from~\eqref{eq:constrain}, it seems quite arbitrary because it essentially amounts to taking the $w\to\infty$ limit at different times in different parts of the very same function (and, as we will see, \textit{in different ways}).
Let us demonstrate this issue by referring to a few steps in the derivation in \cite{Gaberdiel:2023lco}. For instance, the authors encounter the expression $(-1)^{m}\sin(\pi \tfrac{m w}{w+1})$ at finite~$w$.
It is a trigonometric identity that
\begin{equation}\label{identity}
    (-1)^{m}\sin(\pi \frac{m w}{w+1})=\sin(-\pi \frac{m}{w+1})\,,\qquad m,w \in \mathbb{Z}.
\end{equation}
To take limit~\eqref{eq:largewlimit}, the authors single out $q=m/(w+1)$ in the l.h.s., and treat it as a real number; then they argue that the limit of the l.h.s.\ is highly oscillatory, a fact that they use in later manipulations. In reality, the limit was perfectly well-defined to begin with, as it can be seen from the expression on the r.h.s..
To be more explicit: it is always possible to write
\begin{equation}
\label{cheat}
    1=e^{2\pi i m}=e^{2\pi i q (w+1)}\,,\qquad q=\frac{m}{w+1}\,,\qquad m,w\in\mathbb{Z}\,,
\end{equation}
and ``continue'' $q$ to a real number to obtain an oscillatory factor --- essentially taking the limit $w\to\infty$ in $q=m/(w+1)$ but not in $w$ itself. Ref.~\cite{Gaberdiel:2023lco} employs such manipulations repeatedly, see e.g.
\begin{align}
    \lim_{w\to\infty}\frac{\sin(\pi q w)}{\pi(q-\tfrac{n}{w})}=(-1)^n \delta(q-\tfrac{n}{w})\,, &&\text{(5.7) in \cite{Gaberdiel:2023lco}}\,.
\end{align}
Here, the $w,n\to\infty$ limit (with $n/w$ and $m/(w+1)$ fixed) has supposedly been taken but $n$ and $w$ on the right-hand side still appear to be finite. Moreover, the r.h.s.\ is meaningless when $w,n\to\infty$. Another example are the expressions
\begin{align}
    \sqrt{\frac{w+1}{w}}e^{-i\pi \frac{n}{w}}(-1)^{\frac{n(w+1)}{w}}\int dq f_w(q)\phi(q)&&\text{(5.5) in \cite{Gaberdiel:2023lco}}\,,\\
    f_w(q)=\frac{\sin(\pi q w)}{\pi(q-\tfrac{n}{w})}&&\text{(5.6) in \cite{Gaberdiel:2023lco}}\,,
\end{align}
where again $q\in\mathbb{R}$ as in~\eqref{eq:largewlimit} but both $n$ and $w$ remain finite.
The culmination of these manipulations is the claim that the presence of oscillatory factors (which as we saw, are not there when the limit is taken by conventional means) allows one to arbitrarily insert a phase factor in their computation. This leads to their statement that, in the large-$w$ limit~\eqref{eq:largewlimit} the Wick contractions take \textit{either of two values}
\begin{equation}
\label{eq:twolimits}
\wick{ \c{\bar{\alpha}^{1}(q)}\,\c{\alpha^{2}(p)}}=
\wick{ \c{\bar{\psi}^{+}(q)}\,\c{\psi^{-}(p)}}=\delta(p-q)\times \begin{cases}
1\\
e^{-2\pi i p}
\end{cases}\,,\qquad \text{(5.2) in \cite{Gaberdiel:2023lco}},
\end{equation}
whereas \textit{in any consistent limit and regularisation procedure, a given function }(here, the Wick contraction, which is completely fixed at finite~$w$) \textit{should have one limit, not two}.
From there, the authors argue that they can pick one or the other value in various parts of their computation in order to  obtain the desired result known from worldsheet integrability~\cite{Lloyd:2014bsa}.

\paragraph{Physical interpretation and possible solutions.}
In a nutshell, we found that the large-$w$ limit~\eqref{eq:largewlimit} of the expression~\eqref{eq:centr} can be regularised straightforwardly by a standard principal value prescription and it gives zero (as it was the case at finite~$w$). We could not find a consistent regularisation procedure which gives the result of~\cite{Lloyd:2014bsa}, and we regard the procedure of~\cite{Gaberdiel:2023lco} as mathematically not sound.%
\footnote{%
Let us mention that in the more recent work~\cite{Gaberdiel:2024dfw} a different attempt to go off-shell was made, by introducing ``inert'' excitations. These are creation modes acting on the twisted vacuum  which carry $h,\bar{h}$ charge, but on which the symmetry generators do not act (even though in principle they should). This alternative sidestep the problematic manipulations described above, but it unclear whether it is consistent with the orbifold CFT and its perturbation theory.
}
However, \textit{we are not claiming that there is no central extension in the off-shell symmetry algebra of the orbifold CFT!} Rather, we suggest that \textit{the computation of~\cite{Gaberdiel:2023lco} is not able to deal with off-shell states} (those which do not obey the cyclicity constraint), and that is why the central extension vanishes even in the limit~\eqref{eq:largewlimit}. We find two possible reasons for this.
\begin{enumerate}
    \item On the string worldsheet it is easy to relax the level-matching condition; conversely, in the orbifold theory, this is not as easy because orbifold invariance is intimately ``baked into'' the theory by the very definition of the twisted sectors, as it is necessary to sum over permutations, see e.g.~\eqref{eq:811071}. Invariance under the $\mathbb{Z}_w$ cyclic subgroup of the permutation group~$S_N$ enforces level matching. It would be necessary to understand whether it is possible to relax the sum over~$\mathbb{Z}_w$, perhaps by dealing with infinite-length cycles from the get-go (like one does in $\mathcal{N}=4$ SYM).
    \item More technically, the $SL(2)$ invariance condition imposed by the Kronecker~$\delta$ in~\eqref{eq:centr} is itself the result of a regularisation (of the integral over the deforming operator). Since this is the condition which makes the result vanish, it would be interesting to see if this can be avoided by swapping the order of the regularisations: taking the large-$w$ limit~\eqref{eq:largewlimit} in the correlator before integration, and only after perform the integrals. This of course makes the computation technically more challenging.
\end{enumerate}

\paragraph{Comments on finite $w$.}
We would like to reiterate that we find no issue in the results of~\cite{Gaberdiel:2023lco} as far as they deal with on-shell states at finite~$w$.
In particular, the computation of the anomalous dimensions of on-shell magnons in Figure~1 of~\cite{Gaberdiel:2023lco} matches perfectly the expectation from integrability; indeed the plot fits the perturbative expansion of  the dispersion relation found in~\cite{Hoare:2013lja,Lloyd:2014bsa}.
This is remarkable evidence of integrability in the perturbed orbifold, and an important result of~\cite{Gaberdiel:2023lco}! However, this confirmation adds no validity to the large-$w$ ``limit'' of~\cite{Gaberdiel:2023lco}, as it was obtained at finite-$w$ and for on-shell states, without relying on any of the problematic large-$w$ manipulations described above.

\paragraph{Why this matters.}
To find the centrally extended-algebra needed for integrability, we started from a function $\mathcal{C}(n_i,m_i,w)$, which is zero on all positive integers. In a sense we have been asking how to extend it to as a function on the positive real numbers. As always, there are infinitely many ways of doing so, by making use of identities such as~\eqref{cheat} in selected places. The authors of~\cite{Gaberdiel:2023lco} have picked two such possible ways, corresponding to the two ``limits'' in~\eqref{eq:twolimits} and by using either or the other in selected parts of their computation they can reproduce the integrability result of~\cite{Lloyd:2014bsa}. We explained above why this is mathematically inconsistent.
Yet, if they eventually get to the expected result, one may ask: \textit{Where is the harm?}
First of all, we find it necessary to emphasise that a proper derivation of integrability structure from the orbifold CFT is still an important outstanding problem.
Secondly but equally importantly, the issues in the approach of~\cite{Gaberdiel:2023lco} will cast doubt on any future result derived in a similar manner. While the structure of the central extension is very robust and constrained,%
\footnote{See the discussion in~\cite{Borsato:2012ud} where the central extension of a closely-related model (with essentially the same algebra) was derived based on algebraic consistence only.}
other infinite-$w$ results are much more delicate. In particular, the magnon S~matrix is only partially fixed by symmetry but its pre-factors (the ``dressing factors'') are not. Recently, such dressing factors have been proposed based on crossing symmetry and other symmetries such parity, unitarity, etc., as well as based on matching perturbative computations at large string tension~\cite{Frolov:2024pkz,Frolov:2025uwz,Frolov:2025tda}.
To our mind, a computation of the S~matrix and dressing  factors done using the manipulations of~\cite{Gaberdiel:2023lco} could not be trusted, and the same goes for any computation which requires dealing with off-shell states at infinite~$w$.

\end{appendix}
\bibliographystyle{JHEP}
\bibliography{mixing-matrix-AdS3.bib}

@article{Gaberdiel:2023lco,
    author = "Gaberdiel, Matthias R. and Gopakumar, Rajesh and Nairz, Beat",
    title = "{Beyond the tensionless limit: integrability in the symmetric orbifold}",
    eprint = "2312.13288",
    archivePrefix = "arXiv",
    primaryClass = "hep-th",
    doi = "10.1007/JHEP06(2024)030",
    journal = "JHEP",
    volume = "06",
    pages = "030",
    year = "2024"
}

@article{Gaberdiel:2024nge,
    author = "Gaberdiel, Matthias R. and Lichtner, Felix and Nairz, Beat",
    title = "{Anomalous dimensions in the symmetric orbifold}",
    eprint = "2411.17612",
    archivePrefix = "arXiv",
    primaryClass = "hep-th",
    month = "11",
    year = "2024"
}

@article{Gaberdiel:2024dfw,
    author = "Gaberdiel, Matthias R. and Kempel, Dennis and Nairz, Beat",
    title = "{AdS$_3\times$S$^3$ magnons in the symmetric orbifold}",
    eprint = "2412.02741",
    archivePrefix = "arXiv",
    primaryClass = "hep-th",
    month = "12",
    year = "2024"
}

@article{Keller:2019yrr,
    author = "Keller, Christoph A. and Zadeh, Ida G.",
    title = "{Conformal Perturbation Theory for Twisted Fields}",
    eprint = "1907.08207",
    archivePrefix = "arXiv",
    primaryClass = "hep-th",
    doi = "10.1088/1751-8121/ab6b91",
    journal = "J. Phys. A",
    volume = "53",
    number = "9",
    pages = "095401",
    year = "2020"
}

@article{Dei:2019iym,
    author = "Dei, Andrea and Eberhardt, Lorenz",
    title = "{Correlators of the symmetric product orbifold}",
    eprint = "1911.08485",
    archivePrefix = "arXiv",
    primaryClass = "hep-th",
    doi = "10.1007/JHEP01(2020)108",
    journal = "JHEP",
    volume = "01",
    pages = "108",
    year = "2020"
}

@article{Pakman:2009mi,
    author = "Pakman, Ari and Rastelli, Leonardo and Razamat, Shlomo S.",
    title = "{A Spin Chain for the Symmetric Product CFT(2)}",
    eprint = "0912.0959",
    archivePrefix = "arXiv",
    primaryClass = "hep-th",
    reportNumber = "BROWN-HET-1590, YITP-SB-09-41",
    doi = "10.1007/JHEP05(2010)099",
    journal = "JHEP",
    volume = "05",
    pages = "099",
    year = "2010"
}

@article{Arutyunov:1997gi,
    author = "Arutyunov, G. E. and Frolov, S. A.",
    title = "{Four graviton scattering amplitude from S**N R**8 supersymmetric orbifold sigma model}",
    eprint = "hep-th/9712061",
    archivePrefix = "arXiv",
    doi = "10.1016/S0550-3213(98)00326-5",
    journal = "Nucl. Phys. B",
    volume = "524",
    pages = "159--206",
    year = "1998"
}

@article{Pakman:2009zz,
    author = "Pakman, Ari and Rastelli, Leonardo and Razamat, Shlomo S.",
    title = "{Diagrams for Symmetric Product Orbifolds}",
    eprint = "0905.3448",
    archivePrefix = "arXiv",
    primaryClass = "hep-th",
    reportNumber = "BROWN-HEP-1573, YITP-SB-09-11",
    doi = "10.1088/1126-6708/2009/10/034",
    journal = "JHEP",
    volume = "10",
    pages = "034",
    year = "2009"
}

@article{Pakman:2007hn,
    author = "Pakman, Ari and Sever, Amit",
    title = "{Exact N=4 correlators of AdS(3)/CFT(2)}",
    eprint = "0704.3040",
    archivePrefix = "arXiv",
    primaryClass = "hep-th",
    reportNumber = "YITP-SB-07-15, BRX-TH-587",
    doi = "10.1016/j.physletb.2007.06.041",
    journal = "Phys. Lett. B",
    volume = "652",
    pages = "60--62",
    year = "2007"
}

@article{Borsato:2014exa,
    author = "Borsato, Riccardo and Ohlsson Sax, Olof and Sfondrini, Alessandro and Stefanski, Bogdan",
    title = "{Towards the All-Loop Worldsheet S Matrix for $AdS_3\times S^3\times T^4$}",
    eprint = "1403.4543",
    archivePrefix = "arXiv",
    primaryClass = "hep-th",
    reportNumber = "IMPERIAL-TP-OOS-2014-01, HU-MATHEMATIK-2014-05, HU-EP-14-12, SPIN-14-11, ITP-UU-14-10",
    doi = "10.1103/PhysRevLett.113.131601",
    journal = "Phys. Rev. Lett.",
    volume = "113",
    number = "13",
    pages = "131601",
    year = "2014"
}

@article{David:2002wn,
    author = "David, Justin R. and Mandal, Gautam and Wadia, Spenta R.",
    title = "{Microscopic formulation of black holes in string theory}",
    eprint = "hep-th/0203048",
    archivePrefix = "arXiv",
    reportNumber = "TIFR-TH-02-07",
    doi = "10.1016/S0370-1573(02)00271-5",
    journal = "Phys. Rept.",
    volume = "369",
    pages = "549--686",
    year = "2002"
}

@book{polchinski1998string,
  title={String Theory: An introduction to the bosonic string},
  author={Polchinski, J.G.},
  number={v. 1},
  isbn={9780521633031},
  lccn={98004545},
  series={Cambridge monographs on mathematical physics},
  url={https://books.google.it/books?id=jbM3t_usmX0C},
  year={1998},
  publisher={Cambridge University Press}
}

@article{Baggio:2012rr,
    author = "Baggio, Marco and de Boer, Jan and Papadodimas, Kyriakos",
    title = "{A non-renormalization theorem for chiral primary 3-point functions}",
    eprint = "1203.1036",
    archivePrefix = "arXiv",
    primaryClass = "hep-th",
    doi = "10.1007/JHEP07(2012)137",
    journal = "JHEP",
    volume = "07",
    pages = "137",
    year = "2012"
}

@article{Benjamin:2021zkn,
    author = "Benjamin, Nathan and Keller, Christoph A. and Zadeh, Ida G.",
    title = "{Lifting 1/4-BPS states in AdS$_{3}$\texttimes{} S$^{3}$\texttimes{} T$^{4}$}",
    eprint = "2107.00655",
    archivePrefix = "arXiv",
    primaryClass = "hep-th",
    doi = "10.1007/JHEP10(2021)089",
    journal = "JHEP",
    volume = "10",
    pages = "089",
    year = "2021"
}

@article{Fiset:2022erp,
    author = "Fiset, Marc-Antoine and Gaberdiel, Matthias R. and Naderi, Kiarash and Sriprachyakul, Vit",
    title = "{Perturbing the symmetric orbifold from the worldsheet}",
    eprint = "2212.12342",
    archivePrefix = "arXiv",
    primaryClass = "hep-th",
    doi = "10.1007/JHEP07(2023)093",
    journal = "JHEP",
    volume = "07",
    pages = "093",
    year = "2023"
}

@article{Hughes:2023apl,
    author = "Hughes, Marcel R. R. and Mathur, Samir D. and Mehta, Madhur",
    title = "{Lifting of two-mode states in the D1-D5 CFT}",
    eprint = "2309.03321",
    archivePrefix = "arXiv",
    primaryClass = "hep-th",
    doi = "10.1007/JHEP01(2024)183",
    journal = "JHEP",
    volume = "01",
    pages = "183",
    year = "2024"
}

@article{Apolo:2022fya,
    author = "Apolo, Luis and Belin, Alexandre and Bintanja, Suzanne and Castro, Alejandra and Keller, Christoph A.",
    title = "{Deforming symmetric product orbifolds: a tale of moduli and higher spin currents}",
    eprint = "2204.07590",
    archivePrefix = "arXiv",
    primaryClass = "hep-th",
    doi = "10.1007/JHEP08(2022)159",
    journal = "JHEP",
    volume = "08",
    pages = "159",
    year = "2022"
}

@article{Gaberdiel:2015uca,
    author = "Gaberdiel, Matthias R. and Peng, Cheng and Zadeh, Ida G.",
    title = "{Higgsing the stringy higher spin symmetry}",
    eprint = "1506.02045",
    archivePrefix = "arXiv",
    primaryClass = "hep-th",
    reportNumber = "BRX-TH-6297",
    doi = "10.1007/JHEP10(2015)101",
    journal = "JHEP",
    volume = "10",
    pages = "101",
    year = "2015"
}

@article{Gava:2002xb,
    author = "Gava, Edi and Narain, K. S.",
    title = "{Proving the PP wave / CFT(2) duality}",
    eprint = "hep-th/0208081",
    archivePrefix = "arXiv",
    doi = "10.1088/1126-6708/2002/12/023",
    journal = "JHEP",
    volume = "12",
    pages = "023",
    year = "2002"
}

@article{Guo:2022ifr,
    author = "Guo, Bin and Hughes, Marcel R. R. and Mathur, Samir D. and Mehta, Madhur",
    title = "{Universal lifting in the D1-D5 CFT}",
    eprint = "2208.07409",
    archivePrefix = "arXiv",
    primaryClass = "hep-th",
    doi = "10.1007/JHEP10(2022)148",
    journal = "JHEP",
    volume = "10",
    pages = "148",
    year = "2022"
}

@article{Guo:2019ady,
    author = "Guo, Bin and Mathur, Samir D.",
    title = "{Lifting of level-1 states in the D1D5 CFT}",
    eprint = "1912.05567",
    archivePrefix = "arXiv",
    primaryClass = "hep-th",
    doi = "10.1007/JHEP03(2020)028",
    journal = "JHEP",
    volume = "03",
    pages = "028",
    year = "2020"
}

@article{Guo:2020gxm,
    author = "Guo, Bin and Mathur, Samir D.",
    title = "{Lifting at higher levels in the D1D5 CFT}",
    eprint = "2008.01274",
    archivePrefix = "arXiv",
    primaryClass = "hep-th",
    doi = "10.1007/JHEP11(2020)145",
    journal = "JHEP",
    volume = "11",
    pages = "145",
    year = "2020"
}

@article{Hampton:2018ygz,
    author = "Hampton, Shaun and Mathur, Samir D. and Zadeh, Ida G.",
    title = "{Lifting of D1-D5-P states}",
    eprint = "1804.10097",
    archivePrefix = "arXiv",
    primaryClass = "hep-th",
    doi = "10.1007/JHEP01(2019)075",
    journal = "JHEP",
    volume = "01",
    pages = "075",
    year = "2019"
}

@article{Lima:2020boh,
    author = "Lima, A. A. and Sotkov, G. M. and Stanishkov, M.",
    title = "{Microstate Renormalization in Deformed D1-D5 SCFT}",
    eprint = "2005.06702",
    archivePrefix = "arXiv",
    primaryClass = "hep-th",
    doi = "10.1016/j.physletb.2020.135630",
    journal = "Phys. Lett. B",
    volume = "808",
    pages = "135630",
    year = "2020"
}

@article{Lima:2020kek,
    author = "Lima, A. A. and Sotkov, G. M. and Stanishkov, M.",
    title = "{Renormalization of twisted Ramond fields in D1-D5 SCFT$_{2}$}",
    eprint = "2010.00172",
    archivePrefix = "arXiv",
    primaryClass = "hep-th",
    doi = "10.1007/JHEP03(2021)202",
    journal = "JHEP",
    volume = "03",
    pages = "202",
    year = "2021"
}

@article{Lima:2020nnx,
    author = "Lima, A. A. and Sotkov, G. M. and Stanishkov, M.",
    title = "{Correlation functions of composite Ramond fields in deformed D1-D5 orbifold SCFT$_2$}",
    eprint = "2006.16303",
    archivePrefix = "arXiv",
    primaryClass = "hep-th",
    doi = "10.1103/PhysRevD.102.106004",
    journal = "Phys. Rev. D",
    volume = "102",
    number = "10",
    pages = "106004",
    year = "2020"
}

@article{Lima:2020urq,
    author = "Lima, A. A. and Sotkov, G. M. and Stanishkov, M.",
    title = "{Dynamics of R-neutral Ramond fields in the D1-D5 SCFT}",
    eprint = "2012.08021",
    archivePrefix = "arXiv",
    primaryClass = "hep-th",
    doi = "10.1007/JHEP07(2021)211",
    journal = "JHEP",
    volume = "07",
    pages = "211",
    year = "2021"
}

@article{Lima:2021wrz,
    author = "Lima, A. A. and Sotkov, G. M. and Stanishkov, M.",
    title = "{On the dynamics of protected ramond ground states in the D1-D5 CFT}",
    eprint = "2103.04459",
    archivePrefix = "arXiv",
    primaryClass = "hep-th",
    doi = "10.1007/JHEP07(2021)120",
    journal = "JHEP",
    volume = "07",
    pages = "120",
    year = "2021"
}

@article{AlvesLima:2022elo,
    author = "Alves Lima, Andre and Sotkov, G. M. and Stanishkov, M.",
    title = "{Four-point functions with multi-cycle fields in symmetric orbifolds and the D1-D5 CFT}",
    eprint = "2202.12424",
    archivePrefix = "arXiv",
    primaryClass = "hep-th",
    doi = "10.1007/JHEP05(2022)106",
    journal = "JHEP",
    volume = "05",
    pages = "106",
    year = "2022"
}

@article{Sfondrini:2014via,
    author = "Sfondrini, Alessandro",
    title = "{Towards integrability for ${\rm Ad}{{{\rm S}}_{{\bf 3}}}/{\rm CF}{{{\rm T}}_{{\bf 2}}}$}",
    eprint = "1406.2971",
    archivePrefix = "arXiv",
    primaryClass = "hep-th",
    reportNumber = "HU-MATHEMATIK-2014-14, HU-EP-14-24",
    doi = "10.1088/1751-8113/48/2/023001",
    journal = "J. Phys. A",
    volume = "48",
    number = "2",
    pages = "023001",
    year = "2015"
}

@article{Lloyd:2014bsa,
    author = "Lloyd, Thomas and Ohlsson Sax, Olof and Sfondrini, Alessandro and Stefa\'nski, Jr., Bogdan",
    title = "{The complete worldsheet S matrix of superstrings on AdS$_3 \times$ S$^3 \times$ T$^4$ with mixed three-form flux}",
    eprint = "1410.0866",
    archivePrefix = "arXiv",
    primaryClass = "hep-th",
    reportNumber = "IMPERIAL-TP-OOS-2014-04, HU-MATHEMATIK-2014-21, HU-EP-14-34",
    doi = "10.1016/j.nuclphysb.2014.12.019",
    journal = "Nucl. Phys. B",
    volume = "891",
    pages = "570--612",
    year = "2015"
}

@article{Frolov:2023pjw,
    author = "Frolov, Sergey and Sfondrini, Alessandro",
    title = "{Comments on integrability in the symmetric orbifold}",
    eprint = "2312.14114",
    archivePrefix = "arXiv",
    primaryClass = "hep-th",
    doi = "10.1007/JHEP08(2024)179",
    journal = "JHEP",
    volume = "08",
    pages = "179",
    year = "2024"
}

@article{Brollo:2023pkl,
    author = "Brollo, Alberto and le Plat, Dennis and Sfondrini, Alessandro and Suzuki, Ryo",
    title = "{Tensionless Limit of Pure\textendash{}Ramond-Ramond Strings and AdS3/CFT2}",
    eprint = "2303.02120",
    archivePrefix = "arXiv",
    primaryClass = "hep-th",
    doi = "10.1103/PhysRevLett.131.161604",
    journal = "Phys. Rev. Lett.",
    volume = "131",
    number = "16",
    pages = "161604",
    year = "2023"
}

@article{Seibold:2022mgg,
    author = "Seibold, Fiona K. and Sfondrini, Alessandro",
    title = "{Transfer matrices for AdS3/CFT2}",
    eprint = "2202.11058",
    archivePrefix = "arXiv",
    primaryClass = "hep-th",
    reportNumber = "Imperial-TP-FS-2022-01",
    doi = "10.1007/JHEP05(2022)089",
    journal = "JHEP",
    volume = "05",
    pages = "089",
    year = "2022"
}

@article{Arutyunov:2012tx,
    author = "Arutyunov, Gleb and Frolov, Sergey and Sfondrini, Alessandro",
    title = "{Exceptional Operators in N=4 super Yang-Mills}",
    eprint = "1205.6660",
    archivePrefix = "arXiv",
    primaryClass = "hep-th",
    reportNumber = "ITP-UU-12-22, SPIN-12-20, TCD-MATH-12-05, HMI-12-02",
    doi = "10.1007/JHEP09(2012)006",
    journal = "JHEP",
    volume = "09",
    pages = "006",
    year = "2012"
}

@article{Janik:2010kd,
    author = "Janik, Romuald A.",
    title = {{Review of AdS/CFT Integrability, Chapter III.5: L\"uscher Corrections}},
    eprint = "1012.3994",
    archivePrefix = "arXiv",
    primaryClass = "hep-th",
    doi = "10.1007/s11005-011-0511-z",
    journal = "Lett. Math. Phys.",
    volume = "99",
    pages = "277--297",
    year = "2012"
}

@article{Arutyunov:2009kf,
    author = "Arutyunov, Gleb and Frolov, Sergey",
    title = "{The Dressing Factor and Crossing Equations}",
    eprint = "0904.4575",
    archivePrefix = "arXiv",
    primaryClass = "hep-th",
    reportNumber = "ITP-UU-09-17, SPIN-09-17, TCDMATH-09-12, HMI-09-06",
    doi = "10.1088/1751-8113/42/42/425401",
    journal = "J. Phys. A",
    volume = "42",
    pages = "425401",
    year = "2009"
}

@article{Arutyunov:2007tc,
    author = "Arutyunov, Gleb and Frolov, Sergey",
    title = "{On String S-matrix, Bound States and TBA}",
    eprint = "0710.1568",
    archivePrefix = "arXiv",
    primaryClass = "hep-th",
    reportNumber = "ITP-UU-07-50, SPIN-07-37, TCDMATH-07-15",
    doi = "10.1088/1126-6708/2007/12/024",
    journal = "JHEP",
    volume = "12",
    pages = "024",
    year = "2007"
}

@article{Ambjorn:2005wa,
    author = "Ambjorn, Jan and Janik, Romuald A. and Kristjansen, Charlotte",
    title = "{Wrapping interactions and a new source of corrections to the spin-chain/string duality}",
    eprint = "hep-th/0510171",
    archivePrefix = "arXiv",
    reportNumber = "NORDITA-2005-67",
    doi = "10.1016/j.nuclphysb.2005.12.007",
    journal = "Nucl. Phys. B",
    volume = "736",
    pages = "288--301",
    year = "2006"
}

@article{Luscher:1986pf,
    author = "Luscher, M.",
    title = "{Volume Dependence of the Energy Spectrum in Massive Quantum Field Theories. 2. Scattering States}",
    reportNumber = "DESY-86-034",
    doi = "10.1007/BF01211097",
    journal = "Commun. Math. Phys.",
    volume = "105",
    pages = "153--188",
    year = "1986"
}

@article{Luscher:1985dn,
    author = "Luscher, M.",
    title = "{Volume Dependence of the Energy Spectrum in Massive Quantum Field Theories. 1. Stable Particle States}",
    reportNumber = "DESY-85-144",
    doi = "10.1007/BF01211589",
    journal = "Commun. Math. Phys.",
    volume = "104",
    pages = "177",
    year = "1986"
}

@article{Maldacena:1997re,
    author = "Maldacena, Juan Martin",
    title = "{The Large $N$ limit of superconformal field theories and supergravity}",
    eprint = "hep-th/9711200",
    archivePrefix = "arXiv",
    reportNumber = "HUTP-97-A097, HUTP-98-A097",
    doi = "10.4310/ATMP.1998.v2.n2.a1",
    journal = "Adv. Theor. Math. Phys.",
    volume = "2",
    pages = "231--252",
    year = "1998"
}

@article{Giribet:2018ada,
    author = "Giribet, G. and Hull, C. and Kleban, M. and Porrati, M. and Rabinovici, E.",
    title = "{Superstrings on AdS$_{3}$ at $\mathcal{k} =$ 1}",
    eprint = "1803.04420",
    archivePrefix = "arXiv",
    primaryClass = "hep-th",
    reportNumber = "Imperial-TP-2018-CH-01, IMPERIAL-TP-2018-CH-01",
    doi = "10.1007/JHEP08(2018)204",
    journal = "JHEP",
    volume = "08",
    pages = "204",
    year = "2018"
}

@article{Gaberdiel:2018rqv,
    author = "Gaberdiel, Matthias R. and Gopakumar, Rajesh",
    title = "{Tensionless string spectra on AdS$_{3}$}",
    eprint = "1803.04423",
    archivePrefix = "arXiv",
    primaryClass = "hep-th",
    doi = "10.1007/JHEP05(2018)085",
    journal = "JHEP",
    volume = "05",
    pages = "085",
    year = "2018"
}

@article{Eberhardt:2018ouy,
    author = "Eberhardt, Lorenz and Gaberdiel, Matthias R. and Gopakumar, Rajesh",
    title = "{The Worldsheet Dual of the Symmetric Product CFT}",
    eprint = "1812.01007",
    archivePrefix = "arXiv",
    primaryClass = "hep-th",
    doi = "10.1007/JHEP04(2019)103",
    journal = "JHEP",
    volume = "04",
    pages = "103",
    year = "2019"
}

@article{OhlssonSax:2018hgc,
    author = "Ohlsson Sax, Olof and Stefa\'nski, Bogdan",
    title = "{Closed strings and moduli in AdS$_{3}$/CFT$_{2}$}",
    eprint = "1804.02023",
    archivePrefix = "arXiv",
    primaryClass = "hep-th",
    reportNumber = "NORDITA 2018-027, NORDITA-2018-027",
    doi = "10.1007/JHEP05(2018)101",
    journal = "JHEP",
    volume = "05",
    pages = "101",
    year = "2018"
}

@article{Zamolodchikov:1978xm,
    author = "Zamolodchikov, Alexander B. and Zamolodchikov, Alexei B.",
    editor = "Khalatnikov, I. M. and Mineev, V. P.",
    title = "{Factorized s Matrices in Two-Dimensions as the Exact Solutions of Certain Relativistic Quantum Field Models}",
    reportNumber = "ITEP-35-1978",
    doi = "10.1016/0003-4916(79)90391-9",
    journal = "Annals Phys.",
    volume = "120",
    pages = "253--291",
    year = "1979"
}

@article{Minahan:2002ve,
    author = "Minahan, J. A. and Zarembo, K.",
    title = "{The Bethe ansatz for N=4 superYang-Mills}",
    eprint = "hep-th/0212208",
    archivePrefix = "arXiv",
    reportNumber = "UUITP-17-02, ITEP-TH-73-02",
    doi = "10.1088/1126-6708/2003/03/013",
    journal = "JHEP",
    volume = "03",
    pages = "013",
    year = "2003"
}

@article{Demulder:2023bux,
    author = "Demulder, Saskia and Driezen, Sibylle and Knighton, Bob and Oling, Gerben and Retore, Ana L. and Seibold, Fiona K. and Sfondrini, Alessandro and Yan, Ziqi",
    title = "{Exact approaches on the string worldsheet}",
    eprint = "2312.12930",
    archivePrefix = "arXiv",
    primaryClass = "hep-th",
    reportNumber = "NORDITA 2023-083",
    doi = "10.1088/1751-8121/ad72be",
    journal = "J. Phys. A",
    volume = "57",
    number = "42",
    pages = "423001",
    year = "2024"
}

@article{Seibold:2024qkh,
    author = "Seibold, Fiona K. and Sfondrini, Alessandro",
    title = "{AdS3 Integrability, Tensionless Limits, and Deformations: A Review}",
    eprint = "2408.08414",
    archivePrefix = "arXiv",
    primaryClass = "hep-th",
    month = "8",
    year = "2024"
}

@article{Cagnazzo:2012se,
    author = "Cagnazzo, A. and Zarembo, K.",
    title = "{B-field in AdS(3)/CFT(2) Correspondence and Integrability}",
    eprint = "1209.4049",
    archivePrefix = "arXiv",
    primaryClass = "hep-th",
    reportNumber = "NORDITA-2012-67, UUITP-24-12",
    doi = "10.1007/JHEP11(2012)133",
    journal = "JHEP",
    volume = "11",
    pages = "133",
    year = "2012",
    note = "[Erratum: JHEP 04, 003 (2013)]"
}

@article{Beisert:2010jr,
    author = "Beisert, Niklas and others",
    title = "{Review of AdS/CFT Integrability: An Overview}",
    eprint = "1012.3982",
    archivePrefix = "arXiv",
    primaryClass = "hep-th",
    reportNumber = "AEI-2010-175, CERN-PH-TH-2010-306, HU-EP-10-87, HU-MATH-2010-22, KCL-MTH-10-10, UMTG-270, UUITP-41-10",
    doi = "10.1007/s11005-011-0529-2",
    journal = "Lett. Math. Phys.",
    volume = "99",
    pages = "3--32",
    year = "2012"
}

@article{Arutyunov:2009ga,
    author = "Arutyunov, Gleb and Frolov, Sergey",
    title = "{Foundations of the AdS$_{5} \times S^{5}$ Superstring. Part I}",
    eprint = "0901.4937",
    archivePrefix = "arXiv",
    primaryClass = "hep-th",
    reportNumber = "ITP-UU-09-05, SPIN-09-05, TCD-MATH-09-06, HMI-09-03",
    doi = "10.1088/1751-8113/42/25/254003",
    journal = "J. Phys. A",
    volume = "42",
    pages = "254003",
    year = "2009"
}

@article{David:2008yk,
    author = "David, Justin R. and Sahoo, Bindusar",
    title = "{Giant magnons in the D1-D5 system}",
    eprint = "0804.3267",
    archivePrefix = "arXiv",
    primaryClass = "hep-th",
    doi = "10.1088/1126-6708/2008/07/033",
    journal = "JHEP",
    volume = "07",
    pages = "033",
    year = "2008"
}

@article{Arutyunov:1997gt,
    author = "Arutyunov, G. E. and Frolov, S. A.",
    title = "{Virasoro amplitude from the S**N R**24 orbifold sigma model}",
    eprint = "hep-th/9708129",
    archivePrefix = "arXiv",
    reportNumber = "LMU-TPW-97-21",
    doi = "10.1007/BF02557107",
    journal = "Theor. Math. Phys.",
    volume = "114",
    pages = "43--66",
    year = "1998"
}

@article{Dixon:1985jw,
    author = "Dixon, Lance J. and Harvey, Jeffrey A. and Vafa, C. and Witten, Edward",
    editor = "Schellekens, B.",
    title = "{Strings on Orbifolds}",
    reportNumber = "PRINT-85-0616 (PRINCETON)",
    doi = "10.1016/0550-3213(85)90593-0",
    journal = "Nucl. Phys. B",
    volume = "261",
    pages = "678--686",
    year = "1985"
}

@article{Dixon:1986jc,
    author = "Dixon, Lance J. and Harvey, Jeffrey A. and Vafa, C. and Witten, Edward",
    title = "{Strings on Orbifolds. 2.}",
    reportNumber = "PRINT-86-0246 (PRINCETON)",
    doi = "10.1016/0550-3213(86)90287-7",
    journal = "Nucl. Phys. B",
    volume = "274",
    pages = "285--314",
    year = "1986"
}

@article{Dixon:1986qv,
    author = "Dixon, Lance J. and Friedan, Daniel and Martinec, Emil J. and Shenker, Stephen H.",
    title = "{The Conformal Field Theory of Orbifolds}",
    reportNumber = "EFI-86-42-CHICAGO",
    doi = "10.1016/0550-3213(87)90676-6",
    journal = "Nucl. Phys. B",
    volume = "282",
    pages = "13--73",
    year = "1987"
}

@article{Lunin:2000yv,
    author = "Lunin, Oleg and Mathur, Samir D.",
    title = "{Correlation functions for M**N / S(N) orbifolds}",
    eprint = "hep-th/0006196",
    archivePrefix = "arXiv",
    reportNumber = "OHSTPY-HEP-T-00-010",
    doi = "10.1007/s002200100431",
    journal = "Commun. Math. Phys.",
    volume = "219",
    pages = "399--442",
    year = "2001"
}

@article{Frolov:2025tda,
    author = "Frolov, Sergey and Polvara, Davide and Sfondrini, Alessandro",
    title = "{Dressing Factors and Mirror Thermodynamic Bethe Ansatz for mixed-flux AdS3/CFT2}",
    eprint = "2507.12191",
    archivePrefix = "arXiv",
    primaryClass = "hep-th",
    reportNumber = "ZMP-HH/25-12",
    month = "7",
    year = "2025"
}

@article{Brollo:2023rgp,
    author = "Brollo, Alberto and le Plat, Dennis and Sfondrini, Alessandro and Suzuki, Ryo",
    title = "{More on the tensionless limit of pure-Ramond-Ramond AdS3/CFT2}",
    eprint = "2308.11576",
    archivePrefix = "arXiv",
    primaryClass = "hep-th",
    doi = "10.1007/JHEP12(2023)160",
    journal = "JHEP",
    volume = "12",
    pages = "160",
    year = "2023"
}

@article{Cavaglia:2022xld,
    author = "Cavagli{\`a}, Andrea and Ekhammar, Simon and Gromov, Nikolay and Ryan, Paul",
    title = "{Exploring the Quantum Spectral Curve for AdS$_{3}$/CFT$_{2}$}",
    eprint = "2211.07810",
    archivePrefix = "arXiv",
    primaryClass = "hep-th",
    doi = "10.1007/JHEP12(2023)089",
    journal = "JHEP",
    volume = "12",
    pages = "089",
    year = "2023"
}

@article{Eberhardt:2019ywk,
    author = "Eberhardt, Lorenz and Gaberdiel, Matthias R. and Gopakumar, Rajesh",
    title = "{Deriving the AdS$_{3}$/CFT$_{2}$ correspondence}",
    eprint = "1911.00378",
    archivePrefix = "arXiv",
    primaryClass = "hep-th",
    doi = "10.1007/JHEP02(2020)136",
    journal = "JHEP",
    volume = "02",
    pages = "136",
    year = "2020"
}

@article{Gromov:2013pga,
    author = "Gromov, Nikolay and Kazakov, Vladimir and Leurent, Sebastien and Volin, Dmytro",
    title = "{Quantum Spectral Curve for Planar $\mathcal{N} = 4$ Super-Yang-Mills Theory}",
    eprint = "1305.1939",
    archivePrefix = "arXiv",
    primaryClass = "hep-th",
    reportNumber = "IMPERIAL-TP-13-SL-02",
    doi = "10.1103/PhysRevLett.112.011602",
    journal = "Phys. Rev. Lett.",
    volume = "112",
    number = "1",
    pages = "011602",
    year = "2014"
}

@article{Frolov:2010wt,
    author = "Frolov, Sergey",
    title = "{Konishi operator at intermediate coupling}",
    eprint = "1006.5032",
    archivePrefix = "arXiv",
    primaryClass = "hep-th",
    reportNumber = "TCDMATH-10-05, HMI-10-03",
    doi = "10.1088/1751-8113/44/6/065401",
    journal = "J. Phys. A",
    volume = "44",
    pages = "065401",
    year = "2011"
}

@article{Arutyunov:2010gb,
    author = "Arutyunov, Gleb and Frolov, Sergey and Suzuki, Ryo",
    title = "{Five-loop Konishi from the Mirror TBA}",
    eprint = "1002.1711",
    archivePrefix = "arXiv",
    primaryClass = "hep-th",
    reportNumber = "ITP-UU-10-04, SPIN-10-04, TCDMATH-10-01, HMI-10-01",
    doi = "10.1007/JHEP04(2010)069",
    journal = "JHEP",
    volume = "04",
    pages = "069",
    year = "2010"
}

@article{Balog:2010xa,
    author = "Balog, Janos and Hegedus, Arpad",
    title = "{5-loop Konishi from linearized TBA and the XXX magnet}",
    eprint = "1002.4142",
    archivePrefix = "arXiv",
    primaryClass = "hep-th",
    doi = "10.1007/JHEP06(2010)080",
    journal = "JHEP",
    volume = "06",
    pages = "080",
    year = "2010"
}

@article{Lukowski:2009ce,
    author = "Lukowski, T. and Rej, A. and Velizhanin, V. N.",
    title = "{Five-Loop Anomalous Dimension of Twist-Two Operators}",
    eprint = "0912.1624",
    archivePrefix = "arXiv",
    primaryClass = "hep-th",
    reportNumber = "IMPERIAL-TP-AR-2009-4",
    doi = "10.1016/j.nuclphysb.2010.01.008",
    journal = "Nucl. Phys. B",
    volume = "831",
    pages = "105--132",
    year = "2010"
}

@article{Babichenko:2009dk,
    author = "Babichenko, A. and Stefanski, Jr., B. and Zarembo, K.",
    title = "{Integrability and the AdS(3)/CFT(2) correspondence}",
    eprint = "0912.1723",
    archivePrefix = "arXiv",
    primaryClass = "hep-th",
    reportNumber = "ITEP-TH-59-09, LPTENS-09-36, UUITP-25-09",
    doi = "10.1007/JHEP03(2010)058",
    journal = "JHEP",
    volume = "03",
    pages = "058",
    year = "2010"
}

@article{Gromov:2009zb,
    author = "Gromov, Nikolay and Kazakov, Vladimir and Vieira, Pedro",
    title = "{Exact Spectrum of Planar ${\cal N}=4$ Supersymmetric Yang-Mills Theory: Konishi Dimension at Any Coupling}",
    eprint = "0906.4240",
    archivePrefix = "arXiv",
    primaryClass = "hep-th",
    reportNumber = "DESY-09-210",
    doi = "10.1103/PhysRevLett.104.211601",
    journal = "Phys. Rev. Lett.",
    volume = "104",
    pages = "211601",
    year = "2010"
}

@article{Berkovits:1999im,
    author = "Berkovits, Nathan and Vafa, Cumrun and Witten, Edward",
    title = "{Conformal field theory of AdS background with Ramond-Ramond flux}",
    eprint = "hep-th/9902098",
    archivePrefix = "arXiv",
    reportNumber = "IFT-P-012-99, HUTP-99-A004, IASSNS-HEP-99-5",
    doi = "10.1088/1126-6708/1999/03/018",
    journal = "JHEP",
    volume = "03",
    pages = "018",
    year = "1999"
}

@article{Lunin:2001pw,
    author = "Lunin, Oleg and Mathur, Samir D.",
    title = "{Three point functions for M(N) / S(N) orbifolds with N=4 supersymmetry}",
    eprint = "hep-th/0103169",
    archivePrefix = "arXiv",
    reportNumber = "OHSTPY-HEP-T-01-005",
    doi = "10.1007/s002200200638",
    journal = "Commun. Math. Phys.",
    volume = "227",
    pages = "385--419",
    year = "2002"
}

@article{Beisert:2005tm,
    author = "Beisert, Niklas",
    title = "{The SU(2|2) dynamic S-matrix}",
    eprint = "hep-th/0511082",
    archivePrefix = "arXiv",
    reportNumber = "PUTP-2181, NSF-KITP-05-92",
    doi = "10.4310/ATMP.2008.v12.n5.a1",
    journal = "Adv. Theor. Math. Phys.",
    volume = "12",
    pages = "945--979",
    year = "2008"
}

@article{Arutyunov:2006ak,
    author = "Arutyunov, Gleb and Frolov, Sergey and Plefka, Jan and Zamaklar, Marija",
    title = "{The Off-shell Symmetry Algebra of the Light-cone AdS(5) x S**5 Superstring}",
    eprint = "hep-th/0609157",
    archivePrefix = "arXiv",
    reportNumber = "AEI-2006-071, HU-EP-06-31, ITP-UU-06-39, SPIN-06-33, TCDMATH-06-13",
    doi = "10.1088/1751-8113/40/13/018",
    journal = "J. Phys. A",
    volume = "40",
    pages = "3583--3606",
    year = "2007"
}

@article{Hoare:2013lja,
    author = "Hoare, B. and Stepanchuk, A. and Tseytlin, A. A.",
    title = "{Giant magnon solution and dispersion relation in string theory in $AdS_3$x$S^3$x$T^4$ with mixed flux}",
    eprint = "1311.1794",
    archivePrefix = "arXiv",
    primaryClass = "hep-th",
    reportNumber = "IMPERIAL-TP-AS-2013-01, HU-EP-13-56",
    doi = "10.1016/j.nuclphysb.2013.12.011",
    journal = "Nucl. Phys. B",
    volume = "879",
    pages = "318--347",
    year = "2014"
}

@article{Borsato:2012ud,
    author = "Borsato, Riccardo and Ohlsson Sax, Olof and Sfondrini, Alessandro",
    title = "{A dynamic $\mathfrak{su}$(1|1)$^2$ S-matrix for AdS$_3$/CFT$_2$}",
    eprint = "1211.5119",
    archivePrefix = "arXiv",
    primaryClass = "hep-th",
    reportNumber = "ITP-UU-12-46, SPIN-12-43",
    doi = "10.1007/JHEP04(2013)113",
    journal = "JHEP",
    volume = "04",
    pages = "113",
    year = "2013"
}

@article{Dei:2018mfl,
    author = "Dei, Andrea and Sfondrini, Alessandro",
    title = "{Integrable spin chain for stringy Wess-Zumino-Witten models}",
    eprint = "1806.00422",
    archivePrefix = "arXiv",
    primaryClass = "hep-th",
    doi = "10.1007/JHEP07(2018)109",
    journal = "JHEP",
    volume = "07",
    pages = "109",
    year = "2018"
}

@article{Ekhammar:2021pys,
    author = "Ekhammar, Simon and Volin, Dmytro",
    title = "{Monodromy bootstrap for SU(2|2) quantum spectral curves: from Hubbard model to AdS$_{3}$/CFT$_{2}$}",
    eprint = "2109.06164",
    archivePrefix = "arXiv",
    primaryClass = "math-ph",
    reportNumber = "UUITP-44/21, NORDITA 2021-090",
    doi = "10.1007/JHEP03(2022)192",
    journal = "JHEP",
    volume = "03",
    pages = "192",
    year = "2022"
}

@article{Cavaglia:2021eqr,
    author = "Cavagli\`a, Andrea and Gromov, Nikolay and Stefa\'nski, Jr., Bogdan and Jr. and Torrielli, Alessandro",
    title = "{Quantum Spectral Curve for AdS$_{3}$/CFT$_{2}$: a proposal}",
    eprint = "2109.05500",
    archivePrefix = "arXiv",
    primaryClass = "hep-th",
    reportNumber = "DMUS-MP/21-14, DMUS-MP-21/14",
    doi = "10.1007/JHEP12(2021)048",
    journal = "JHEP",
    volume = "12",
    pages = "048",
    year = "2021"
}

@article{Frolov:2021bwp,
    author = "Frolov, Sergey and Sfondrini, Alessandro",
    title = "{Mirror thermodynamic Bethe ansatz for AdS3/CFT2}",
    eprint = "2112.08898",
    archivePrefix = "arXiv",
    primaryClass = "hep-th",
    doi = "10.1007/JHEP03(2022)138",
    journal = "JHEP",
    volume = "03",
    pages = "138",
    year = "2022"
}

@article{Frolov:2024pkz,
    author = "Frolov, Sergey and Polvara, Davide and Sfondrini, Alessandro",
    title = "{Dressing factors for mixed-flux AdS3\texttimes{}S3\texttimes{}T4 superstrings}",
    eprint = "2402.11732",
    archivePrefix = "arXiv",
    primaryClass = "hep-th",
    doi = "10.1103/PhysRevD.111.L081901",
    journal = "Phys. Rev. D",
    volume = "111",
    number = "8",
    pages = "L081901",
    year = "2025"
}

@article{Frolov:2025uwz,
    author = "Frolov, Sergey and Polvara, Davide and Sfondrini, Alessandro",
    title = "{Massive dressing factors for mixed-flux AdS$_{3}$/CFT$_{2}$}",
    eprint = "2501.05995",
    archivePrefix = "arXiv",
    primaryClass = "hep-th",
    reportNumber = "ZMP-HH/25-1",
    doi = "10.1007/JHEP07(2025)171",
    journal = "JHEP",
    volume = "07",
    pages = "171",
    year = "2025"
}

@article{Borsato:2014hja,
    author = "Borsato, Riccardo and Ohlsson Sax, Olof and Sfondrini, Alessandro and Stefanski, Bogdan",
    title = "{The complete AdS$_{3} \times$ S$^3 \times$ T$^4$ worldsheet S matrix}",
    eprint = "1406.0453",
    archivePrefix = "arXiv",
    primaryClass = "hep-th",
    reportNumber = "IMPERIAL-TP-OOS-2014-03, HU-MATHEMATIK-2014-11, HU-EP-14-19, SPIN-14-15, ITP-UU-14-17",
    doi = "10.1007/JHEP10(2014)066",
    journal = "JHEP",
    volume = "10",
    pages = "066",
    year = "2014"
}

@article{Baggio:2018gct,
    author = "Baggio, Marco and Sfondrini, Alessandro",
    title = "{Strings on NS-NS Backgrounds as Integrable Deformations}",
    eprint = "1804.01998",
    archivePrefix = "arXiv",
    primaryClass = "hep-th",
    doi = "10.1103/PhysRevD.98.021902",
    journal = "Phys. Rev. D",
    volume = "98",
    number = "2",
    pages = "021902",
    year = "2018"
}

@article{Frolov:2021fmj,
    author = "Frolov, Sergey and Sfondrini, Alessandro",
    title = "{New dressing factors for AdS3/CFT2}",
    eprint = "2112.08896",
    archivePrefix = "arXiv",
    primaryClass = "hep-th",
    doi = "10.1007/JHEP04(2022)162",
    journal = "JHEP",
    volume = "04",
    pages = "162",
    year = "2022"
}

@article{Burrington:2012yq,
    author = "Burrington, Benjamin A. and Peet, Amanda W. and Zadeh, Ida G.",
    title = "{Operator mixing for string states in the D1-D5 CFT near the orbifold point}",
    eprint = "1211.6699",
    archivePrefix = "arXiv",
    primaryClass = "hep-th",
    doi = "10.1103/PhysRevD.87.106001",
    journal = "Phys. Rev. D",
    volume = "87",
    number = "10",
    pages = "106001",
    year = "2013"
}

@article{Burrington:2012yn,
    author = "Burrington, Benjamin A. and Peet, Amanda W. and Zadeh, Ida G.",
    title = "{Twist-nontwist correlators in $M^N/S_N$ orbifold CFTs}",
    eprint = "1211.6689",
    archivePrefix = "arXiv",
    primaryClass = "hep-th",
    doi = "10.1103/PhysRevD.87.106008",
    journal = "Phys. Rev. D",
    volume = "87",
    number = "10",
    pages = "106008",
    year = "2013"
}

\end{document}